\documentclass[aps,twocolumn,eprint,pre,superscriptaddress,nofootinbib]{revtex4-2}
\usepackage{setspace}

\usepackage[utf8]{inputenc}
\usepackage{ dsfont }
\usepackage{hyperref}
\usepackage{amsmath,amssymb,color,enumerate}
\usepackage[margin=1in]{geometry}

\usepackage{physics}
\usepackage{tensor}
\usepackage{graphicx}
\usepackage{breqn}
\usepackage{ amssymb }
\usepackage{setspace}
\usepackage{comment} % to comment
\usepackage{xspace} % for additional spaces

\usepackage[most]{tcolorbox}

\usepackage{etoolbox}   % For conditional checks
\usepackage{xcolor}     % For color customization
\usepackage{natbib}     % For standard citation formatting

\newcommand{\emptycitecolor}{red}

\newcommand{\Poincare}{Poincar\'e\xspace}

\makeatletter
\renewcommand{\cite}[1]{%
  \ifstrempty{#1}{%
    \textcolor{\emptycitecolor}{[?]}% If \cite{} is empty, display [?] in red
  }{%
    \citep{#1}% If \cite has content, process it normally with \citep
  }%
}
\makeatother
\makeatletter
\patchcmd{\normalsize}{10pt}{14pt}{}{}
\patchcmd{\large}{12pt}{16pt}{}{}
\patchcmd{\footnotesize}{8pt}{12pt}{}{}

\makeatother

\makeatletter
\let\cat@comma@active\@empty
\makeatother

\begin{document}

\title{Transition to chaos with conical billiards}
\author{Lara Braverman}
\affiliation{John A. Paulson School of Engineering and Applied Sciences, Harvard University, Cambridge, MA 02138, USA}
\author{David R. Nelson}
\affiliation{John A. Paulson School of Engineering and Applied Sciences, Harvard University, Cambridge, MA 02138, USA}
\affiliation{Department of Physics, Harvard University, Cambridge, MA 02138, USA}
\date{\today}

\begin{abstract}

We adapt ideas from geometrical optics and classical billiard dynamics to consider particle trajectories with constant velocity on a cone with specular reflections off an elliptical boundary formed by the intersection with a tilted plane, with tilt angle $\gamma$.  We explore the dynamics as a function of $\gamma$ and the cone deficit angle $\chi$  that controls the sharpness of the apex, where a point source of positive Gaussian curvature is concentrated.   We find  regions of the ($\gamma, \chi$) plane where, depending on the initial conditions, either (A) the trajectories  sample the entire cone base and avoid the apex region; (B)  sample only a portion of the base region while again avoiding the apex; or (C) sample the entire cone surface much more uniformly, suggestive of ergodicity. The special case of an untilted cone displays only type A trajectories which form a ring caustic at the distance of closest approach to the apex. However, we observe an intricate transition to chaotic dynamics dominated by Type (C) trajectories for sufficiently large $\chi$ and $\gamma$.  A \Poincare map that summarizes trajectories decomposed into the geodesic segments interrupted by specular reflections provides a powerful method for visualizing the transition to chaos.   We then analyze the similarities and differences of the path to chaos for conical billiards with other area-preserving conservative maps.

\end{abstract}
\maketitle
\section{Introduction}

The ergodic hypothesis, which lies at the core of much of statistical mechanics, states that time averages of one trajectory from a single initial condition and phase space averages or equivalently averages over many initial conditions are the same~\cite{sethna2021statistical}. This implies that a single trajectory, in say, an ideal gas, is expected to reach almost all points in an appropriately defined phase space. In addition to ergodicity, related features such as chaos--where trajectories from nearby initial conditions diverge exponentially fast; and strongly mixing--where trajectories uniformly sample almost all of phase space ~\cite{adams2023ergodicity}), see Sec.\ref{sec:chaos}, often justify the use of statistical mechanics to replace time averages with phase space averages for systems in thermodynamic equilibrium~\cite{sethna2021statistical}.

However, many physical systems are neither ergodic nor chaotic, and instead exhibit alternative complex behavior. For systems for which ideas from equilibrium statistical mechanics do not apply, the emergent order can be sensitive to boundary conditions and to the geometric and topological properties of the environment~\cite{wioland2013confinement,ben2022disordered,keber2014topology,sun2023geometric}. For example, quenched random fluctuations of the boundary can destroy the order that is constructed through spontaneous motility-induced phase separation of active Brownian particles~\cite{ben2022disordered}. Similarly, the curvature of a surface and activity can influence biological processes, physical organization, and behavior of particles confined to the surface~\cite{bausch2003grain,keber2014topology,sun2023geometric}. Recent investigations have shown that even curvature at a distance, such as when Gaussian curvature is concentrated at the apex of a cone, can attract topological defects of like sign and repel defects of opposite sign~\cite{turner2010vortices,zhang2022fractional,vafa2025defect}. The concentrated curvature contributes to a breakdown of ideas from equilibrium states in the presence of activity~\cite{vafa2024periodic}. 

In this paper, we explore the interplay of curvature and boundary conditions in mathematical billiards, a class of dynamical systems where a particle travels along a geodesic on a surface and reflects specularly off its boundaries, as if it were a ball on a pool table or a light ray interacting with a reflecting boundary. Similar to the physical systems described above, mathematical billiards have been shown to be sensitive to both  Gaussian curvature and  boundary conditions ~\cite{kourganoff2016uniform,tyc2022spherical,carmo2024mixing,bunimovich1979ergodic,wojtkowski2020principles,Tabachnikov1995,lynch2019integrable,stachel2022motion,stachel2021geometry,koiller1996static,tabanov1994separatrices,dietz2022intermediate,lenz2007classical,lenz2009evolutionary,stone2010chaotic,koiller1996static}. Early studies of mathematical billiards suggested that truly ergodic and mixing billiard trajectories are possible only on manifolds with either negative Gaussian curvature~\cite{Tabachnikov1995} or concave boundaries~\cite{lazutkin1973existence,wojtkowski2020principles}, such as in the classic Sinai problem~\cite{sinai1970dynamical,wojtkowski2020principles}. However, later work demonstrated that non-smoothly defined convex boundaries, like those in the Bunimovich stadium (billiards on a rectangle capped by semicircles on the opposite ends) can also produce ergodic and mixing trajectories~\cite{bunimovich1979ergodic,wojtkowski2020principles}. In fact, recent forays into  dynamical systems theory have been able to identify analytically tractable features of billiards on polygons~\cite{mcmullen2023billiards}. This work has been extended into three dimensions for non-smooth closed surfaces embedded in $\mathbb{R}^3$~\cite{tyc2022spherical}.

Billiard problems, beyond being a model system for the interplay between curvature and boundary conditions, have broad implications throughout physics and have recently been extended to the field of soft and active matter~\cite{spagnolie2017microorganism,dutta2020system,albers2024billiards,badeau2024statistical}. One recent study models micro-swimmers as billiards near boundaries as their tumbling is suppressed due to hydrodynamic interactions~\cite{spagnolie2017microorganism}. Similarly, billiards with spatial memory have been used to explain complex patterns observed in single particle systems~\cite{albers2024billiards}. Outside of soft matter, chaotic dynamical billiards have been used in developing statistical wave field theory~\cite{badeau2024statistical} and as a physical basis for information theory~\cite{dutta2020system}. 

\begin{figure}
    \centering
    \includegraphics[width = \columnwidth]{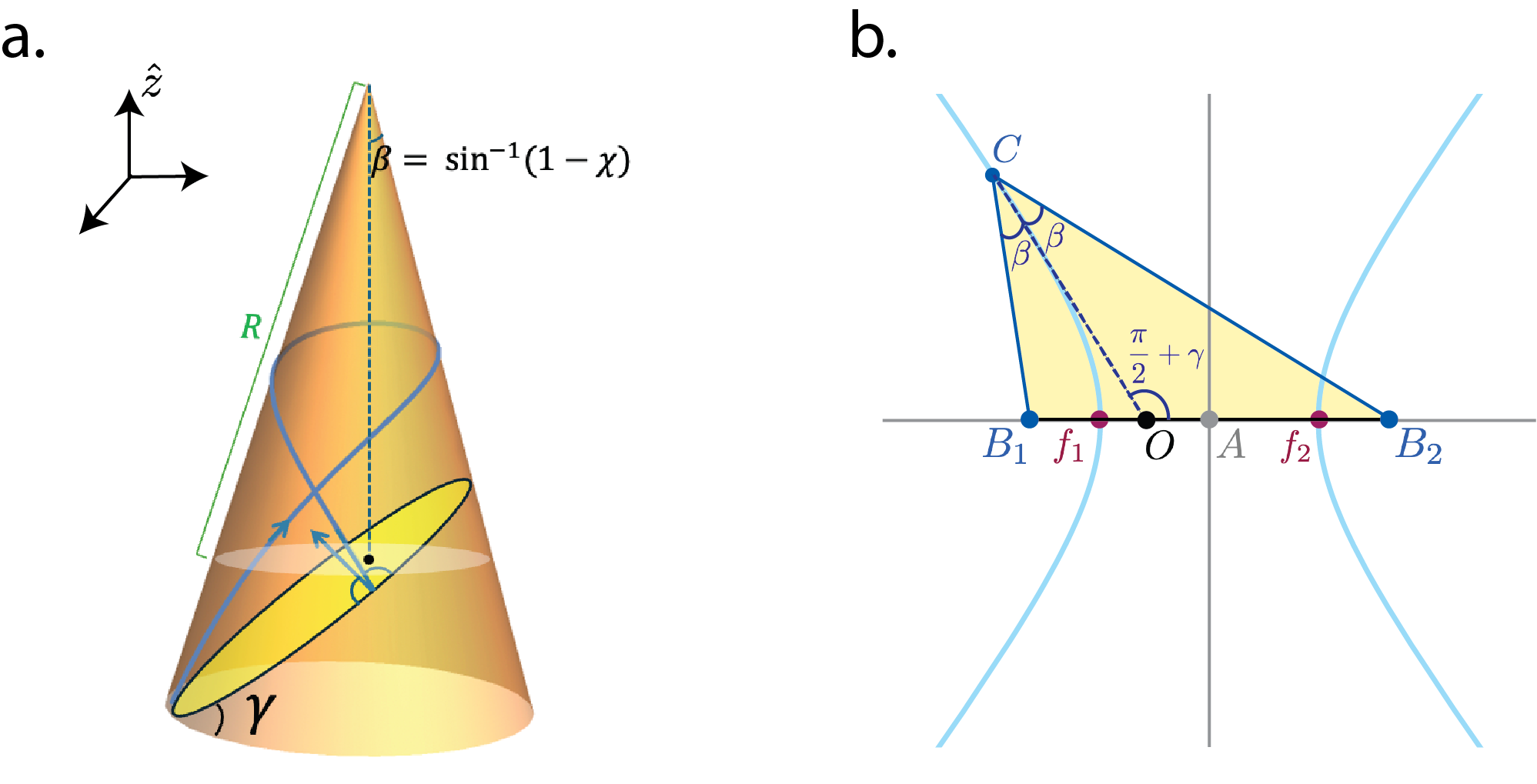}
    \caption{ (a) Schematic set up for the billiard problem. A cone and its reflecting boundary condition at the base are defined by two variables $\chi$ and $\gamma$ where $\beta = \sin^{-1}(1-\chi)$ is the cone half-angle and $\gamma$ is the angle the plane that cuts through the cone to form an elliptical base makes with a horizontal. The deficit angle associated with Gaussian curvature at the apex is $2\pi\chi$. The distance $R$, which is kept the same for all simulations of the paper, is the flank distance corresponding to an untilted cone whose base intersects the base of the tilted cone at $x=0$ and $y=0$. Here the coordinate axes are defined so that the apex is at $(0,0,0)$ and the height of the cone is parallel to the $z$-axis. The intersection at  $x=0$ and $y=0$, between the bases of the tilted and untilted cones and the $z-axis$, marked with a black dot, is \emph{not} one of the foci of the tilted cone base ellipse. (b) In fact, if any cone constructed in the manner described in (a) is tilted so that the major axis of the elliptical base is placed on the horizontal axis with the center of the major axis placed at (0,0), then the cross section, through the major axis and the apex of the cone, will look like the yellow region on this panel. The apex of the cone (C), will lie on a hyperbola (in light blue) whose focal points are the ends of the major axis of the ellipse ($B_1$ and $B_2$) and whose vertices lie on the focal points of the ellipse ($f_1$ and $f_2$). Note that the point $O$ corresponds to the black point in panel (a) and is not a focal point of the ellipse. Trigonometric manipulations show that $O$ lies in a line with endpoints $f_1$,and $A$. $O$ will only coincide with $f_1$ in the extreme case when $C = f_1$ and with $A$ in the extreme case when $f_1 =A$ (billiards on an untilted cone).  }
    \label{fig:setupFirst}
\end{figure}
Moreover, billiards on planar conic sections, specifically ellipses and their perturbations, have been extensively studied~\cite{Tabachnikov1995,lynch2019integrable,stachel2022motion,stachel2021geometry,koiller1996static,tabanov1994separatrices,dietz2022intermediate}. Well-studied problems such as billiards on ellipses have been used to understand more broadly prevalent and experimentally realizable physical systems such as those with particles interacting in-elastically with a boundary~\cite{lynch2019integrable,kroetz2016dynamical,nagler2007leaking} or behaving like quantum particles~\cite{kudrolli1994signatures,waalkens1997elliptic,blomquist2002geometry}. In addition, rich billiard dynamics have also been observed on surfaces with conic section boundaries that are time-dependent, driven, or involve energy loss during collisions~\cite{lenz2007classical,lenz2009evolutionary,stone2010chaotic,koiller1996static}. While billiards on ellipses are integrable~\cite{Tabachnikov1995,lynch2019integrable,stachel2022motion,stachel2021geometry}, this integrability is not robust to boundary deformations~\cite{koiller1996static,tabanov1994separatrices}. Yet, properties such as ergodicity and mixing are not guaranteed, and often require very large boundary deformations~\cite{koiller1996static}. In fact, the non-ergodicity of elliptical billiards has been shown to be more robust to perturbations than billiards on a disk. For example, while the aforementioned Bunimovich stadium model, a perturbation of billiards on a disk, is ergodic and mixing, a similar perturbation of an ellipse with large eccentricity is neither ergodic nor mixing~\cite{koiller1996static}.

Motivated by these recent discoveries, in this work we present an alternative perturbation to classic billiard systems (billiards confined to a circle or an ellipse) via a simple escape into a third dimension.  Our model considers billiards on the surface of a tilted cone, defined by its tilt angle and by the deficit angle $2 \pi \chi$ at the apex, where $\chi = 1-\sin(\beta)$ is the angular fraction of a flat disk removed to make the cone and $\beta$ is the half-angle of the cone. The \emph{boundary} of the cone is defined by an ellipse i.e., the conic section of the cone produced by an intersection with a plane at an angle $\gamma$ (see Fig.~\ref{fig:setupFirst}). By combining a cone apex singularity (which is somewhat like the corner in a polygon) and a curved boundary, conical  billiards provide an intriguing mix of features of both polygonal billiards and billiards with boundaries that are simple planar conic sections~\cite{mcmullen2023billiards}. Despite the surface possessing non-negative curvature and a static smooth convex (in three dimensions from the perspective of an observer outside the surface, but concave from the perspective of a light ray bouncing inside the surface) boundary, we present evidence here that the combination of a delta function of positive Gaussian curvature and an asymmetric boundary can be sufficient for ergodic, mixing, and chaotic orbits.

\begin{figure*}[t!]
    \centering
    \includegraphics[width = \linewidth]{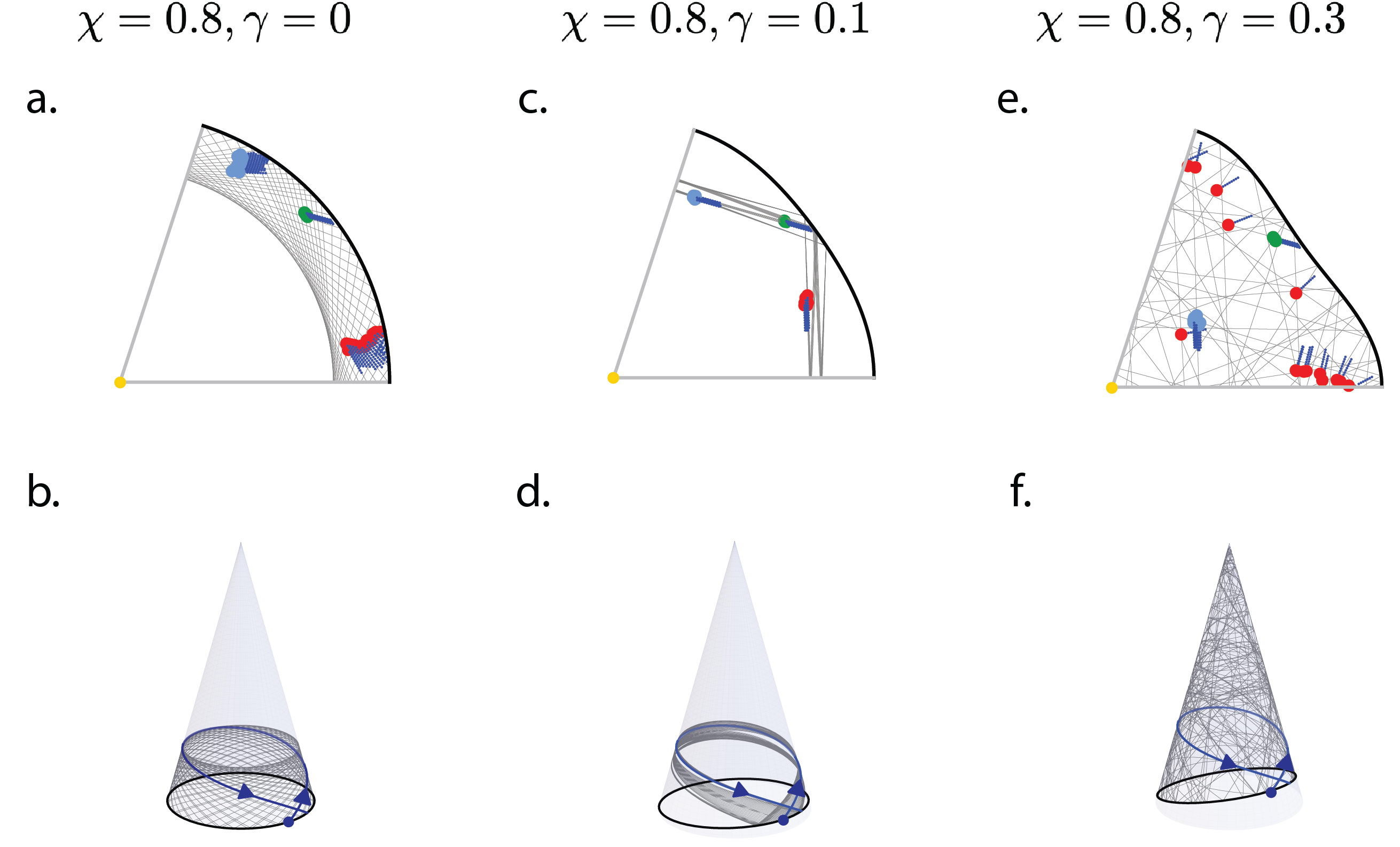}
        \caption{Representation of paths from a particular initial condition at the base for cones with $\chi = 0.8$, and increasing $\gamma = 0 (a, b), 0.1 (c, d),$ and $0.3 (e, f)$. (a) Unwrapped untilted cone with $\chi = 0.8$, and $\gamma = 0$. The black boundary line marks the base of the cone. The unwrapping is done in all cases by cutting a tilted cone along the longest geodesic running from the base to the apex, and then unrolling. The two light gray edges are the boundary created by the cut. When trajectories hit a gray edge of an unrolled cone, they reappear at the other gray edge at an identical distance from the apex (yellow dot) but rotated by $2\pi \chi$. Large green, blue, and red dots mark the positions of particles beginning from nearby initial conditions along the base at 10, 510, and 1010 time steps respectively, where we assume particles moving with unit velocity. The "stems" formed by the small dark blue dots show the position of the path at time steps (0-10), (500-510), and (1000-1010).  The thin dark gray line is the full path for one of the initial conditions for approximately the first 1200 time steps or 30 bounces off the cone base. Note the accumulation of segments of the trajectory creates a caustic boundary when $\gamma =0$. (b) Representation of the gray path from (a) in three dimensions. The heavy black line is the base of the cone, and the blue line is the path between the initial condition (dark blue dot) and the first intersection with the boundary. (c,d) Same figures as in panels (a) and (b) respectively, but for $\gamma = 0.1$. (e,f) Same figures as in panels (a) and (b) respectively, but in a more chaotic regime reached when $\gamma = 0.3$. Note how after 1010 time steps paths which begin at nearby initial conditions diverge in this case. } 
    \label{fig:intro}
\end{figure*}

Individually, neither breaking azimuthal symmetry by deforming a circular billiard table into an ellipse, nor adding curvature, via erecting a cone apex over a circular base with a delta function of Gaussian curvature, are sufficient to produce ergodicity or chaotic dynamics. However, we show here that their combination leads to a relatively simple billiard system that nonetheless exhibits a rich spectrum of complex behaviors, ranging from near-integrability to chaos to ergodicity and mixing. Examples of the wide range of trajectories which we observe are shown in Fig.~\ref{fig:intro}. Notably, for a large region of $(\chi, \gamma)$-parameter space, we find that conical billiards exist at the interface between integrability and chaos--``the edge of chaos''--where small perturbations in initial conditions can shift trajectories from integrable to chaotic. At such interfaces dynamical systems can optimize complex computations~\cite{bertschinger2004real,toyoizumi2011beyond}, consequently garnering widespread interest in fields such as neural networks~\cite{bertschinger2004real}.  

In regions near the edge of chaos, we observe a transition from integrable orbits to strongly mixing orbits that parallels the transition to chaos in area-preserving maps such as the Chirikov-Taylor kicked rotor map~\cite{chirikov2008chirikov,weissert1997kolmogorov,shenker1982critical}. This transition is consistent with the Kolmogorov, Arnold, and Mosner (KAM) theorem, which predicts  that under small perturbations from an integrable limit, most orbits will be deformed but will persist, and the ones that will not will form a chaotic sea, the volume of which in phase space grows with the size of the perturbation~\cite{weissert1997kolmogorov,shenker1982critical}. The deformed integrable orbits are called KAM tori. The transition to chaos in our system occurs in two steps: (1)  the formation of a new type of KAM tori absent in the integrable limit of $\gamma =0$; and (2) the dissolution of these tori into a chaotic sea. Both steps occur in a similar, though not identical manner to the Chirikov-Taylor kicked oscillator map~\cite{chirikov2008chirikov}. A feature of our system not present in previously studied transitions to chaos is that the ergodic region of parameter space is  trapped between two limits, that of an untilted cone with $\gamma=0$ and one that approximates a flat ellipse. These limits are summarized in Fig. \ref{fig:setupFirst}(b).

This paper is organized as follows. In Sec.~\ref{sec:integrable}, we provide a brief review of integrable limits of conical billiard problems. We then define our main problem in Sec.~\ref{sec:boundary}, and discuss three different types of trajectories-- rim, hourglass, and mixing--which are illustrated for tilted cones in Sec.~\ref{sec:transition to chaos} (Fig.\ref{fig:intro} shows an example of each type).  We introduce a \Poincare map that tracks the slope and cone base intercept of trajectories interrupted by specular reflections, which provides a powerful way to classify various billiard paths. In Sec.~\ref{sec:chaos}, we discuss in more detail the space-filling chaotic trajectories that we observed in Sec.~\ref{sec:transition to chaos} and demonstrate that they are mixing in some regions of the $(\chi, \gamma)$ parameter space. We conclude in Sec.~\ref{sec:conclusion} by reviewing our results and suggesting future directions of research, including extending our analysis to hyperbolic cones, with a negative delta-function Gaussian curvature at the origin, and investigating the semiclassical regime of quantum conical billiards.  We present a heuristic argument for our results in Appendix ~\ref{sec:Appendix heuristic}, discuss \Poincare maps for elliptical billiards in  ~\ref{sec: appendix ellipse}, discuss conical geodesics as a function of $\chi$, and shows that the winding number divergesas $\chi \rightarrow 1$ in Appendix ~\ref{sec: appendix winding number}, and expand on the phase diagram presented in the main text in Appendix~\ref{sec:Appendix full phase diagram}. Technical details are relegated to Appendices~\ref{sec: appendix Coordtransform}, ~\ref{sec: appendix caustic}, ~\ref{sec:Appendix coneboundary}, ~\ref{sec:Appendix thetamax}, and ~\ref{sec: appendix distribution}.

\section{Related integrable systems}\label{sec:integrable}
A particle, with mass $m$, moving on a two dimensional surface has four degrees of freedom: two each for the position coordinate $\vec r$ and momentum coordinate, $\vec p = m \vec v$. In a mathematical billiard problem, the collisions with the wall are elastic and thus conserve energy $\frac{|\vec p|^2}{2m}$,  a constraint that leads to three independent degrees of freedom. According to the \Poincare-Bendixson theorem, any trajectory in a dynamical system with a bounded phase space and fewer than three degrees of freedom must eventually settle into a fixed point, a closed orbit, or a quasi-periodic orbit~\cite{strogatz2018nonlinear}. Thus, any billiard system with an additional conserved quantity beyond energy is integrable, and thus neither ergodic nor chaotic.

To set the stage for the more intricate conical billiard dynamics that is the focus of this paper, we begin by briefly discussing two extreme limits, both of which are integrable: billiards on flat ellipses and on conventional cones with an untilted base. 
\subsection{Billiards on an ellipse}
\begin{figure}[t!]
    \centering
    \includegraphics[width=\columnwidth]{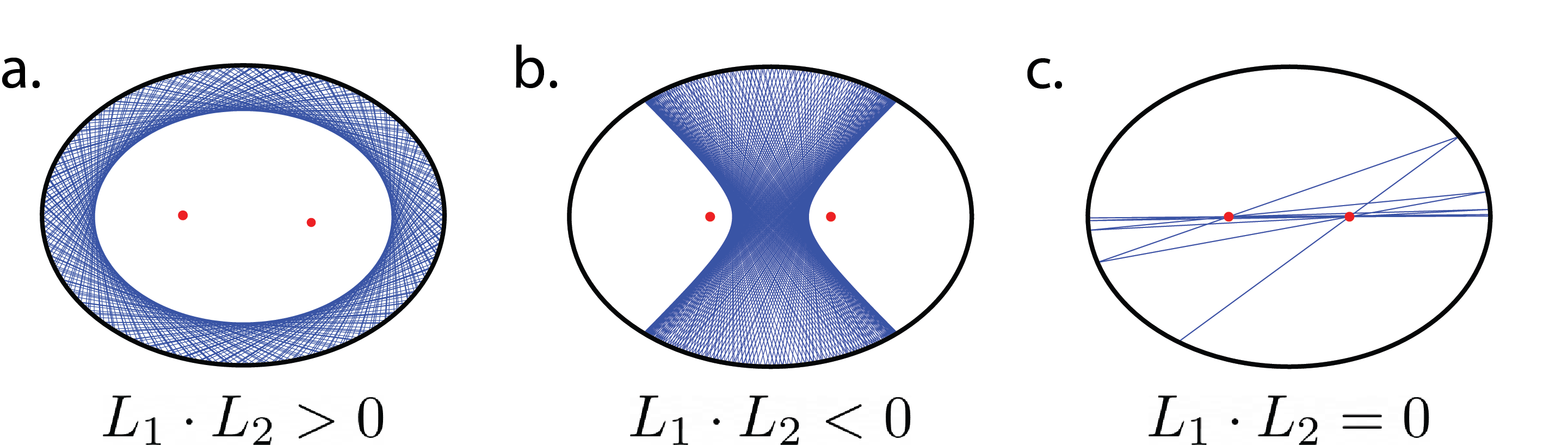}
    \caption{Three different types of trajectories on an ellipse with specular reflections on the boundary: (a) rim trajectory, and (b) hourglass trajectory, (c) homoclinic trajectory. The focal points are labeled in red. Fig.~\ref{fig: ellipse Poincare} in Appendix~\ref{sec: appendix ellipse} shows the \Poincare maps, similar to those defined in Fig.~\ref{fig:poincareMap}, that describe elliptical billiards with various eccentricities.}.
    \label{fig:ellipse}
\end{figure}
On a flat ellipse, in addition to the kinetic energy, the product of the angular momenta ($L_1, L_2$) about the two focal points is conserved~\cite{Tabachnikov1995,lynch2019integrable}. Billiards on an ellipse exhibit three distinct non-periodic types of integrable trajectories: (a) rim trajectories, where $L_1\cdot L_2 >0$ for which the trajectory never crosses between the two focal points and reflects off almost all points along the boundary (Fig.~\ref{fig:ellipse} (a)); (b) hourglass trajectories, where $L_1\cdot L_2 <0$ for which the trajectory always remains between the two focal points, densely sampling points only on a portion of the boundary (Fig.~\ref{fig:ellipse} (b)); and (c) homoclinic trajectories, where $L_1\cdot L_2 =0$, for which each segment of the trajectory intersects a focal point in an alternating fashion (Fig.~\ref{fig:ellipse} (c))~\cite{lynch2019integrable}. These homoclinic trajectories eventually approach a cycle, corresponding to the major axis of the ellipse, for long times. The first two types of trajectories are bounded in the ellipse interior by curves in real space called catacaustics,  a type of caustic singularity formed by billiard or ray-optic trajectories which reflect off boundaries~\cite{lynch2019integrable,avendano2010caustics}. Caustics are curves tangent to a collection of rays where the lines pile-up. When the caustics are formed by rays of light (for example, at the bottom of a swimming pool on a sunny day), they are the regions of concentrated light intensity~\cite{avendano2010caustics}. Rim and hourglass trajectories are named here for the shape of their catacaustic, which are an ellipse or a hyperbola confocal to the original ellipse, respectively~\cite{Tabachnikov1995,lynch2019integrable}. While the interior of a planar ellipse is not a simple limit of our conical billiard construction, we present it here, because qualitatively, these three types of trajectories can also be seen in conical billiards, as shown in Fig.~\ref{fig:3 types}(a-c). In addition, as discussed in Appendix~\ref{sec:Appendix heuristic}, a cone in the limit that the tilt angle $\gamma \rightarrow \pi/2-\beta$ (see Fig.\ref{fig:setupFirst}(a)) is nearly an ellipse. We note that, in the limiting case of a disk,an ellipse with eccentricity $0$ (or a cone with $\chi =0, \gamma =0$), where the two foci coincide at the center, only trajectory types (a) and (c) are possible. 

\begin{figure*}[t!]
    \centering
    \includegraphics[width=\linewidth]{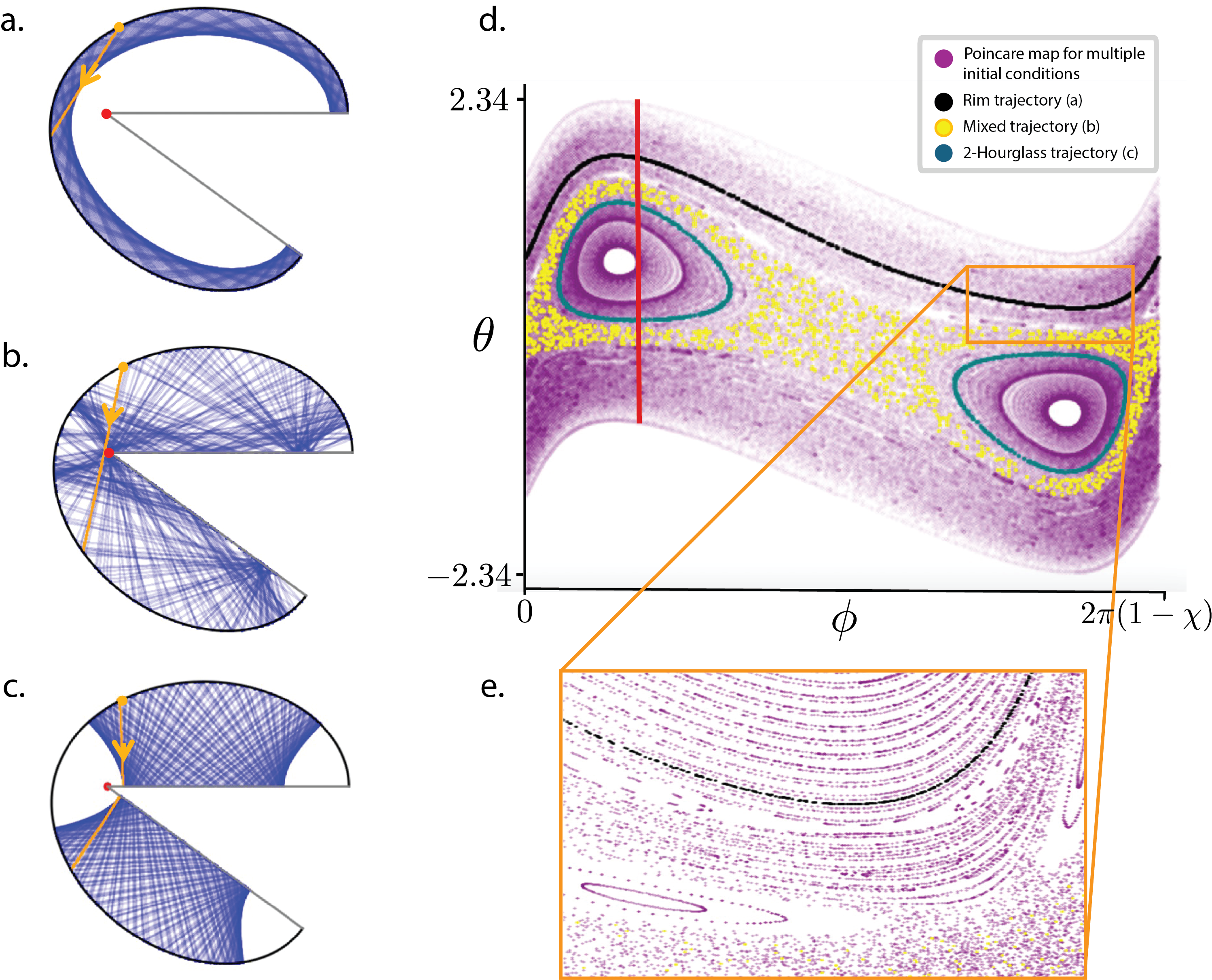}
    \caption{(a-c) Three qualitatively different types of trajectories on an unrolled tilted cone with an elliptical base for $\chi = 0.1$, $\gamma = 0.3$. The orange line indicates the initial condition that created the trajectories drawn in blue. The red point signifies the cone apex. (a) shows a rim trajectory, (b) a mixed trajectory, and (c) an hourglass trajectory. Note how the trajectories (a) and (c) look qualitatively similar to those in Fig.~\ref{fig:ellipse} (a,b). (d) The Poincare map for multiple initial conditions chosen uniformly along the red line, again for $\chi = 0.1$, $\gamma = 0.3$. A \Poincare map plots each trajectory as a sequence of geodesic line segments on the unrolled cone labeled by the angles ($\phi, \theta$), defined on Fig.\ref{fig:poincareMap}. The three trajectories in (a), (b), and (c) are marked in black, yellow, and blue respectively. Although the angles $(\phi_n, \theta_n)$ jump around with $n$, these points can eventually form smooth curves, such as the black line and blue loops. Note that the yellow points arising from the mixed trajectory in (b) do not settle down to a smooth curve, and are much more chaotic. For a billiard on an elliptical table, hourglass and rim trajectories look qualitatively similar in \Poincare space as illustrated in Appendix~\ref{sec: appendix ellipse}(e). A zoom in on a section of the \Poincare map to more easily see the three types of trajectories. Notice how small loops that signify a hourglass trajectory are also visible in this region of \Poincare space. Unlike the blue loops in (c) this hourglass trajectory will cover more than two regions of the real space boundary as it orbits a fixed point of period larger than two as discussed in Sec.\ref{Sec:box}.}.
    \label{fig:3 types}
\end{figure*}

As an important illustration of sensitivity to initial conditions, we note that two nearby trajectories that graze opposite sides of a focal point of an ellipse will exhibit markedly different behaviors, belonging to type (a) ($L_1\cdot L_2 >0$) or type (b) ($L_1 \cdot L_2 <0$), respectively. Despite this sharp contrast, the system behaves smoothly because as trajectories approach a focal point from either side, their corresponding catacaustic approaches the line connecting the two focal points~\cite{lynch2019integrable, Tabachnikov1995}. A somewhat similar phenomena occurs on a cone, where trajectories that graze the cone apex on either side have qualitatively different behaviors. However, on a cone, the limit of trajectories that pass increasingly closer to the apex from opposite sides lead to markedly different predictions for the trajectory. Trajectories  passing exactly through the cone apex are undefined. Analogous behavior has been observed in billiards on polygons which have corners~\cite{mcmullen2023billiards}. The effect of this ambiguity is explored in some detail in Appendix~\ref{sec:Appendix heuristic}.

\subsection{Billiards for $\gamma=0$}\label{sec.gamma0}

\begin{figure}
    \centering
    \includegraphics[width=\linewidth]{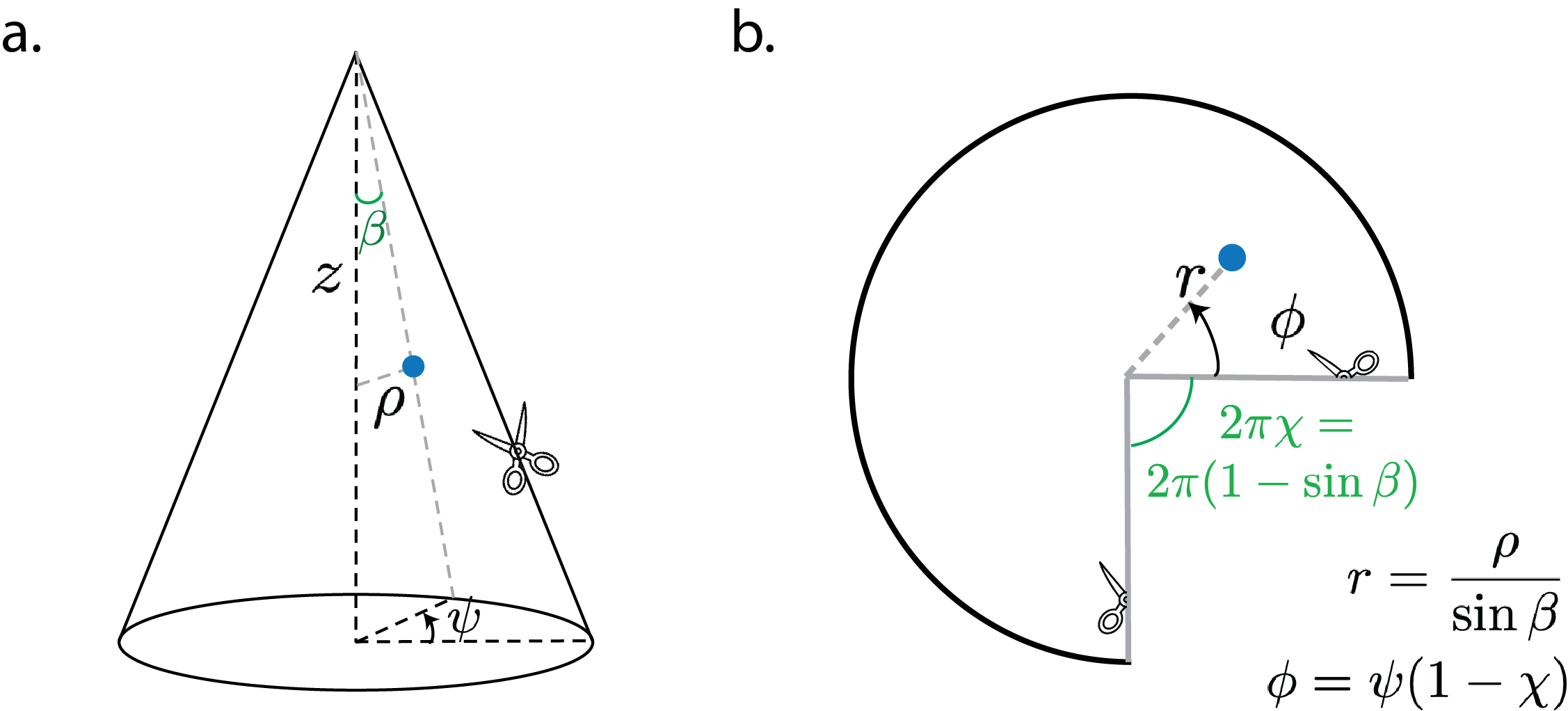}
    \caption{(a) Coordinates for a point (in blue) on the surface of a cone. In order to convert to flat space in two dimensions, we cut along a geodesic which hits the apex. The path along which we cut is marked with the scissors. (b) Unrolled coordinates for the cone for the same blue point. When reconstructing the cone back into three dimensions, the gray lines with broken scissors at $\phi = 0$ and $\phi = 2\pi(1-\chi)$ are identified.}
    \label{fig:coordinatesY}
\end{figure}

The second integrable system we consider is a special untilted conical billiard problem with $\gamma =0$. Since a cone is locally flat (has zero Gaussian curvature) everywhere except at the apex, it can be mapped onto a flat 2D plane with all distances and angles preserved. In three dimensions, the surface of a cone with half angle $\beta$ (see Fig.~\ref{fig:setupFirst}(a)) can be described as all points with cartesian coordinates 
\begin{align}
    \vec r(\rho, \psi)= (x=\rho \cos \psi , y= \rho \sin \psi, z = \frac{\rho}{\tan(\beta)})
\end{align}, 
with $0<\psi <2\pi$, as shown in Fig.~\ref{fig:coordinatesY}(a). To implement unrolling, we then equate a point on the cone in $\mathbb R^3$ to a point in $\mathbb R^2$ with cartesian coordinate 
\begin{align}
    &(x= r\cos \phi,y = r\sin \phi) =\nonumber \\
    & \quad \Big(\frac{\rho}{\sin(\beta)} \cos(\psi (1-\chi)), \frac{\rho}{\sin(\beta)} \sin(\psi (1-\chi))\Big)
\end{align}
 where we have periodic boundary conditions in the sense that $\phi = \phi +2\pi (1-\chi)$. This mapping corresponds to cutting the cone along its flank at $\psi = 0$ and unrolling, placing one side of the cut on the $x-$axis as shown on Fig.~\ref{fig:coordinatesY}(b). A similar transformation will be carried out when $\gamma \neq 0$. For $\gamma \neq 0$, we will always position the longest flank distance from the apex to the base at the location of the cut, which we set at $\psi=0$. The primary advantage of these coordinates is that geodesics on the unrolled cone become straight lines in the plane, as discussed in more detail in Appendix~\ref{sec: appendix Coordtransform},  simplifying our analysis.

For any geodesic on the cone, the angle of incidence, $\theta$,  equal to the angle at which the path intersects the base of the cone, is conserved in the $\gamma =0$ billiard problem (see Fig.\ref{fig:setup}a in Appendix \ref{sec: appendix Coordtransform} for illustration). It follows that each segment of the trajectory reaches a minimum flank distance of $R \sin(\theta)$ from the apex (See Fig. \ref{fig:setup}). A similar perspective has been used to analyze billiards on a circular disk, which correspond to the $\chi = 0$ limit of our $\gamma =0$ system~\cite{berry1981regularity}. On a disk, this limit reduces the problem to the special case where the two focal points coincide, making $L_1=L_2$. Conservation of $L_1\cdot L_2 = L_1^2$ in this limit is equivalent to the conservation of the angular momentum about the center of the disk~\cite{berry1981regularity}. For an untilted cone with arbitrary $\chi$, this additional conservation law ensures that the system is not chaotic, as implied by the \Poincare-Bendixson theorem. 

For the remainder of this section, we will introduce a \Poincare map by identify trajectories by the sequence of encounters they make with the untilted cone base, see Fig.~\ref{fig:setup}(a). At each reflection, we record two quantities: the polar angle,  $\phi_i$, $0 \leq \phi_i < 2\pi(1-\chi)$, specifying the location of the intersection on the boundary, and the angle $\theta_i$, $-\pi < \theta_{max}(\phi_i) - \pi< \theta_i< \theta_{max}(\phi_i) <\pi$, ($\theta_{max}(\phi_i)$ is defined in Appendix \ref{sec:Appendix thetamax}), which the trajectory makes with the vector from its initial position to the apex. These quantities are illustrated for an unrolled cone in Fig.~\ref{fig:setup} (a). The first coordinate plays the role of a $y-$intercept, while the latter is an analog of the slope, $m$, in the usual $y = mx+b$ representation of a line in a plane. Similar coordinates have been used in conventional billiard systems~\cite{tyc2022spherical,Tabachnikov1995}. This approach allows us to represent trajectories in terms of a \Poincare map--a two dimensional mapping which records trajectories' intersections and slopes at the one dimensional boundary at the untilted cone base~\cite{strogatz2018nonlinear}. The sequence of ($\phi_n, \theta_n$)-coordinates obtained in this way,  jump discontinuously after each bounce, but nevertheless represent a kind of phase space with $\phi$ playing the role of position, and $\theta$ playing the role of the ratio of the two components of momentum for a particle with mass $m=1$ and constant speed $|\vec v|=1$. We will call this space the \Poincare space. A simple example of such a map for $\gamma =0$ can be seen in Fig.~\ref{fig:setup} (b).  Further examples of  \Poincare maps for ellipses of different eccentricities are provided in Appendix~\ref{sec: appendix ellipse}. 
\begin{figure*}[t!]
    \centering
    \includegraphics[width = \linewidth]{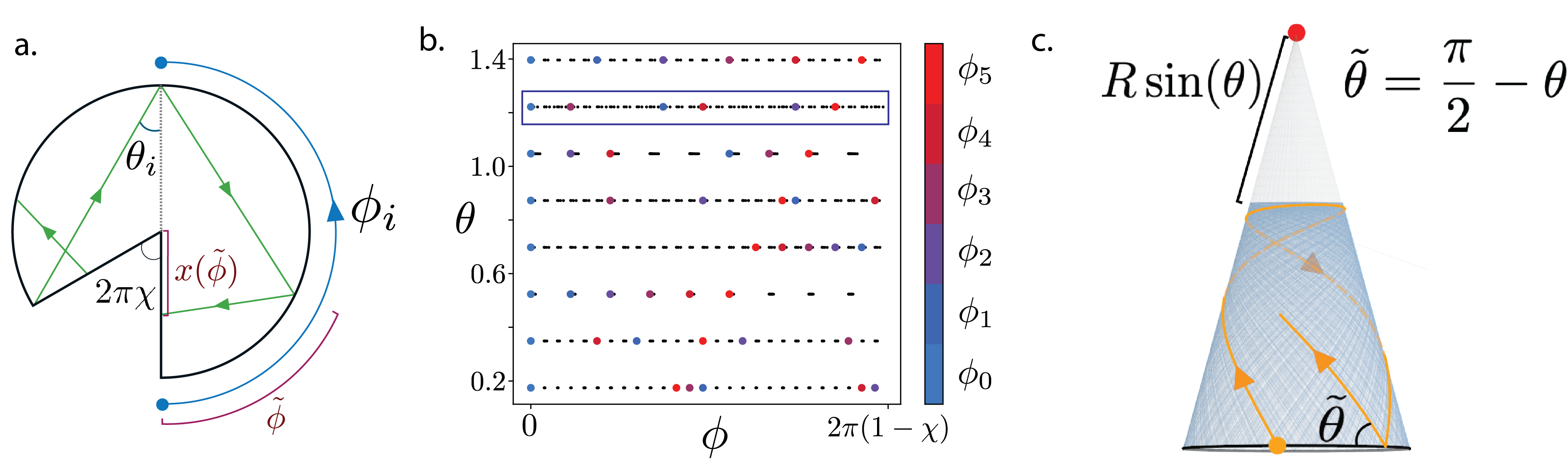}
        \caption{(a) Definition of coordinates $\theta_i$ and $\phi_i$ in the unrolled geometry for an untilted cone, which can be viewed as an analog of a line slope and intercept, respectively. (b) Poincare map for $\gamma = 0$ and $\chi =0.8$ for eight distinct initial conditions illustrating the conservation of the bounce angle $\theta_n$. Each line of black point cllusters corresponds to a different initial condition. The first six points of the sequence defining each trajectory are marked by larger points. They are colored such that the first point $(\phi_0, \theta_0)$ is in blue and $(\phi_5,\theta_5)$ is in red and all the intermediate points colors lie on a gradient from blue to red. Note that some of the black points cluster together to look like dashes. (c) The trajectory boxed in blue on panel (b) reconstituted on a three dimensional cone. Orange line plots highlights the trajectory from its initial condition to the first bounce off the base. The minimal distance to the apex, $R \sin \theta$ is labeled. Also labeled at the position of the first bounce off the base is $\tilde \theta = \pi/2-\theta$. } 
    \label{fig:setup}
\end{figure*}
% \textcolor{red}{I introduce this here and also in the next section. Not sure which is better.}
As illustrated in Fig.~\ref{fig:setup} (b), for $\gamma =0$, $\theta_i = \theta_0$ and $\phi_{i+1} - \phi_{i} = 2 (\pi/2 -\theta_0)$ will both be constants of motion preserved by bounces off the base. By exploiting the azimuthal symmetry of the untilted cone, we can set the initial cone base coordinate $\phi_0 = 0$ without loss of generality.  

In \Poincare space, any trajectory on an untilted cone can thus be represented via the \Poincare map as a sequence of ordered pairs ($\phi_n, \theta_n$) with: 
\begin{align}
    &\phi_n =2 n (\pi/2 -\theta_n)\mod(2 \pi (1-\chi)) \\
    &\theta_n = \theta_0
\end{align}
where $n = \{0,1,2,...\}$, and the circumferential arc-length around the cone base is given by $s_n = R\phi_n$. Note that $s_n$ can in principle be longer than the length of the cone base, thus causing the geodesic to loop around the cone apex. The winding number of such geodesics (number of loops the geodesic makes around the apex as it gets turned back to the base) diverges in the limit of extremely pointed cones such that $\chi \rightarrow 1$, see Appendix~\ref{sec: appendix winding number}. Fig.~\ref{fig:setup}(b) highlights this simple \Poincare map for various choices of the conserved initial angle $\theta_0$. If $\frac{2 (\pi/2 -\theta_0)}{2 \pi (1-\chi)}\in \mathbb{Q}$, the set of rational numbers, the trajectory will eventually form a closed cycle of finite length (such periodic trajectories are also discussed in more detail in Appendix~\ref{sec: appendix winding number}). Otherwise, the untilted cone trajectory will be incommensurate, eventually uniformly covering the base of the cone. 

However, the trajectory will not uniformly cover the entire cone surface, as shown in Fig.~\ref{fig:setup} (c). In fact, there is a ring caustic, an accumulation of trajectory points at a distance of $x = R\sin(\theta)$ from the apex. We now determine the functional form of this accumulation for the incommensurate case, without loss of generality, along the geodesic $\phi =0$. 

All path segments crossing $\phi=0$ will next intersect the base somewhere between $\phi=0$ and $\phi=2 (\pi/2 -\theta)$ on the unrolled cone. Upon assuming that the trajectory uniformly covers the base, as will be the case for incommensurate trajectories, we expect the probability density $\tilde P(\phi)$ for a point on the path being found at position $\phi$ on the boundary will be uniform:
\begin{align}
     \tilde P(\phi) = \frac{1}{2 (\pi/2 -\theta)} ,
\end{align}
where $\tilde P(\phi)$ is the normalized probability density for $\phi \in [0, 2 (\pi/2 -\theta)]$. 

\begin{figure}
     \centering
     \includegraphics[width=\columnwidth]{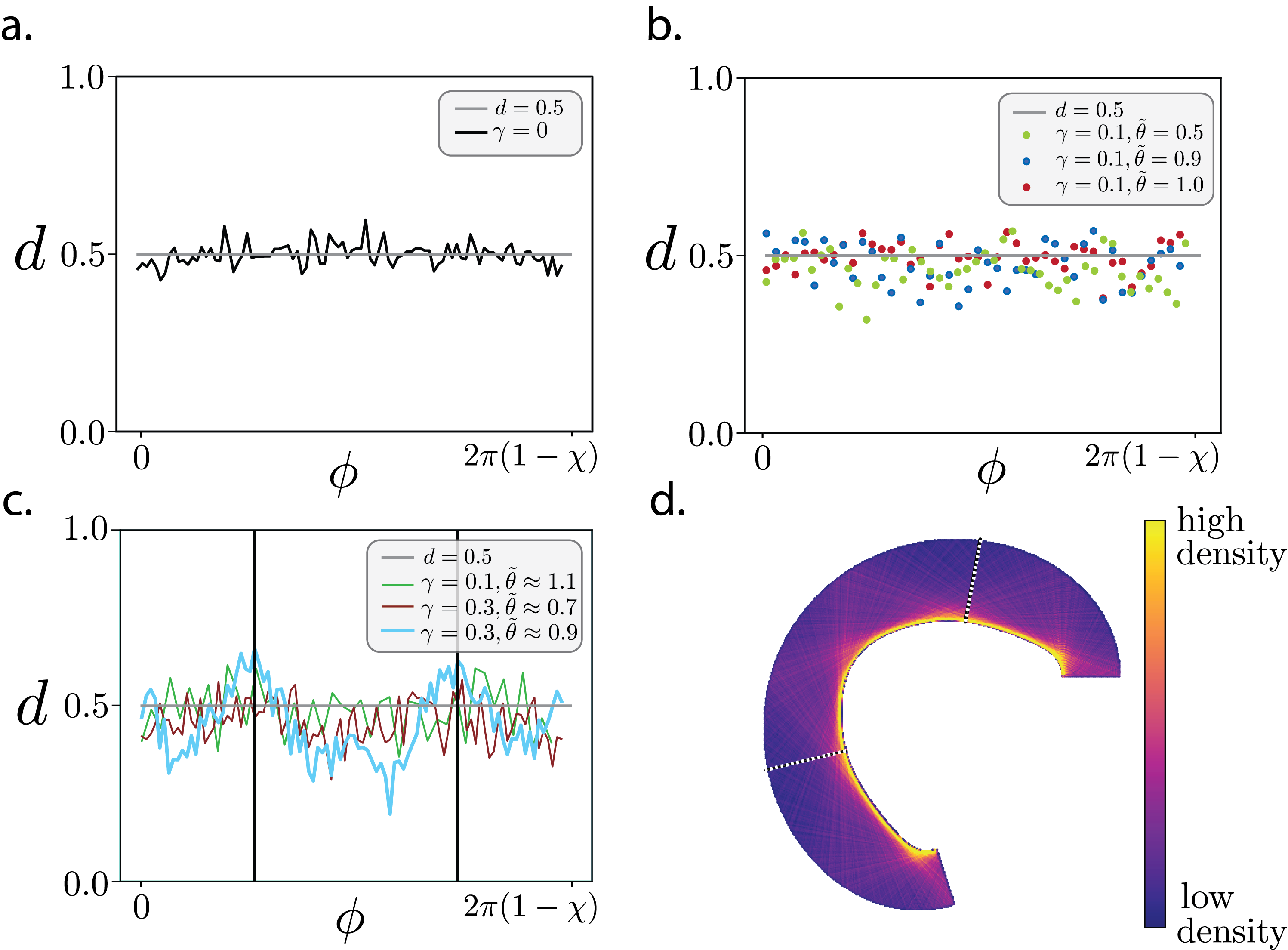}
     \caption{(a-c) The critical exponent, $d$, for the density pile-up at the catacaustic for $\chi=0.2$, and various values of  $\gamma$ and $\tilde \theta = \theta_{max}(\phi) -\theta$ fitted to $\frac{b}{(x-a)^d}$ as a function of $\phi$ using the protocol discussed in Appendix~\ref{sec: appendix caustic}. The horizontal gray line marks an exponent $d = 0.5$. (a) $\gamma = 0$, notice how the critical exponent hovers around $d = 0.5$ as expected from Eq.\ref{Eq.critexp}. We use this untilted cone as a control to confirm that the numerical method for calculating critical exponents is accurate. Slight deviations are expected due to the method used to coarse grain the data, the numerical determination of the catacaustic, and the simulation being run for less than infinite time. (b) Tilt parameter $\gamma = 0.1$ with three different values of $\tilde \theta$, as defined in Eq.~\ref{eq.tildetheta}. Note that even though $\tilde \theta$ is as large as $1.0$ we still see that the critical exponent hovers around $d=0.5$ for all values of $\phi$. (c) Same plot but for larger values of $\tilde \theta$ and for $\gamma = 0.1$ and $0.3$. Note how the critical exponent seems to depend on $\phi$. Two vertical black lines are drawn in regions of higher critical exponent. (d) Density heat map corresponding to the light blue line in (c), $\gamma = 0.3$, $\tilde \theta = 0.9$. The two white dotted lines show the normals to the catacaustic at the locations marked with black lines in panel (c).}
     \label{fig:caustic}
 \end{figure}

We now let $x(\tilde \phi)$ denote the distance from the apex at $\phi=0$ for a segment of the path that previously intersected the base at $\phi = \tilde \phi$ (labeled in Fig.~\ref{fig:setup}(a)). It can be shown straight forwardly that
\begin{align}
    x(\tilde \phi) = \frac{R \sin(\theta)}{\sin(\tilde \phi+\theta)} .
\end{align}
Thus, since $\tilde P(\phi) d\phi = P(x) dx$, where $P(x)$ is probability the trajectory being at apex distance $x$, we have
\begin{align}\label{Eq.critexp}
     &\lim_{x \rightarrow R\sin(\theta)} P(x) \nonumber \\& \qquad =\lim_{x \rightarrow R\sin(\theta)} \frac{R \sin(\theta)}{(\frac{\pi}{2} -\theta) x^2 \sqrt{1-\frac{R^2\sin^2(\theta)}{x^2}}} \nonumber \\& \qquad \propto \frac{1}{\sqrt{x-R\sin(\theta)}} .
\end{align}
This result reveals a square root caustic-like divergence in the density of points visited as $x$ approaches $R\sin(\theta)$. We numerically confirm this behavior using the method described in Appendix~\ref{sec: appendix caustic}, with the results shown in Fig.~\ref{fig:caustic}(a). Note that the amplitude of the square root of the divergence itself diverges as $\theta_0 \rightarrow \pi/2$ and the trajectory points directly at the apex. Notably, $\chi$ does not appear in the diverging probability density Eq.\ref{Eq.critexp} describing this catacaustic, and only influences the order of bounces along the cone base and whether the path is periodic or incommensurate. In Sec.\ref{sec: rim}, we will present numerical evidence that this square root power law divergence in the density is approximately correct for small values of $\gamma, \chi$ and $\theta$. 

\section{Tilted Cone Parametrization}\label{sec:boundary}

\begin{figure}
    \centering
    \includegraphics[width=\columnwidth]{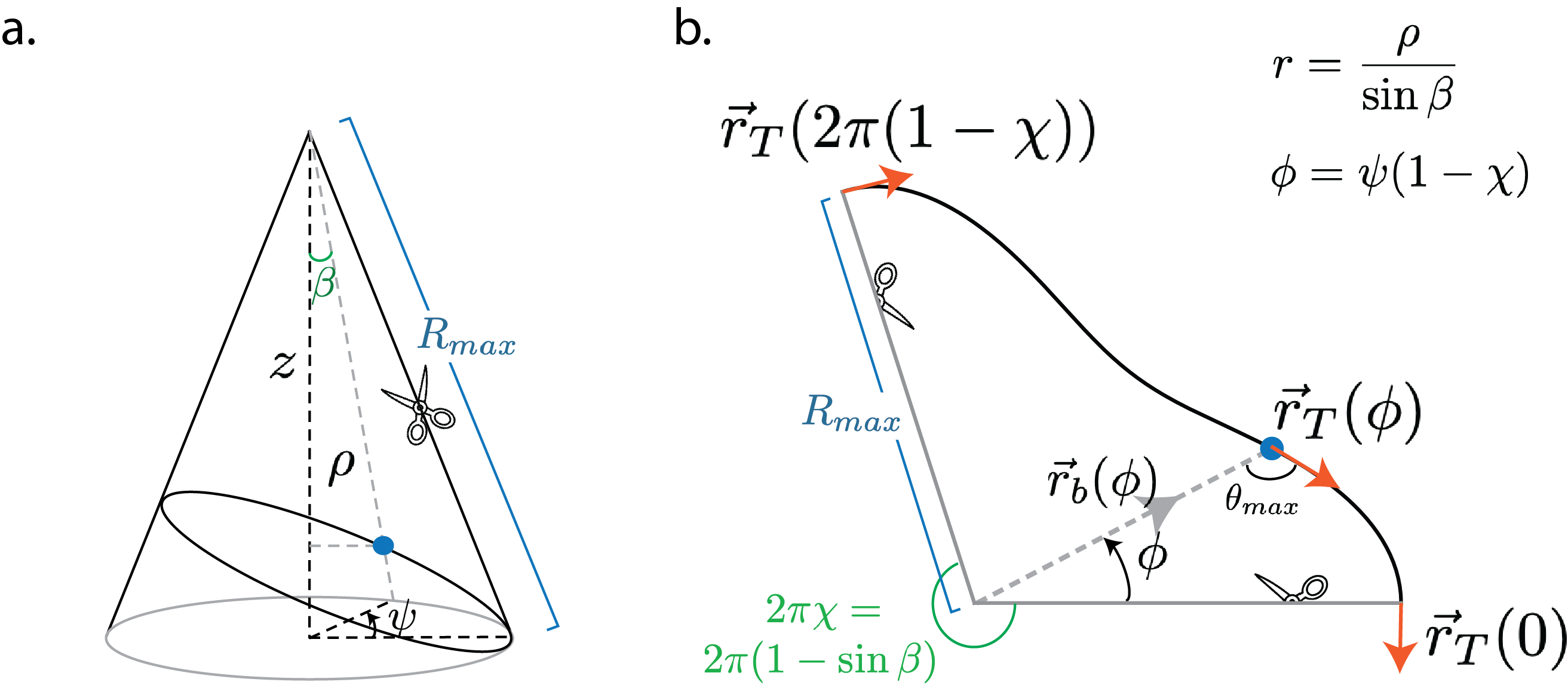}
    \caption{(a) Coordinates for a point (in blue) on the surface of a cone for non-zero $\gamma$. Just like for $\gamma = 0$, in order to convert to two dimensions, in a similar method to that illustrated in Fig.~\ref{fig:coordinatesY}, we cut along a geodesic which hits the apex. We choose to cut along the longest flank distance. The path along which we cut is marked with the scissors. (b) Unrolled coordinates for a cone with $\gamma = 0.6$ and $\chi = 0.7$ with the same blue point on the rim. When the unrolled cone is converted back into three dimensions the gray lines with broken scissors at $\phi = 0$ and $\phi = 2\pi(1-\chi)$ are identified. We mark the vector from the apex to the boundary, $\vec r_b(\phi)$ in dashed gray, and $\vec r_T(\phi)$ the unit tangent at the boundary at $\phi$ in red at the blue point. We also mark $\vec r_T(2\pi(1-\chi))$ and $\vec r_T(0)$. $\theta_{max}(\phi)$, the maximum angle between $\vec r_b(\phi)$ and $r_T (\phi)$, is labeled as well. Notice how the black boundary line is concave near $\phi = \pi(1-\chi)$, half way along the base.}
    \label{fig:tiltedConeBound}
\end{figure}
\begin{figure}[t!]
    \centering
    \includegraphics[width = \columnwidth]{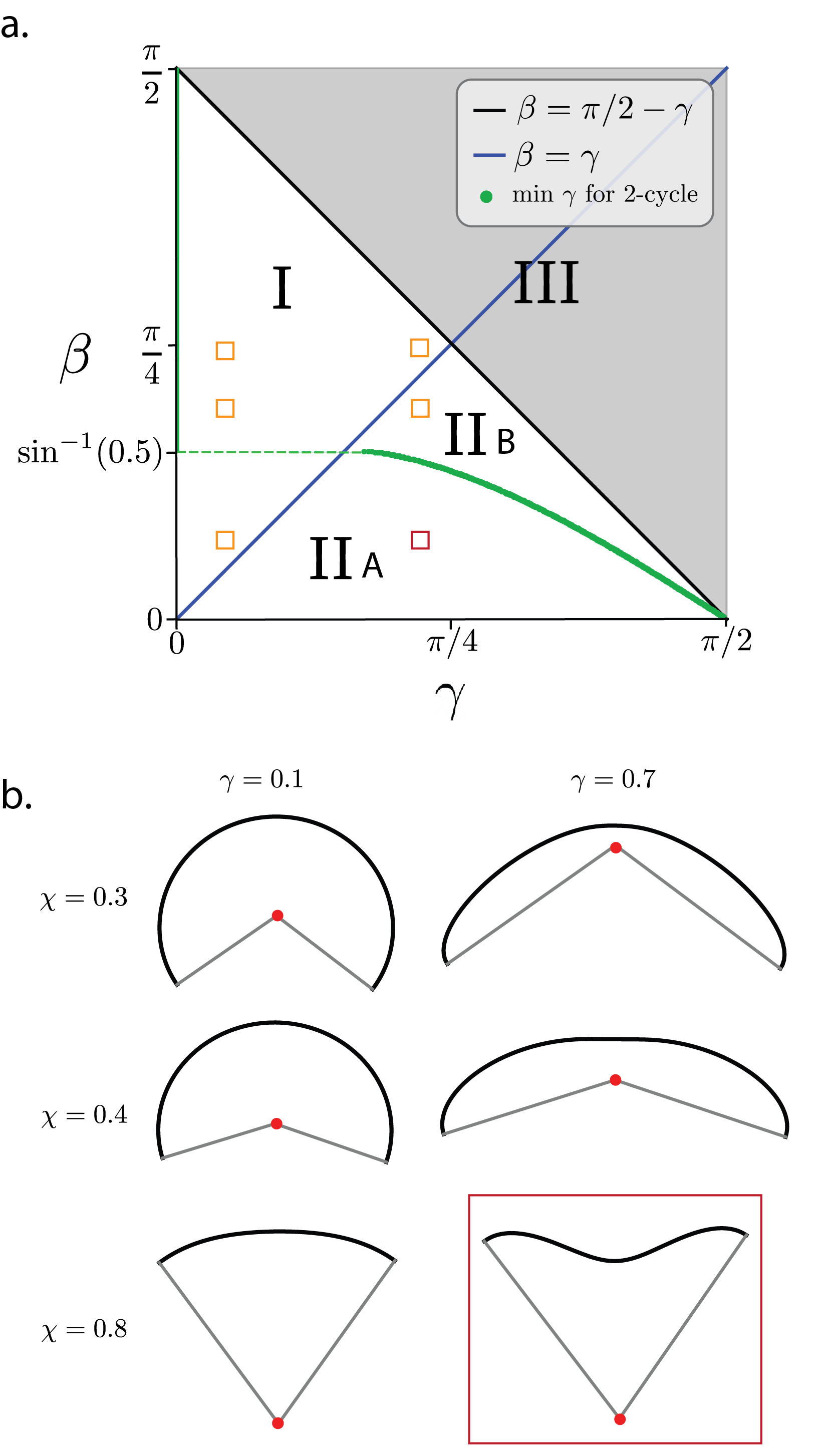}
    \caption{a. Plot of ($\gamma$,$\beta$)-plane  with the blue line marking the transition between convex and partially concave unrolled cone bases (regions I and II respectively) for the unrolled cone. Note that $\sin\beta = 1-\chi$. The black line marks the maximum allowed value of $\gamma$ for a given $\beta$ before the elliptical cone base becomes a hyperbola. The green line marks the numerically calculated minimal $\gamma$ where the unrolled cone base is parallel to the cut, separating regions IIA and IIB. Region III, shown in gray, is the region of disallowed values of $\gamma$. The orange and red squares mark the unrolled cones drawn in panel (b). Sample unrolled boundaries for six different combinations of $\chi$ and $\gamma$ are shown. Note that all the unrolled cones outside of region $I$ ($\gamma = 0.7$ and $\chi = 0.4$ and $ 0.8$) contain a concavity in the unrolled cone base. Sec.\ref{sec:transition to chaos} and Sec.\ref{sec: rim} rule out ergodicity in regions $IIB$ and $I$ respectively. Thus, of the unrolled cones represented in (b) only the one boxed in red ($\chi = 0.8$, $\gamma = 0.7$) could contain ergodic trajectories.}.
    \label{fig:boundary}
\end{figure}

We will now move on to the more interesting case of tilted cones, i.e., $\gamma \neq 0$. As before, we unroll the cone, this time along the longest geodesic passing from the base to the apex, and identify the boundaries in plane polar coordinates $(r, \phi)$ at $\phi=0$ and $\phi = 2\pi(1-\chi)$, as shown in Fig.~\ref{fig:tiltedConeBound}. The boundary of the unrolled tilted cone in flat space, $\vec r_b = \big(r_b(\phi) \cos(\phi),r_b(\phi) \sin(\phi) \big)$, calculated in Appendix \ref{sec:Appendix coneboundary}, is given by
\begin{align}
    r_b(\phi) = \frac{R\cos(\gamma)\sqrt{\chi (2-\chi)}}{\sqrt{\chi(2-\chi)}\cos(\gamma) - \cos(\frac{\phi}{1-\chi})(1-\chi)\sin(\gamma)} 
\end{align}
where $\sin^{-1} (1-\chi) = \beta$ and 
\begin{align}
    R = \frac{R_{\max}}{1+\frac{(1-\chi)\sin(\gamma)}{\cos(\sin^{-1}(1-\chi) + \gamma)}} 
\end{align}
is the flank distance of the conventional cone with the same apex,$\chi$ value, and height directly below the apex,as is shown graphically in Fig.~\ref{fig:setupFirst}. For consistency, we choose $R$ to be the same for all simulations in this paper. Here, $R_{\max}$ is the length of the longest flank geodesic distance of our cone which we take to be the line along which the cone is cut (see Fig.\ref{fig:tiltedConeBound}). While the boundary in the three dimensional embedding of our cone is convex, when unrolled this need not always be the case. This indentation of the unrolled cone boundary, already evident in Fig.~\ref{fig:tiltedConeBound}, allows for more complex dynamics than when the unrolled cone base is convex. In Appendix \ref{sec:Appendix coneboundary}, we analytically determine that the concavity first appears when 
\begin{align}
    \gamma >\tan^{-1}(\frac{1-\chi}{\sqrt{\chi(2-\chi)}}) = \beta.\label{eq.phasediagramboundaries}
\end{align}
These observations allow us to identify three important regions, I, II, and III in the ($\gamma, \beta$)-plane, as shown in Fig.~\ref{fig:boundary}: 
\begin{align}
     \begin{cases} 
      \text{I}: 0<\gamma <\beta & \text{boundary is convex} \\
      \text{II}: \beta<\gamma <\frac{\pi}{2} - \beta & \text{boundary is partially concave}\\
      \text{III}: \frac{\pi}{2} - \beta < \gamma & \text{our problem is ill-defined}
   \end{cases}
\end{align}
Because $\sin(\beta) = 1-\chi$ and $\cos(\beta) = \sqrt{\chi(2-\chi)}$, these regions could also be delineated in the $(\gamma,\chi)$ plane, but we find it easier to work with the $(\gamma, \beta)$ plane.

\section{Transition to Chaos}\label{sec:transition to chaos}

\begin{figure}[t!]
    \centering
    \includegraphics[width=\columnwidth]{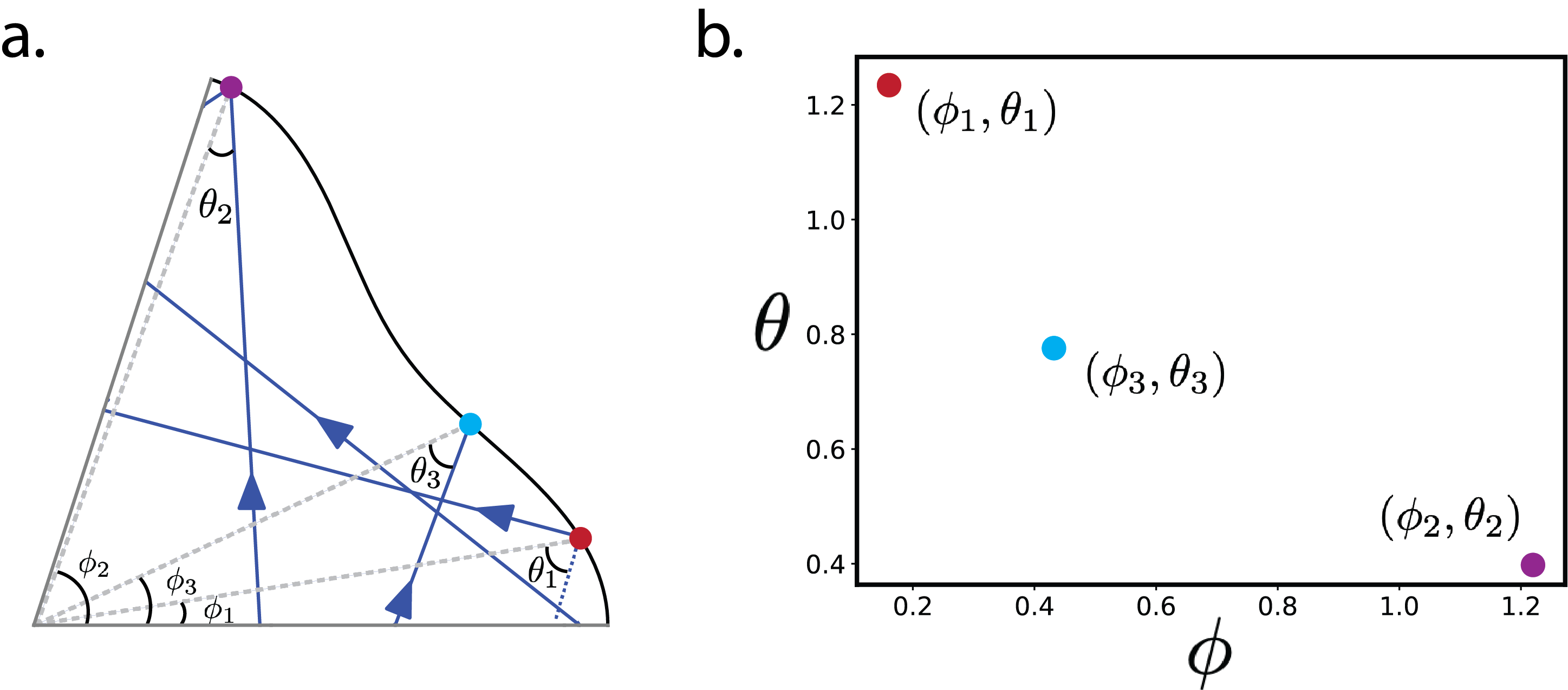}
    \caption{Illustration of the connection between continuous trajectories on an unrolled cone and their Poincare mappings. A typical trajectory on a cone with a concavity is chosen.  (a) A trajectory on a cone with $\chi = 0.8$, $\gamma =0.5$ is shown in blue. Note that as for $\gamma =0$, when a trajectory hits one of the cuts (solid gray lines), it reappears at the other cut, at the same distance from the apex but rotated clockwise by an angle $2\pi\chi$.The first three intersections it makes with the unrolled cone base are marked in order by a red, purple, and blue point. At each intersection $\phi_i$, the azimuthal position along the cone, and $\theta_i$, the line segment slope relative to the cone apex, are labeled. These discrete points are used to create a sequence of $\phi$ and $\theta$ values that is then plotted as a Poincare map. (b) The positions and orientations of geodesic segments labeled in (a) as points in Poincare space. The points are labeled and color coordinated with their base intersections in (a). Note that the trajectory bounces about Poincare space discontinuously and that the third point in the sequence is closer to the first point than the second point is. }.
    \label{fig:poincareMap}
\end{figure}

As we did for $\gamma =0$ we proceed by examining our continuous trajectories via \Poincare maps (See Fig.\ref{fig:setup}(b)). An illustration of this analysis can be seen in Fig.~\ref{fig:poincareMap}. We first summarize the transition to chaos qualitatively and then go into more detail in various subsections.

\begin{figure*}[ht!]
    \centering
    \includegraphics[width=\linewidth]{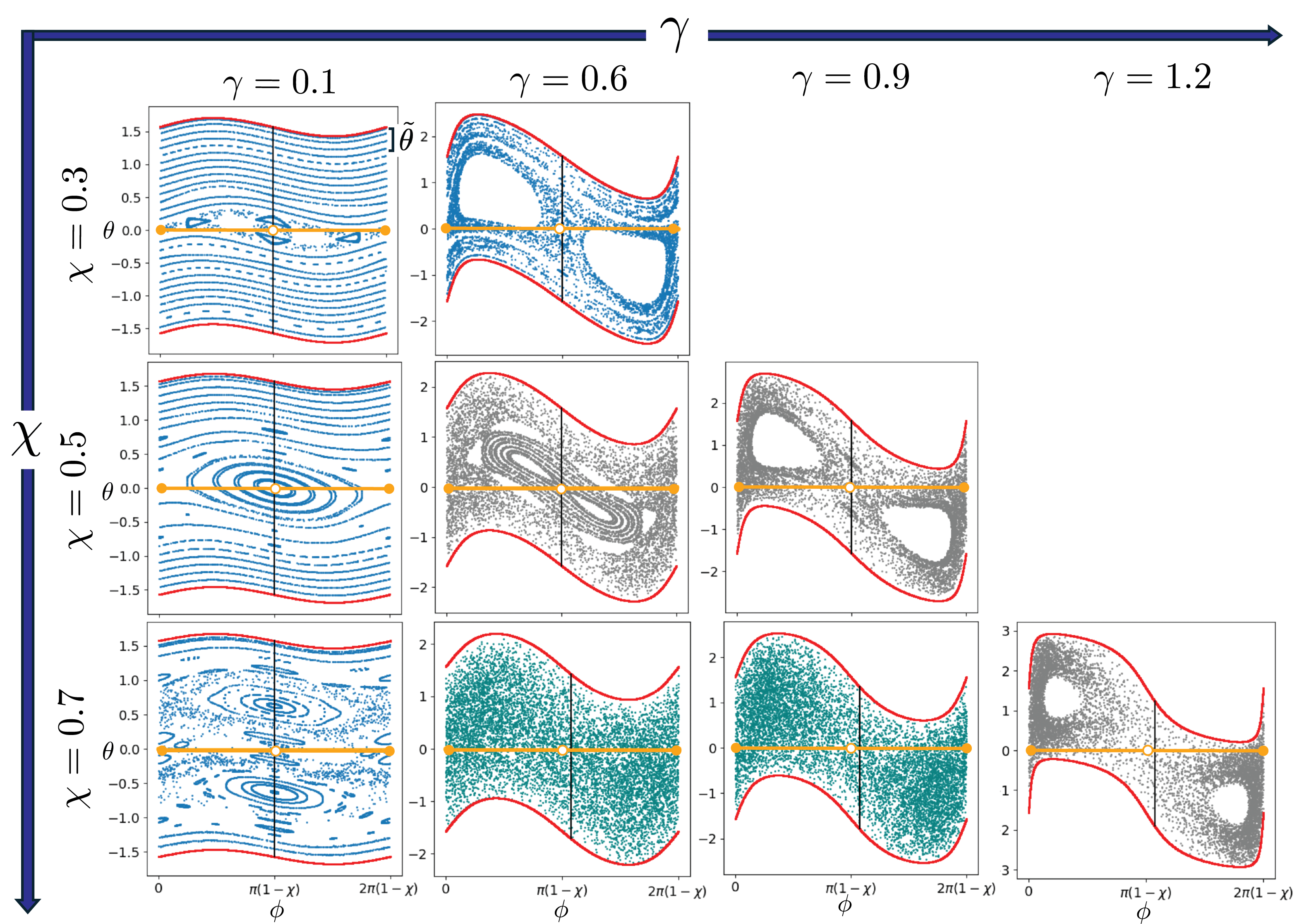}
    \caption{Selected \Poincare maps for varying $\chi$ and $\gamma$. The horizontal yellow lines mark the points in \Poincare space that correspond to trajectories directly hitting the apex. The \Poincare maps are periodic in $\phi$. The wavy red lines mark the boundaries of the \Poincare space. A procedure to determine the equation for these lines is given in Appendix \ref{sec:Appendix thetamax}. The black vertical line marks the initial conditions that generate these \Poincare maps. The empty white spaces are regions that are not reached by trajectories beginning on those initial conditions. Blue plots are in region I, teal ones are in region IIA, and gray ones are in region IIB as defined in Sec.~\ref{sec:boundary} and Fig.~\ref{fig:boundary}. All \Poincare maps  in the blue region generate rim trajectories when periodicity in $\phi$ is taken into account. There are no rim trajectories on the gray and teal \Poincare maps, and all the gray plots and the blue plots for $\chi <0.5$ have a white region corresponding to two loop hourglass trajectories as discussed in the text. While being in region IIA of Fig.~\ref{fig:boundary} does not guarantee that trajectories are mixing and ergodic, we see here that many are. Note that in the top left figure, the relative coordinate $\tilde \theta = \theta_{max}-\theta$ is labeled. A more densely populated version of this figure can be see in Fig.\ref{fig:full phase diagram} of Appendix \ref{sec:Appendix full phase diagram}, which also tabulates the effect of snipping off the top of the cone by including an additional boundary near the apex.} 
    \label{fig:phasediagram}
\end{figure*}

Fig.~\ref{fig:phasediagram} uses Poincare maps to reveal three distinct types of trajectories present as $\chi$ and $\gamma$ are varied, also labeled in Fig.~\ref{fig:3 types}. We will refer to the two non-chaotic trajectory classes as rim and hourglass trajectories, emphasizing their approximate similarity with the rim and hourglass trajectories on an ellipse, see Fig.~\ref{fig:ellipse} and Appendix~\ref{sec: appendix ellipse}. More chaotic trajectories, which we will call mixed trajectories, appear in between regions of rim and hourglass trajectories. The latter trajectories are analogs to the homoclinic trajectories passing through the foci on an elliptical billiard table, but occupy a larger region of phase space and, unlike homoclinic trajectories on an ellipse, do not approach a two cycle at long times. As $\gamma$ and $\chi$ increase, the mixed trajectories occupy an increasingly larger area in \Poincare space, reflecting the growing complexity of the conical dynamics.

Note that hourglass trajectories appear on \Poincare diagrams as collections of loops. Trajectories can jump from loop to loop, where each loop spans a disjoint range of $\phi$ values, as shown in Fig.~\ref{fig:3 types}.  These trajectories come in groups of progressively smaller nested loops. The centers of these loops are $n$-cycle trajectories where $n$ is the number of loops in \Poincare space that the trajectory jumps between. Henceforth, we will say that a trajectory is \emph{n-hourglass} if it surrounds an n-cycle. 

\begin{figure}
    \centering
    \includegraphics[width=\columnwidth]{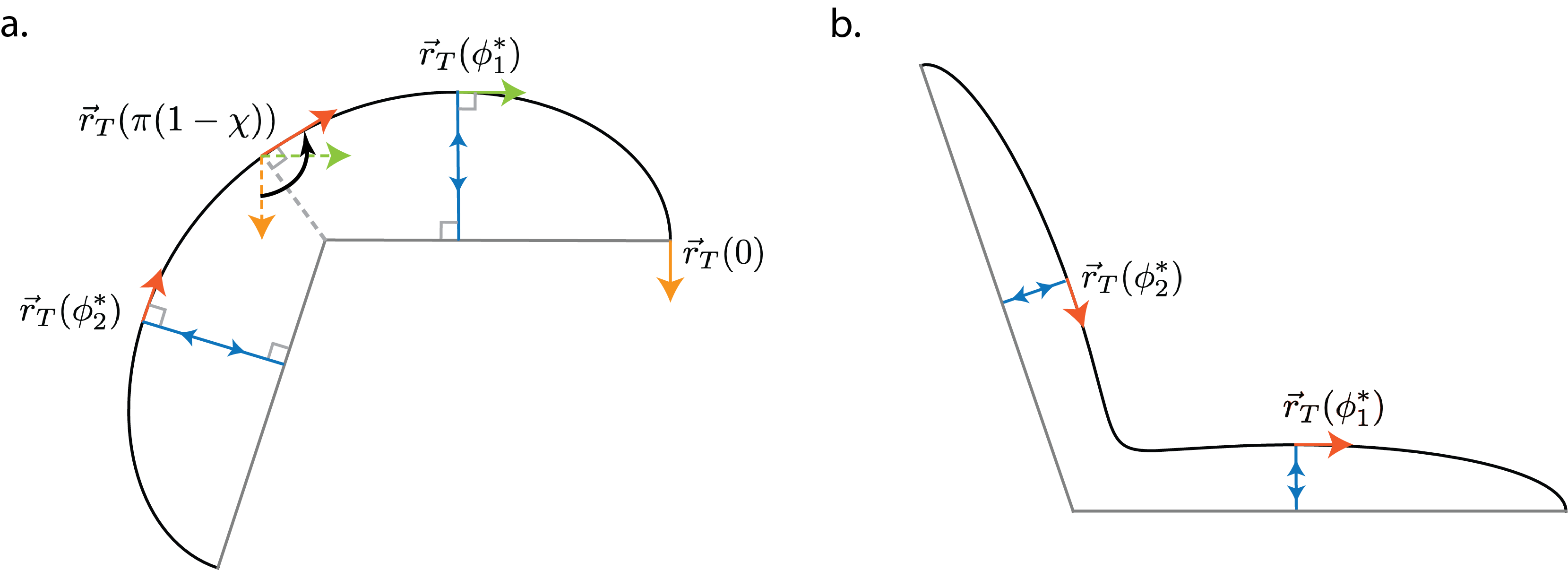}
    \caption{(a) An unrolled cone for $\chi = 0.3$, $\gamma = 0.4$. Four tangents at different points along the boundary are labeled: $\vec{r}_T(0)$ , $\vec{r}_T(\phi_1^*)$ , $\vec{r}_T(\pi(1-\chi))$ , and $\vec{r}_T(\phi_2^*)$. The angles $\phi_1^*$ and $\phi_2^*$ correspond to points at which the line tangent to the unrolled cone base is parallel to the cut. These will be the intersection points of a two cycle trajectory. The angle $\phi$ is the polar angle as labeled in Fig.\ref{fig:coordinatesY}(b) and ranges form $0$ to $2\pi(1-\chi)$.   To illustrate the continuous rotation of the tangents, the green and light orange lines have been translated to $\phi = \pi(1 - \chi)$. This helps visualize how, when moving counterclockwise along the boundary from $\phi = 0$ (orange) to $\phi = \pi(1 - \chi)$ (red), the tangent must pass through the orientation of $\vec{r}_T(\phi_1^*)$ (green). Since the tangents vary continuously, this guarantees the existence of $\phi_1^*$ where $\vec{r}_T(\phi_1^*)$ is parallel to the cut. By symmetry, a similar point $\phi_2^*$ must also exist.     A two-cycle trajectory starting at $\phi_1^*$ and perpendicular to the boundary follows the blue line, reflects off the boundary at $\phi_2^*$, and returns to $\phi_1^*$. Note that the convex shape of the boundary near $\phi_1^*$ acts like a concave spherical mirror, focusing nearby trajectories and preventing them from diverging. This focusing effect creates a region of 2-hourglass trajectories near the 2-cycle.     (b) The points $\vec{r}_T(\phi_1^*)$ and $\vec{r}_T(\phi_2^*)$ are now shown for the more extreme values $\chi = 0.7$, $\gamma = 1.15$. As in (a), a trajectory starting at $\phi_1^*$ perpendicular to the boundary follows the blue path, reflects at $\phi_2^*$, and returns to $\phi_1^*$. The convex boundary near $\phi_1^*$ again acts like a concave spherical mirror, focusing nearby trajectories and preventing them from spreading apart. This structure again results in a region of 2-hourglass trajectories near the 2-cycle.}
    \label{fig:2cycle}
\end{figure}

In the case of a planar billiard table with an elliptical boundary, all hourglass trajectories envelope the two-cycle that coincides with the minor axis of the ellipse. In the case of a tilted cone, we would expect such a two cycle to exist if the unrolled boundary contains a region where the tangent to the boundary is parallel to the cut of the cone, as illustrated in Fig.~\ref{fig:2cycle}. A trajectory that begins perpendicular to the boundary at such a point will intersect the cut at a right angle. Since the cone boundary is symmetric about the line bisecting the unrolled cone (dashed line in Fig.\ref{fig:2cycle}(a)), the trajectory will reflect off the boundary again at a location where the boundary is also parallel to the cut, perpetually bouncing between those two points as shown in blue in Fig.~\ref{fig:2cycle}(a-b). If the boundary is convex, i.e. bulging outward, at these two points, this configuration will correspond to a two cycle, and will be surrounded by 2-hourglass trajectories, similar to those observed for an elliptical billiard table \cite{lynch2019integrable}, and displayed in Fig.~\ref{fig:ellipse}(b). 

 The tangent to the unrolled cone base is a continuous function. At $\phi = 0$, the tangent is at an angle of $\pi/2$ radians to the boundary, and at $\phi = \pi(1-\chi)$, i.e. half way around the cut of the cone, it is perpendicular to the line $\phi = \pi(1-\chi)$ (see Fig. \ref{fig:2cycle}(a)). Thus, by the intermediate value theorem, there must exist at least one point along the boundary where the tangent is perpendicular in the range $0$ to $\pi(1-\chi)$. For all tilt angles $\gamma \in (0, \pi/2 - \beta)$ and $\chi < 0.5$, this implies that there exists a point where the tangent is perpendicular to the line $\phi = \pi/2$ and thus parallel to the line $\phi =0$ and thus parallel to the cut. Thus there exists a trajectory that is perpendicular both to the boundary and the cut. Since this point will be between $\phi = 0$ and $\phi = \pi/2$ the boundary will be convex at this position. Thus, all cones with $\chi <0.5$  will contain a two-cycle. Because the boundary is convex near the two cycle, we expect the boundary to act as a focusing mechanism and nearby trajectories to remain nearby. Hence, the two-cycle will not be unstable and will be surrounded by 2-hourglass trajectories. Hence, there will be a region in which trajectories are not space-filling and thus the dynamical system cannot be ergodic.

For $\chi > 0.5$, the boundary is not necessarily parallel to the cut for all $\gamma$. However, by the same argument as above, if this condition holds for some $\gamma'$, it must also hold for all $\gamma >\gamma'$. We numerically determine the minimal $\gamma$ for which the boundary is parallel to the cut at some point and plot the threshold (in green) in Fig.~\ref{fig:boundary}(a).  This graph divides region II into two sub-regions $A$ and $B$, where in subregion $B$ a region of 2-hourglass trajectories must exist. As discussed in the following sections, region I cannot have ergodic dynamics due to the convexity of the boundary. Thus, the only region where we can hope to see ergodic and mixing trajectories is subregion II$A$ of Fig.\ref{fig:boundary}(a). An example of a cone with $\chi>0.5$ and a point where the boundary is parallel to the cut is shown in Fig.~\ref{fig:2cycle}(b). The subregion IIA in Fig.~\ref{fig:boundary}(a) could of course be further subdivided by looking for three-cycles, four-cycles etc..

\subsection{Rim Trajectories}\label{sec: rim}

Rim trajectories, for tilted cones, are those that visit the entire range of rim coordinates $\phi$ while remaining confined to a narrow range of $\theta$ values. Such rim trajectories are appear as blue lines for small $\gamma$ and $\chi$ in the \Poincare plots of Fig.\ref{fig:phasediagram}. They are analogs of the rim trajectories on an elliptical billiard table with $L_1\cdot L_2>0$ and represent small perturbations of trajectories with $\gamma =0$ (untilted cones), for which all trajectories are rim trajectories. The Kolmogorov-Arnold-Moser (KAM) theorem predicts that perturbations on an integrable billiard system will produce non-chaotic trajectories similar to those of the original system, such as the rim trajectories we observe for tilted cones, that are called KAM tori\cite{weissert1997kolmogorov}. As has been observed for other dynamical systems, we expect for small tilt angles these KAM tori to have similar properties to the trajectories on $\gamma =0$, such as uniformly covering the base\cite{shenker1982critical}. 

To better understand rim trajectories, it is helpful to define a new relative coordinate: 
\begin{align}
    \tilde \theta = \min[\theta_{\max}(\phi) - \theta, \theta-\theta_{\max}(\phi) +\pi] ,\label{eq.tildetheta}
\end{align}
 which ranges from $0$ to $\pi$. Here, $\theta_{\max}$ is the maximum allowed value of the slope variable $\theta$ for a given $\phi$ value, as shown on the top left \Poincare plot of Fig.~\ref{fig:phasediagram}. In \Poincare space, rim trajectories populate one-dimensional lines and exist only for initial conditions where $\tilde \theta$ is below a critical value. 
 % Note that all trajectories for $\gamma =0$ are rim trajectories. 

 For any billiard system with a convex boundary, a rim trajectory must exist~\cite{Tabachnikov1995}. We can understand this assertion by the following qualitative argument valid for both elliptical and conical billiards: Consider a trajectory with an initial value of $\tilde \theta\approx 0$, so that the trajectory is nearly tangent to the boundary.  Since $\tilde \theta$ can be taken to be infinitesimally small, it is always possible to choose initial conditions such that the trajectory closely hugs the boundary, thus sampling virtually all values of $\phi$ of the boundary. Thus no billiard trajectory in two dimensions with convex boundaries (such as that on an ellipse or an unrolled conical billiard) can truly reach all points of the surface or in \Poincare space. For a cone the later is true as a non-rim trajectory will inevitably be limited to only points in \Poincare space with sufficiently large $\tilde \theta$. Thus, ergodicity is ruled out. However, this restriction no longer holds for a cone with an unrolled cone base with a concave indentation. Thus, truly ergodic trajectories can only exist when $\gamma>\beta$, i.e. below the blue line in Fig.\ref{fig:boundary}.

 However, rim trajectories do not only constrain the transition to chaos and ergodicity. Similar to trajectories on a cone with $\gamma =0$, rim trajectories produce caustics, i.e. pile-ups of trajectories. These caustics again bound the approach of the trajectory to the  cone apex. For sufficiently small $\tilde \theta$, the rim trajectories in \Poincare space are nearly parallel to the trajectory along the boundary in \Poincare space,(see blue curvilinear plots in Fig.~\ref{fig:phasediagram}). In analogy with $\gamma=0$, and inspired by the rim trajectories for small $\tilde \theta$ on the blue plots in Fig.~\ref{fig:phasediagram}, we assume that the trajectory visits all $\phi$ values along the boundary with approximately equal probability and that $\tilde \theta$ is constant throughout the trajectory. This is a reasonable assumption for small values of $\gamma$ and , as discussed previously, the rim trajectories are analogous to KAM tori and represent small perturbations on rim trajectories for $\gamma =0$. Similar to the case when $\gamma =0$, we choose a particular $\phi = \phi_0$, draw a perpendicular to the caustic at $\phi_0$ as shown by the dashed white line in Fig.\ref{fig:caustic}(d). We are then interested in the probability density of a trajectory hitting this perpendicular at different distances $\tilde x$ from the caustic. For $\gamma =0$ this perpendicular can be extended to intersect the apex. For an untilted cone $\tilde x = x - R\sin(\theta)$ where $x$ is defined in Fig.\ref{fig:setupFirst}(a). Given that $\tilde \theta$ is small, $\tilde \phi$, defined as the angular distance between last intersection of the trajectory with the base prior to intersecting the perpendicular and $\phi_0$, is also small. Thus, we can approximate the distance $x(\tilde\phi)$:
 \begin{align}
     x(\tilde \phi) \approx  a_0(\chi, \gamma, \tilde \theta) + a_1(\chi, \gamma, \tilde \theta)\tilde \phi +a_2(\chi, \gamma, \tilde \theta)\tilde \phi^2
 \end{align}
Upon solving for $\tilde \phi(x)$, we find that
 \begin{align}
     &\tilde \phi(x) = \frac{-a_1 \pm \sqrt{a_1^2 - 4a_2(a_0 - x)}}{2 a_2}  \nonumber \\ &\qquad \implies d\tilde \phi = \frac{1}{\sqrt{a_1^2 - 4a_2(a_0 - x)}}dx .
 \end{align}
Provided we can approximate $P(\tilde \phi)$ by a constant, we find
\begin{align}
    P(x) \propto \frac{1}{\sqrt{x-x_{\text{crit}}}} ,
\end{align} 
as for an untilted cone, where $x_{\text{crit}}$ is a complicated function of $(\chi, \gamma, \tilde \theta)$. Thus, we expect that the divergence of the probability near the catacaustic of a rim trajectory to be similar to an untilted cone for most rim trajectories where the tilt $\gamma$ is small. 
In Fig.~\ref{fig:caustic}(b), we show numerically that, for $\chi = 0.2$, $\gamma=0.1$, the power law divergence seems to persist even for relatively large $\tilde \theta$. 

However, when the rim trajectories transition into hourglass trajectories and can no longer be approximated as either uniformly covering all values of $\phi$ or having a constant $\tilde \theta$, we expect this approximation to break down. An example of this break down is shown in Fig.~\ref{fig:caustic} (c,d). Specifically, while the average $d \approx 0.5$, the critical exponent seems to become weakly $\phi$-dependent for rim \Poincare sets that appear near hourglass \Poincare sets. 

Fig.~\ref{fig:caustic} (c,d) hints at the way rim trajectories evolve into hourglass trajectories as $\gamma$ and the initial slope between the boundary and the trajectory, $\tilde \theta$ (marked in Fig.\ref{fig:phasediagram}), increase. Unlike for small $\gamma$ and $\tilde \theta$, for large $\gamma$ and $\tilde \theta$, the critical exponent $d$, plotted in Fig.\ref{fig:caustic}(c), depends on  the polar angle,$\phi$, around the cone. If we plot the density distribution heatmap on the cone in this case, we see that for the $\phi$ values where the fitted critical exponent appears to be large, the density decreases to nearly zero (dark purple) near the cone boundary while having a high concentration (yellow) near the catacaustic. In contrast, in the $\phi$ regions with low critical exponent, such as those labeled in red on Fig.~\ref{fig:caustic} (d), we see a region of higher density near the boundary. As $\tilde \theta$ increases, these regions will become the only regions along the boundary reached by the trajectory. In the case of $\gamma = 0.3$, $\chi = 0.2$, and $\tilde \theta$ slightly greater than $0.9$, the trajectory will become of 5-hourglass type, i.e. surrounding a 5-cycle trajectory and composed of five separate regions in \Poincare space.

\subsection{Hourglass Trajectories}\label{Sec:box}
Generic n-hourglass trajectories on tilted cones only partially populate the base with their reflections. As discussed in the introduction to this section, they appear on \Poincare maps as collections of n loops. Trajectories bounce from loop to loop, where each loop spans a disjoint range of $\phi$ values, as shown in Fig.~\ref{fig:3 types}(c).  
In the remainder of this section, we discuss three qualitatively different sources of hourglass trajectories. The first of these sources has been previously discussed in Sec.~\ref{sec:boundary}.
For sufficiently high values of $\chi$ a 2-hourglass trajectory which is an analog of the 2-hourglass trajectory on an elliptical billiard table appears on a tilted cone, the origin of which is summarized in Fig.\ref{fig:2cycle}. This type of hourglass trajectory formation occurs in region IIA of Fig.~\ref{fig:boundary}. 
\begin{figure}[t!]
    \centering
    \includegraphics[width=\columnwidth]{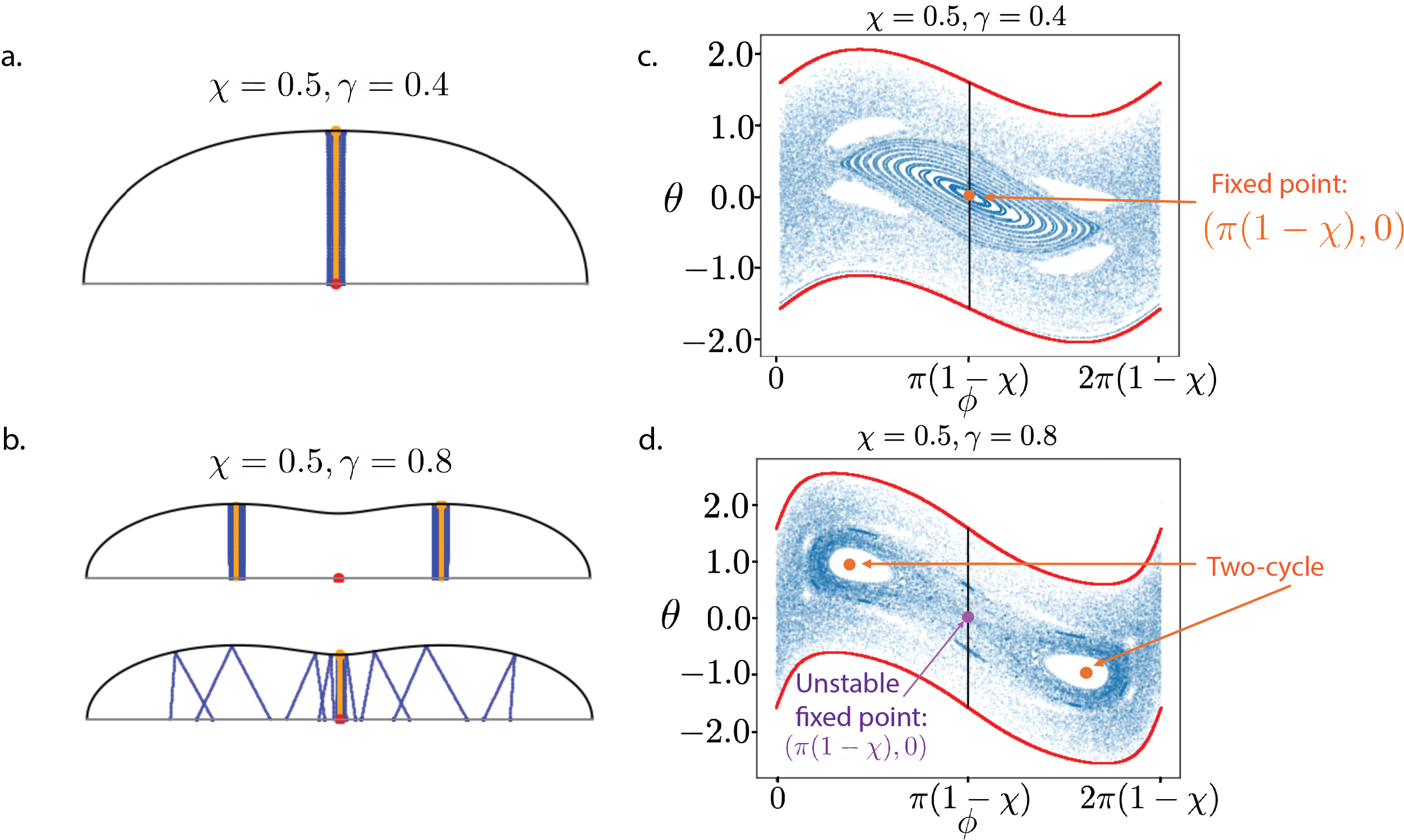}
    \caption{Visual representation of a Hopf Bifurcation-like transition leading from 1-hourglass to 2-hourglass trajectories in both real space of unrolled cones ((a) and (b)) and \Poincare space ((c) and (d)). One center fixed point transitions to an unstable fixed point and stable limit cycle. Panels (a) and (b) show trajectories close to the fixed points (panel (a) and bottom of panel (b)) and 2-cycle (top of panel (b)) for $\chi =0.5$ and $\gamma=0.4$ and $0.8$ respectively. Note how in panel (a) and top of panel (b) the trajectory (marked in blue) remains close to its initial path (marked in orange). The bottom part of panel (b) shows the instability of the trajectory in (a) caused by the concave portion of the boundary centered on $\phi  =\pi(1-\chi)$. Panels (c) and (d) show the Poincare map for $\chi =0.5$ and $\gamma=0.4$ and $0.8$ respectively with the fixed points and limit cycles shown in panels (a) and (b) labeled. Initial conditions are taken along the line (in black) $\phi = \pi(1-\chi)$.}.
    \label{fig:change in fixed points}
\end{figure}

As an example of a transition within this region, we present in Fig.\ref{fig:change in fixed points} a bifurcation occurring for $\chi = 0.5$, as the concavity threshold is crossed with increasing $\gamma$. For $\gamma$ below the concavity threshold, of the unrolled cone boundary, in the \Poincare map, a region of 1-hourglass trajectories appears as a loop surrounding the point $\phi = \pi(1-\chi)$ and $\theta = 0$, with $\chi = 0.5$. This point corresponds to a trajectory that begins at the boundary position closest to the apex and shoots straight at the apex. Slightly shifting the initial value of $\phi$ sends a trajectory to the cut of the cone at nearly a right angle. As such, it will emerge on the other side of the cut also at nearly a right angle and at the boundary nearly equidistant from $\phi = \pi(1-\chi)$ as the initial trajectory.  Note that trajectories at $\theta=0$ (ones shooting directly at the apex) are not well defined in general, since trajectories approaching them from different directions will have different limits (see Sec.~\ref{sec:integrable}). However, in the case of the trajectory starting at $\phi = \pi(1-\chi)$ and $\theta = 0$, trajectories from all nearby initial conditions, regardless on which side of the point they start, will form increasingly smaller loops on the \Poincare map, and will approach a fixed point as initial conditions are taken closer and closer to $\phi = \pi(1-\chi)$ and $\theta = 0$. This is a special feature for cone parameters with $\chi = 0.5$ and $\gamma < \beta =\pi/6$. Thus,  on the \Poincare map, the point at $\phi = \pi(1-\chi)$ and $\theta = 0$ is well defined, a fixed point, and not unstable. In fact, as illustrated in the \Poincare diagram in Fig.\ref{fig:change in fixed points}(c), any trajectory starting on one of the loops surrounding the fixed point will remain at a finite distance from the fixed point, typically jumping around the loop in a quasi-periodic manner. This behavior is analogous to  continuous dynamical systems which exhibit a transition from a fixed point to a fixed point called a center in which continuous trajectories remain at a finite distance about a fixed point forming nested loops surrounding it~\cite{strogatz2018nonlinear}. 

\begin{figure*}[t!]
    \centering
    \includegraphics[width=\linewidth]{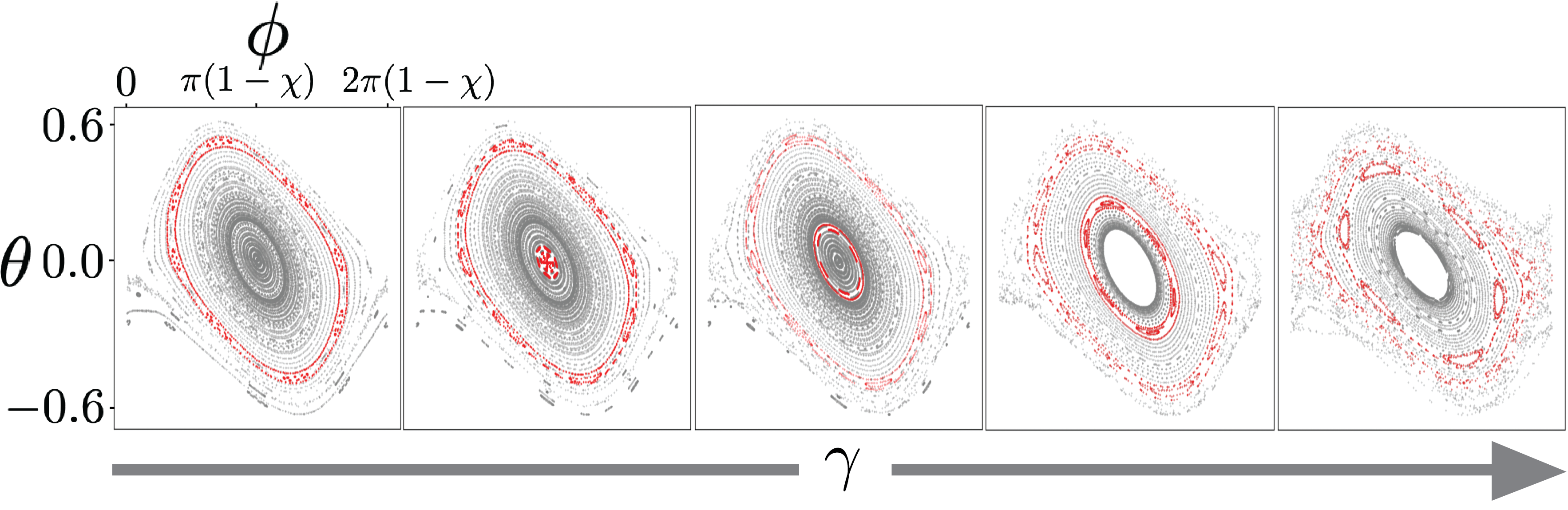}
    \caption{Evolution of the central region of Poincare space for $\chi = 0.5$ and from left to right, $\gamma = 0.114, 0.115, 0.118,0.129, 0.169$. Ranges of $\phi$ and $\theta$ shown in all the panels of the figure are labeled for the leftmost panel. A few trajectories near two rings of hourglass loops are marked in red. As $\gamma$ increases we see the outer ring merge with the mixed trajectory, and the inner ring appear from the original fixed point and move out to the bulk of the region of 1-hourglass trajectories. Initial conditions on all panels are taken along the vertical line $\phi = \pi(1-\chi)$. However a higher density of initial conditions is taken near the highlighted  (in red) regions in order to better see the structure of trajectories near them. Some initial conditions near the horizontal line $\theta = 0$ are omitted for the two rightmost panels for clarity.}.
    \label{fig:bif2}
\end{figure*}

However, for the $\chi = =0.5$ cone in Fig.\ref{fig:bif2}, when the boundary becomes partially concave, Fig.\ref{fig:bif2}(b), it does so at precisely the $\phi$ value where we see the fixed point ($\phi = \pi(1-\chi)$). At the onset of concavity, the slope of the tangent line at $\phi = \pi(1-\chi)$ remains the same and the trajectory with the initial condition $(\phi = \pi (1-\chi), \theta = 0)$  remains a fixed point, as shown in orange the lower part of Fig.~\ref{fig:change in fixed points}(b). However, it is now unstable, with the concave segment of the boundary acting as a convex mirror in optics, sending trajectories near the fixed point farther away from it with every bounce, as shown in blue on the lower part of Fig.~\ref{fig:change in fixed points}(b) for a sample trajectory. As this happens two new fixed points are formed at the two points where the boundary is parallel to the cut of the cone. These are precisely the two points that form the 2-cycle described in Fig.\ref{fig:2cycle}. In real space, the two cycle corresponding to this trajectory is shown in Fig.~\ref{fig:change in fixed points}(b). We can extend the analogy with conventional dynamical systems by noticing the similarity between this transition and a Supercritical Hopf Bifurcation where a stable fixed point becomes an unstable fixed point with a stable limit cycle forming around it~\cite{strogatz2018nonlinear}.

\begin{figure*}[t!]
    \centering
    \includegraphics[width=\linewidth]{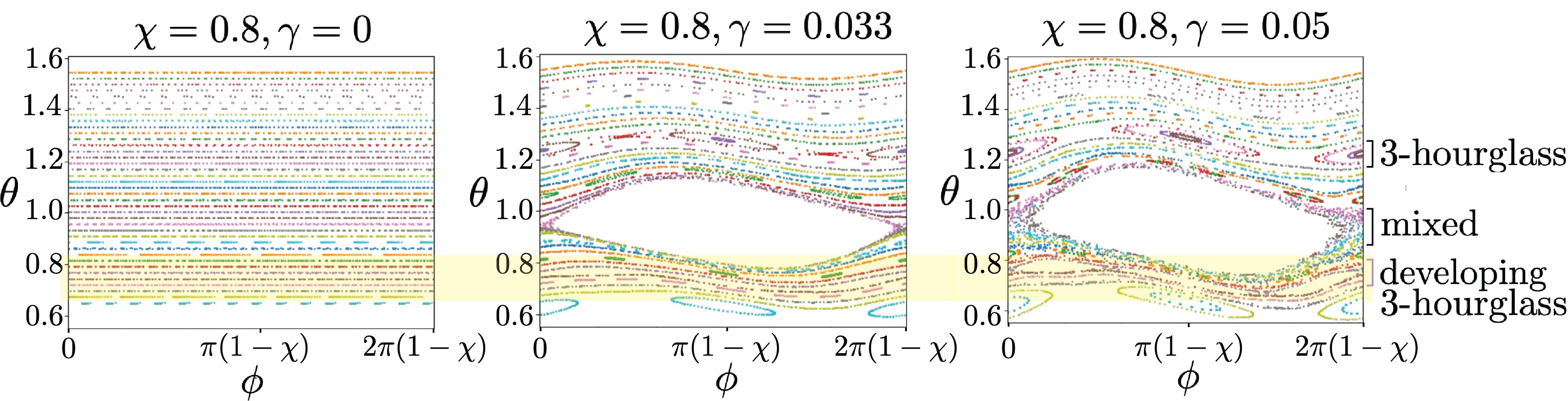}
    \caption{Appearance of mixed and hourglass trajectories with increasing $\gamma$ for $\chi =0.8$ on a segment of a Poincare map from $\theta = 0.6$ to $1.6$. Here, initial conditions are chosen at a range of $\theta$ values for $\phi =0$. The large white region in the middle and right panel is a region in which hourglass trajectories which never hit $\phi=0$ would appear if different initial conditions, e.g., along a vertical line with $\phi = \pi(1-\chi)$, were chosen.  Different colors mark \Poincare sequences with different initial conditions.  The region in which mixed trajectories appear is labeled for $\gamma = 0.05$, on the right hand side of the figure. A region of 3-hourglass trajectories, i.e., three families of nested loops, near $\theta = 1.2$ is labeled right hand side of the figure as well. Note how the red trajectory near $\theta = 1.2$ and the pink and gray trajectories near $\theta = 0.9$ in the middle panel occur on either side of the hourglass trajectory. As $\gamma$ increases, the region in space spanned by these trajectories increases, leading to a mixed trajectory. In the rightmost panel we see how the pink and gray trajectories become the first mixed trajectories to appear for this range of initial conditions.  Another 3-hourglass trajectory begins forming in the yellow highlighted region. Notice how the brown and purple trajectories in the yellow highlighted region begin parallel for $\gamma =0$, appear to approach each other at three separate $\phi$ values at $\gamma = 0.033$, and pinch off forming a 3-hourglass  trajectory at $\gamma = 0.05$. }.
    \label{fig:transition}
\end{figure*}

The second mechanism for formation of hourglass trajectories occurs as $\gamma$ slightly increases from zero for a fixed $\chi$. As we summarized in Fig.\ref{fig:setup}, at $\gamma=0$ all trajectories are rim trajectories. When $\gamma$ increases, nearby rim trajectories distort in such a way that they approach each other in \Poincare space and pinch off, creating an $n$-loop hourglass trajectory surrounding an $n$-periodic fixed point as $\gamma$ increases from zero. Such a process is visible in the purple and gray trajectories in the highlighted region of Fig.~\ref{fig:transition}. Note that in Fig.~\ref{fig:transition} the vertical line of initial conditions are chosen at $\phi = 0$ in order to emphasize this transition. Because of these initial conditions, 1-hourglass trajectories which form near the fixed point at $\phi = \pi (1-\chi)$ will not appear on the figure as these trajectories never reach $\phi=0$. This is the origin of the large white space in the center of the second and third panels. 

Another interesting transition not visible in many less symmetric area preserving dynamical systems, is the appearance of hourglass trajectory loops that form a ring which emerges from a singular fixed point as shown in Fig. \ref{fig:bif2}. As $\gamma$ increases, the ring of loops moves further away from the original fixed point until it merges into a more chaotic mixed \Poincare sequence. In this way the mixed trajectory region grows in area, and gains islands of hourglass trajectories inside of it, giving it a fractal like structure. This kind of transformation has been observed in other highly symmetric billiard problems such as on an oval~\cite{berry1981regularity}.

\subsection{Mixed Trajectories}\label{sec: mixed}
While both hourglass and rim trajectories can be classified as quasi periodic, the emergence of mixed trajectories signifies the beginning of a transition from integrability to chaos. The Kolmogorov-Arnold-Moser (KAM) theory, for Hamiltonian area preserving systems such as this one, suggests that as this transition occurs the set of remaining quasi periodic trajectories will be a fractal in phase space (similar to a Cantor set) with shrinking fractal dimension as the system is further perturbed from the integrable limit (in this case the limit of $\gamma=0$)~\cite{weissert1997kolmogorov, dumas2014kam}. Correspondingly, the non-integrable trajectories themselves will form a fractal with growing fractal dimension as the system is further perturbed from the integrable limit.

One way to track this transition would be to study the fractal dimension of mixed trajectories. For the purpose of this analysis we will focus on $\chi = 0.7$ and $0.8$ for which one mixed trajectory forms and takes up increasingly more space as $\gamma$ increases from zero. As discussed in the Introduction, our system naturally exists in a three dimensional space (two position dimensions, and one corresponding to the orientation of the velocity). However, by tracking the intersection of the conical billiard trajectories with the base of the cone, we isolate a two-dimensional cross section of the phase space, corresponding to the intersection point along the cone base, $\phi$, and the trajectory angle at this point,$\theta$, (See Fig.\ref{fig:poincareMap}).  We would naturally expect an ergodic trajectory to take up the whole cross section of this two-dimensional parameter space, and an integrable trajectory to occupy a one-dimensional subspace. The \Poincare maps summarized in Fig.\ref{fig:phasediagram} distort this two-dimensional cross section in phase space so that it can be represented in a flat plane, $\mathbb R^2$.

 A commonly used way to numerically determine fractal dimensions of sets of points is to calculate the correlation dimension, which is a lower bound for the fractal dimension and often coincides with the fractal dimension, and is significantly easier to calculate in a complicated system~\cite{strogatz2018nonlinear}~\cite{grassberger1983measuring}. This is the approach we will take here, as we study numerically the distribution of chaotic trajectories in \Poincare space.
 
\begin{figure}[t!]
    \centering
    \includegraphics[width=\columnwidth]{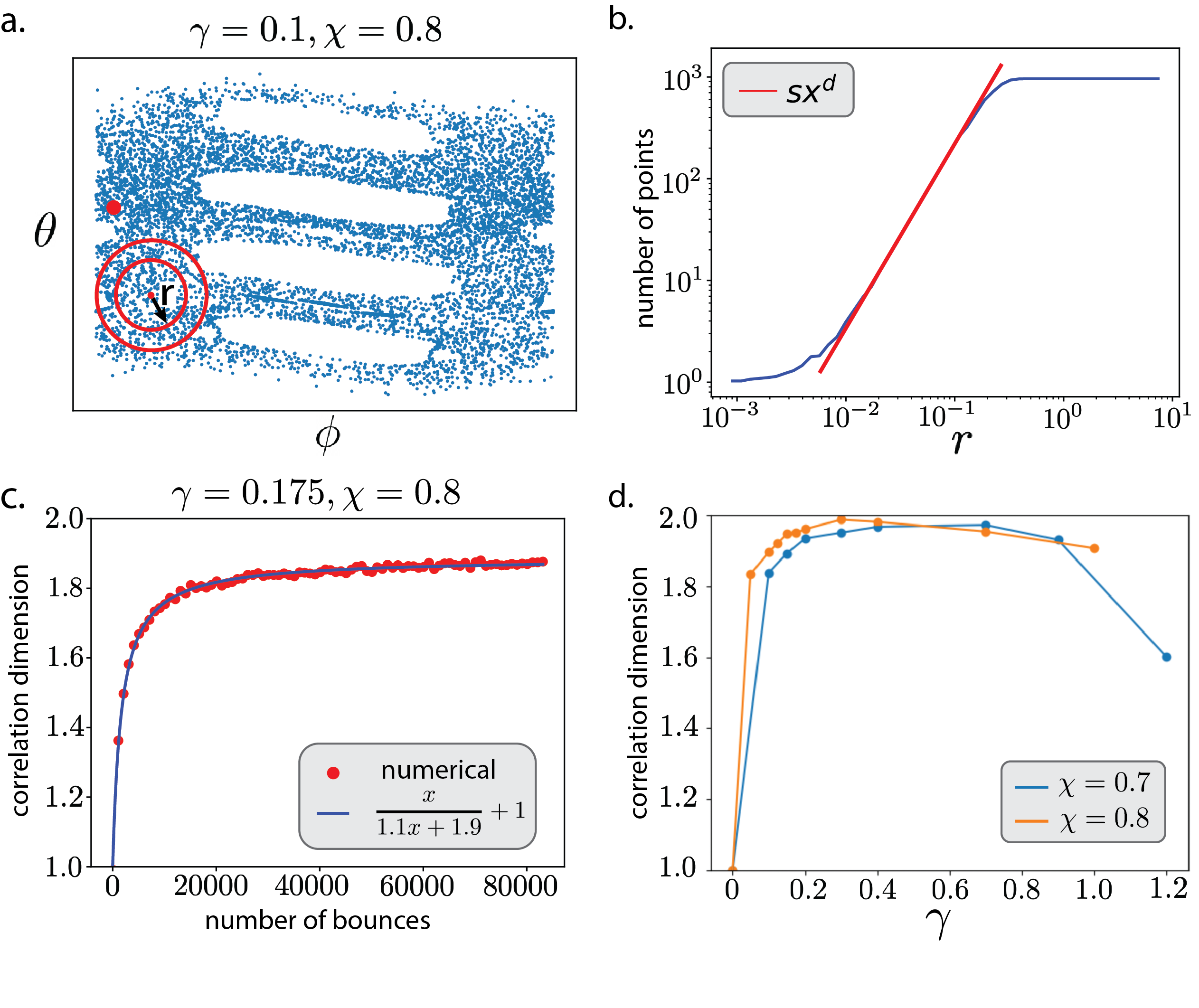}
    \caption{Calculation of correlation dimension for mixed trajectories in \Poincare space. Panel (a) shows the first 8000 bounces of a mixed trajectory for $\gamma = 0.1$ and $\chi = 0.8$, with the initial condition marked with a large red dot. For a set of randomly selected points in the trajectory (such as the small red point marked on the figure) we draw circles of radius $r$ and count the number of points encircled by them and average across the other randomly chosen centers. We then plot the  number as a function of $r$ on a log-log plot in panel (b). We next find a best fit power law segment of the curve, to obtain the correlation dimension, $d$. (c) For each $\gamma$, we truncate the trajectory at different numbers of bounces and perform the above procedure at each of them, plotting $d$ as a function of the number of bounces. We then fit the data for each $\gamma$ to the blue line of the form $y=\frac{x}{a(\gamma) x +b(\gamma)}+1$ as discussed in the main text. (d) We plot the correlation dimension, $d(\gamma) = \lim_{x\rightarrow \infty} y(x) = 1+\frac{1}{a(\gamma)}$. Note how the maximum achieved correlation dimension for both $\gamma=0.8$ and $\gamma=0.7$ is nearly two, indicating ergodicity for the relevant region in \Poincare space.}.
    \label{fig:8 chaos}
\end{figure}

 The correlation dimension is the power with which the number of points inside a circle of radius $r$, in, say, \Poincare space, grows with the radius~\cite{strogatz2018nonlinear}. Note that this dimension should not be affected by mild distortions of the shape inside which the number of points is calculated. For example, if we assume that the fractal dimension is constant in a region of space, and we determine that the number of points inside a circle in that region of radius $r$ scales as $r^d$, then the number of points inside an ellipse with major axis $n*r$ and minor axis $r$ would scale as $r^d*n^k$ where $k$ is a constant. Provided $n$ remains constant in our scaling procedure, we expect that changing the circle of sampled points to an ellipse will not alter the power law that connects the number of points in a region and the size of the region. For conical billiards it seems reasonable to assume that, even though our \Poincare map is a distorted version of phase space, as long as regions of sufficiently small radii are taken, the correlation dimension of a set of points in \Poincare space approximates its analog in the original phase space.

 Our method of determining the correlation dimension in \Poincare space is illustrated in Fig.~\ref{fig:8 chaos}(a-c). For a given cone tilt angle, $\gamma$, we calculate $d$ for circles of radius $r$, centered on a randomly chosen set of one hundred points of the mixed trajectory region of \Poincare space independently, while respecting the periodic boundary conditions on the $\phi$ axis for plots like those in Fig.~\ref{fig:phasediagram}. We then average the correlation dimension $d$ for all such points for the given value of $\gamma$ \cite{strogatz2018nonlinear, grassberger1983measuring}. To check that this averaging scheme is accurate, we repeat it for multiple realizations of randomly chosen points, and find nearly identical results. 

As the radius $r$ becomes comparable to the dimensions of the cone, we would expect the  number of points inside a disk of radius $r$ in \Poincare space and $r$ to no longer be approximated well by a power law for two reasons,: First, for large radii the map between phase space and \Poincare space is not well approximated as a linear distortion. Second, once we reach the boundary of \Poincare space increasing the radius will no longer engulf any more points. Deviations from a power law will also occur if the radius so small it is comparable to the distance between the points for a given number of bounces off the cone base. We can see this effect in Fig.~\ref{fig:8 chaos}(b).

 Thus, the correlation dimension is only relevant for intermediate radii of the circles in Fig.\ref{fig:8 chaos}. In this regime, the number of points encircled by a ring of radius $r$ grows as a power law with $r$, as shown for example in Fig.~\ref{fig:8 chaos}(b). If the trajectory uniformly covered \Poincare space, one would expect the correlation dimension calculated in this way to grow with the number of points in \Poincare space (number of bounces the conical billiard trajectory experiences) until it approaches the full dimension of the two-dimensional \Poincare space. However, if the mixed trajectory is a fractal, one would expect the correlation dimension to asymptote to some value $d<2$. 
 
  To estimate the asymptotic correlation dimension, $d$, for each $\gamma$, we track  this quantity as a function of the number of bounces, $x$, in the trajectory. To model the asymptotes, we fit the x and y coordinates of Fig.$\ref{fig:8 chaos}$(c) to
 \begin{align}
     y=\frac{x}{a(\gamma) x +b(\gamma)}+1
 \end{align}
 and record $d=\frac{1}{a(\gamma)} +1$ as the correlation dimension which seems to fit the data quite well. 

Fig.\ref{fig:8 chaos}(d) suggests that for $\chi =0.8$ and $\chi = 0.7$, the correlation dimension of the space taken up by a mixed trajectory is at first monotonically increasing, and thus the correlation dimension characterizing integrable elliptical and hourglass trajectories is decreasing, as suggested by the KAM theorem. However, for larger $\gamma$, the system begins to resemble the limit of flat elliptical billiards, and hourglass trajectories reappear, leading to an eventual drop in correlation dimension of the mixed trajectories as predicted in Appendix~\ref{sec:Appendix heuristic}. In this regime, trajectories that appear to be mixed evantually approach hourglass trajectories after long times, leading to a correlation dimension of one after many bounces. However, before dropping to one, the correlation dimension, $d$, grows with the number of bounces of the trajectory. To create the plot on Fig.~\ref{fig:8 chaos}(d), we calculated the asymptote for the mixed trajectory only in the region where it increases with the number of bounces, because we are concerned about numerical error when the number of bounces is large.

Note that for both $\chi = 0.8$ and $\chi = 0.7$ the maximum correlation dimension in \Poincare space is approximately 2, is reached for $\gamma > \beta = \sin^{-1}(1-\chi)$ for both values of $\chi$ as expected for systems which are approximately ergodic.

\section{Emergence of Ergodicity}\label{sec:chaos}
Having understood the three types of trajectories on tilted cones as a function of $\chi$ and $\gamma$, we now analyze the seemingly ergodic behavior observed in region IIA of Fig.\ref{fig:boundary} and suggested in Sec.\ref{sec: mixed}. In this section, we will show that mixed trajectories are chaotic, and as such, when they become space filling (have fractal dimension of two, in most of \Poincare space), exhibit chaotic dynamics. We also show that in some regions of $\chi$ and $\gamma$ space, almost all trajectories exhibit a strong mixing property and are thus ergodic.

Chaotic systems are deterministic, but with sensitive dependence on initial conditions; almost all trajectories exhibit long-term aperiodic behavior~\cite{strogatz2018nonlinear}. Our system is inherently deterministic. As shown in the previous section, as $\chi$ and $\gamma$ approach region IIA in Fig.\ref{fig:boundary}, the fractal dimension in \Poincare space of hourglass and rim trajectories decreases, and the majority of trajectories become mixed and aperiodic. Thus, to show that in this region of  $\chi$ and $\gamma$ space the system is indeed chaotic, we must show sensitive dependence on initial conditions. One way to verify sensitive dependence on initial conditions is to determine whether the distance between nearby trajectories on the unrolled cone diverges exponentially over time. The rate of this divergence is quantified by the Lyapunov exponent, where a positive Lyapunov exponent indicates sensitive dependence on initial conditions~\cite{strogatz2018nonlinear}. For our billiard problem two slightly misaligned trajectories will diverge linearly until they intersect the cone base and bounce off of it. Hence we compare the distance between nearby trajectories at the moment of impact.

\begin{figure}[t!]
    \centering
    \includegraphics[width=\columnwidth]{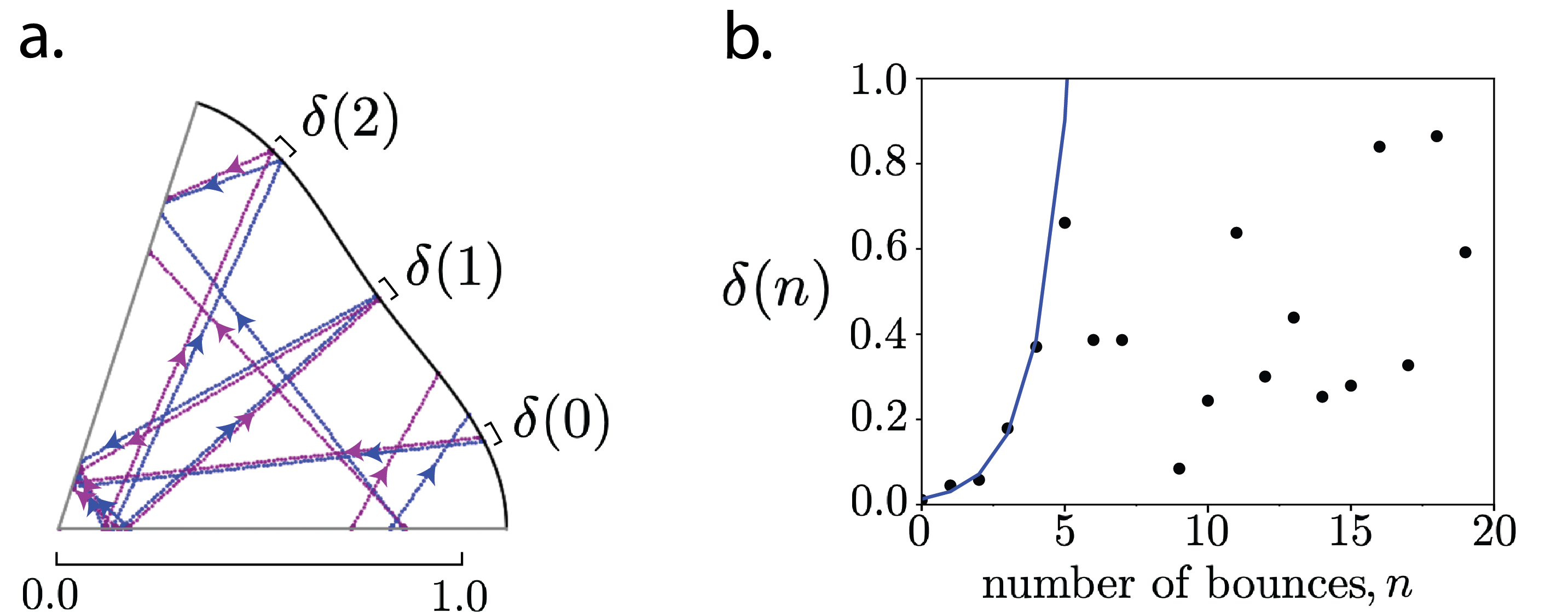}
    \caption{(a) Nearby purple and blue trajectories on a cone with $\chi = 0.8$ and $\gamma = 0.3$, which start at $\theta = 0.1$ and $\phi = \frac{\pi}{2}(1-\chi) + .105 \pm .05$. On the bottom of the panel note the scale bar (this same scale is used in panel (b)); we are working with cones whose macroscopic dimensions are of order unity. Over time the two trajectories move further apart from each other along the cone base, as measured by $\delta(n)$. (b)The real space distance between the two trajectories calculated at every bounce. Note how for large numbers of bounces the trajectories no longer seem to be correlated. However, for small numbers of bounces, the distance seems to grow exponentially. The blue line is an exponential fit for the first five bounces with $\delta(n) \approx e^{0.84n}$, so the Lyapunov exponent here is $\approx 0.84$.}.
    \label{fig:lyapunov}
\end{figure}

We measured the Lyapunov exponent for two nearby trajectories on a cone with $\chi = 0.8$ and $\gamma = 0.3$, which start at $\theta = 0.1$ and $\phi = \frac{\pi}{2}(1-\chi) + .105 \pm .05$. This cone was chosen because of the particularly high correlation dimension of the mixed trajectory, see Fig.~\ref{fig:8 chaos}(d). On the unrolled cone, the trajectories for their first three bounces are shown in Fig.\ref{fig:lyapunov}(a). At the $n^{th}$ bounce, we record the real space distance, $\delta(n)$ between the two trajectories. We then plot these  distances as a function of the number of bounces. Note that when the $\delta(n)$ becomes comparable to half the boundary length (roughly 0.4 in this case), the trajectories can no longer move away from each other exponentially due to finite size effects as can be see in Fig.~\ref{fig:lyapunov}(b). Prior to this point, we fit $\delta(n)$ to an exponential, also shown in  Fig.~\ref{fig:lyapunov}(b). The power of the exponential growth is the Lyapunov exponent. From Fig.~\ref{fig:lyapunov}(b) we find $\delta(n) \approx e^{0.84n}$, i.e, a positive Lyapunov exponent of $0.84$ as expected for a chaotic system.  Note that, in the non-ergodic region, we expect that the results are dependent on the initial condition. However, in the region where the fractal dimension of mixed trajectories approaches $2$, we expect the Lyapunov exponent to be nearly the same for all of space as has been proven for other chaotic billiard systems \cite{datseris2019estimating}. 

In addition to a sensitive dependence on initial conditions, we find that conical billiard trajectories in the chaotic regime display an even more powerful property: strong mixing. To define a strong mixing criterion for our conical billiard system, consider two regions of $\theta$ and $\phi$ values in \Poincare space, denoted by $A$ and $B$. A billiard transformation $T(A)$ that maps a point in \Poincare space to its next point in \Poincare space along the billiard trajectory, is strongly mixing if the following condition holds:
\begin{equation}\label{eq. mixing definition}
    \lim_{n\rightarrow \infty} \mu(T^{-n}(A) \cap B) = \mu(A)\mu(B) ,
\end{equation}
where $\mu(C)$ is the area in \Poincare space occupied by the set $C$ and $C \cap D$ means the intersection of the sets $C$ and $D$. The $T^{-n}(A)$ operation appears because it is conventional to focus on where the points in $A$ came from in this reservable system. Eq.~\ref{eq. mixing definition} means that after sufficiently many iterations, it is impossible to determine from where in \Poincare space a trajectory originated, as it is statistically independent of its initial location ~\cite{Tabachnikov1995}. 

To better understand this statement, which is a stronger property than mere ergodicity, assume the set $T^{-n}(A)$ is a small region of area $\tilde A$ of adjacent points in the \Poincare space. Suppose, for concreteness, $\tilde A = 1/10$. Let $B$ be a distinct region with the same area as $T^{-n}(A)$. Since the billiard transformation $T$ is volume preserving~\cite{Tabachnikov1995}, we know that $\mu(A) = \mu(T^{-n}(A))$. Thus $\mu(A)\mu(B) = 1/100$. If the system is strongly mixing, Eq.~\ref{eq. mixing definition} implies that $1/10$th of $\mu(T^{-n}(A))$ will intersect $B$, meaning $\mu(T^{-n}(A))$ will be uniformly distributed~\cite{wikiMix}.

\begin{figure}[t!]
    \centering
    \includegraphics[width=\columnwidth]{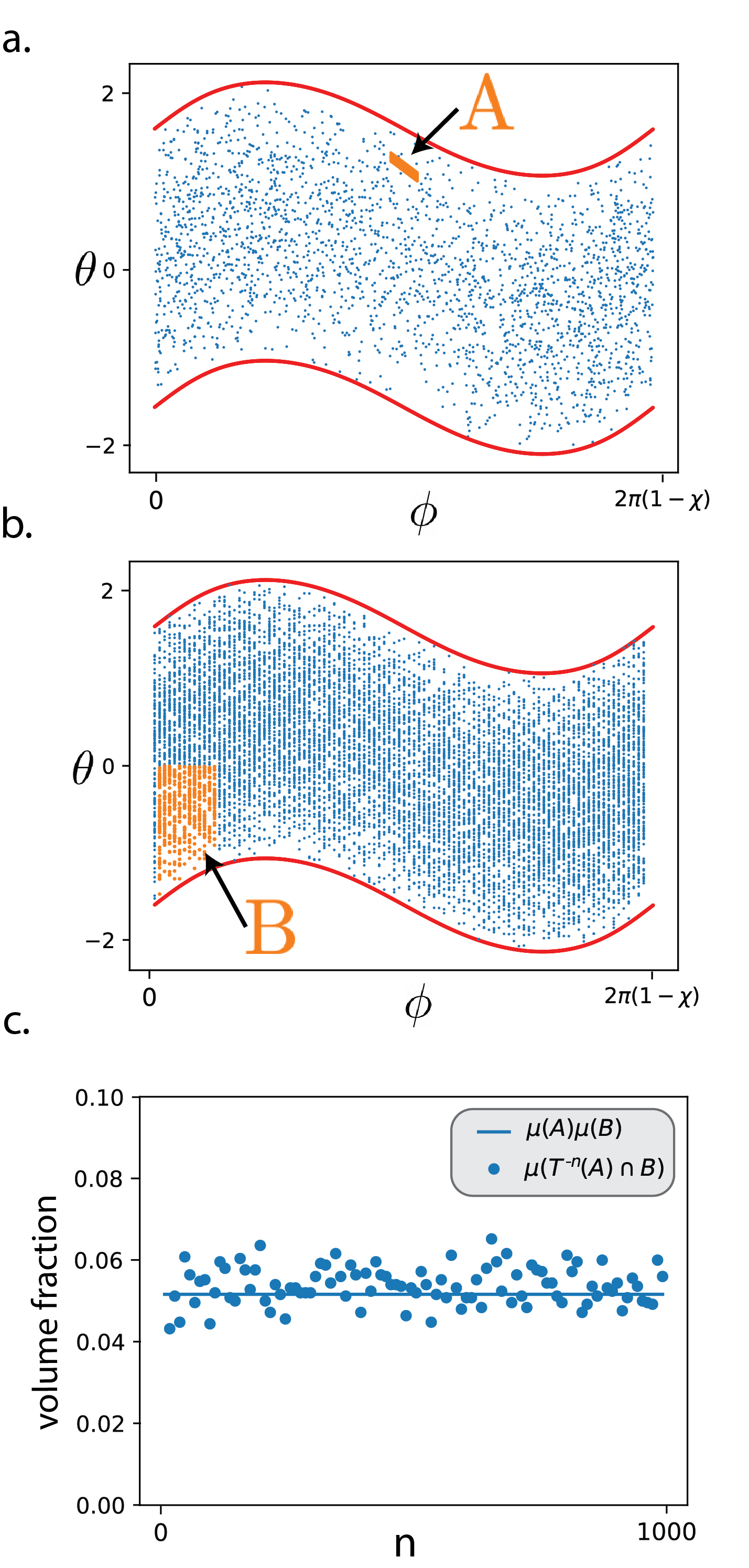}
    \caption{ (a) In blue, 2500 points that make up $T^{-n}(A)$ for $n = 1000$. In orange, $2500$ more densely packed points that make up $A$.(b) In blue, random distribution of $\phi$ and $\theta$ points. In orange, an area containing the set $B$ is highlighted. (c) For the choice of $B$ and $A$ in panels (a) and (b) the volume fraction of $T^{-n}(A)\cap B$ for different choices of $n$, blue dots, and the product of area fractions of $A$ and $B$, solid blue line.}.
    \label{fig:9 chaos}
\end{figure}

To check that our system  can beis strongly mixing in the sense discussed above, we choose test $\gamma$ and $\chi$ values deep in the chaotic regime, i.e. in the teal-colored region of Fig.\ref{fig:phasediagram}. We focus here on $\gamma = 0.5, \chi = 0.7$, but find similar results for other chaotic values of $\chi$ and $\gamma$, such as $\gamma = 0.3$, $\chi = 0.8$. We take a small region of \Poincare space for this particular $\gamma$ and $\chi$ (marked in orange on Fig.~\ref{fig:9 chaos}(a)) which has the area of $\approx 0.01$ and evenly populate it with 2500 initial condition points. This will be our set $A$. We then run the simulation for 1000 iterations. Since running the trajectories forwards in time is the same as running them backwards in time with the velocities reversed, the resultant set in \Poincare space produces $T^{-n}(A)$, shown in blue on Fig.~\ref{fig:9 chaos}(a). 
We determine what a random distribution of geodesic segments would look like in \Poincare space, Fig.~\ref{fig:9 chaos}(b), using a procedure delineated in Appendix~\ref{sec: appendix distribution}. We then choose set $B$, marked in orange on , Fig.~\ref{fig:9 chaos}(b). 

As a proxy for the volume of a set $B$ in \Poincare space, we measure the ratio of the points in set $B$ to the total number of randomly distributed points. Because we are looking at ratios, we only need the relative numbers of points in different regions of the \Poincare map, provided the number of points is large.

We choose $B$  to be sufficiently large so that a large number of points of the model \Poincare space are inside. Similarly, we determine the area of $\mu(T^{-n}(A) \cap B)$ by taking the number of points of $\mu(T^{-n}(A))$ inside the region $B$ and dividing it by the total number of points of $A$. As illustrated for a specific region $B$ in Fig.~\ref{fig:9 chaos}, we find that for even very small values of $n$, $\mu(T^{-n}(A) \cap B) \approx  \mu(A)\mu(B)$ so evidently the billiard map is strongly mixing. Strong mixing implies ergodicity, which means that for these values of $\chi$ and $\gamma$ our system is ergodic \cite{tabanov1994separatrices}. 

\section{Conclusion}\label{sec:conclusion}

In this paper, we introduced and extensively investigated dynamical billiards on the surface of a cone with a tilted base. Upon varying the cone angle $\beta$, corresponding to a deficit angle $2\pi \chi = 2\pi(1-\sin(\beta))$, and tilt angle $\gamma$, we identified three distinct types of trajectories with associated \Poincare map for conical billiards: rim, hourglass, and mixed. By combining numerics and analytics, we  delineated three distinct regions in $(\gamma, \chi)$ -parameter space: Region I, where \Poincare space consists of rim, hourglass, and mixed trajectories; Region IIB, where \Poincare space consists of only hourglass and mixed trajectories; and Region IIA, in which we find choices of $\gamma$ and $\chi$ for which almost all trajectories are strongly mixing. We then studied the $\gamma$-dependence of the transitions between these three regions and presented evidence that the correlation dimension of mixed trajectories initially grows and then decreases with increasing $\gamma$. We also developed a scheme for identifying \emph{strongly mixing} trajectories~\cite{adams2023ergodicity}.

Our work provides additional insights into billiard problems that contains both corners and curved boundaries. We were able to systematically analyze the role of varying the corner angle (changing $\chi = 1-\sin\beta$, where $\beta$ is the cone half-angle) and varying the curvature of the boundary (changing the tilt angle $\gamma$). In doing so, we also demonstrated the influence of Gaussian curvature at a distant cone apex on ballistic particle motion on a surface even in the absence of long range interactions with the apex or the tilted cone boundaries. Furthermore, we were able to show that a dynamical billiard on a surface with exclusively convex and positive Gaussian curvature in three dimensions can still exhibit ergodic behavior in certain parameter regimes.

We anticipate that similar ideas about strong mixing will be relevant with to the addition of soft boundary interactions or long-range interactions~\cite{lynch2019integrable}, and perhaps to related problems on cones with topological defects in active nematics~\cite{vafa2024periodic}. In addition, we believe the results of this paper can be viewed as a first step in understanding the behavior of crystal defects, such as dislocations which can move via glide motion along geodesic lattice lines on surfaces of cones\cite{amir2013theory}. 

A particularly intriguing feature of this system is that by tuning $\chi$ and $\gamma$, nearly all points in $(\theta,\phi)$ \Poincare space describing conical line segments in between bounces can be placed at the edge between chaotic and integrable dynamics. Thus this work highlights the potential of conical billiards as a model system for exploring intriguing problems inspired by neural networks at the  ``edge of chaos''~\cite{bertschinger2004real, toyoizumi2011beyond}.

Future directions for this work include extensions to ballistic dynamics on surfaces with a negative delta-function of Gaussian curvature, where we expect negative curvature to enhance the extent of chaotic behavior. Another natural extension of this work involves investigating the semiclassical billiard regime. The well-characterized yet complex behavior of classical conical billiards, combined with the relatively simple geometric structure of this problem, provides an ideal framework for exploring quantum scars both theoretically and experimentally~\cite{lepore2003diffraction,ge2024direct}. For example, it would be interesting to manufacture cones whose walls act as conical waveguides and inject laser light. 

To summarize, we hope that conical billiards will serve as a versatile and stimulating model system for a wide variety of complex behaviors which have been observed in a range of dynamical systems. By presenting a system that is experimentally accessible (for example, by sending laser beams through the walls of a cone created using fiber-optic-like technology), conceptually simple, yet able to capture a variety of complex behaviors, we hope this work can stimulate further research into both classical and quantum chaos.

\begin{acknowledgements}
The authors would like to thank Grace Zhang for discussions on cone geodesics, Jonathan Bauermann for discussions on random distributions of segments in a confined space and suggesting the elliptical cone boundary, and Farzan Vafa for helpful comments on the manuscript. We also thank Curtis McMullen and Jayadev Athreya for helpful insights on elliptical billiards. L. Braverman would also like to acknowledge NSF GRFP Grant No.DGE 2140743. This research was funded in part by the Harvard University Materials Research Science and Engineering Center through NSF grant DMR-2011754. 

\end{acknowledgements}

\appendix
\section{Analytic expression for Conical Billiards}\label{sec:Appendix heuristic}
 Because conical billiards are a deterministic system, one can determine the location at any given time of a particle following a single trajectory given its initial condition. Similarly, one could in principle calculate the location and orientation of the particle at the next intersection the trajectory makes with the boundary given the location and orientation of the previous intersection. This calculation however, proves to be analytically challenging.  However, the problem can be simplified by realizing that each trajectory is composed of a sequence of geodesics. Instead of using a \Poincare map (see Fig.~\ref{fig:phasediagram}), each geodesic on a cone can be described by its distance of closest approach to the apex, $H$, and azimuthal angle,$\phi$, at which a particle traveling along the geodesic in the direction of the trajectory intersects the boundary.

We can thus represent each trajectory as a sequence $(H_n, \phi_n)$. We determine that $H_n$ can be recursively defined as a function of $H_{n-1}$ and $\phi_{n-1}$ in the following way:

\begin{widetext}
\begin{align}
    H_n = r_b(\chi, \phi_{n-1}, \gamma)\left(\frac{H_{n-1}}{r_b(\chi, \phi_{n-1}, \gamma)} \cos(2\tilde\gamma(\chi, \phi_{n-1}, \gamma)) \pm \sqrt{1-\frac{H_{n-1}^2}{r_b(\chi, \phi_{n-1}, \gamma)^2}}\sin(2\tilde \gamma(\chi, \phi_{n-1}, \gamma))\right) \label{eq:T(h)definition}
\end{align}
\end{widetext}
where $\chi$ and $\gamma$ are the deficit and tilt angles of the cone, $\vec r_b(\chi, \phi_n, \gamma)$ is the vector from the apex to the unrolled cone base at $\phi_n$,  and $\tilde \gamma(\phi_n)$ is the angle between the normal vector to the boundary at $\phi_n$ and $\vec r_b(\chi, \phi_n, \gamma)$. 

\begin{figure}[t!]
    \centering
    \includegraphics[width=\columnwidth]{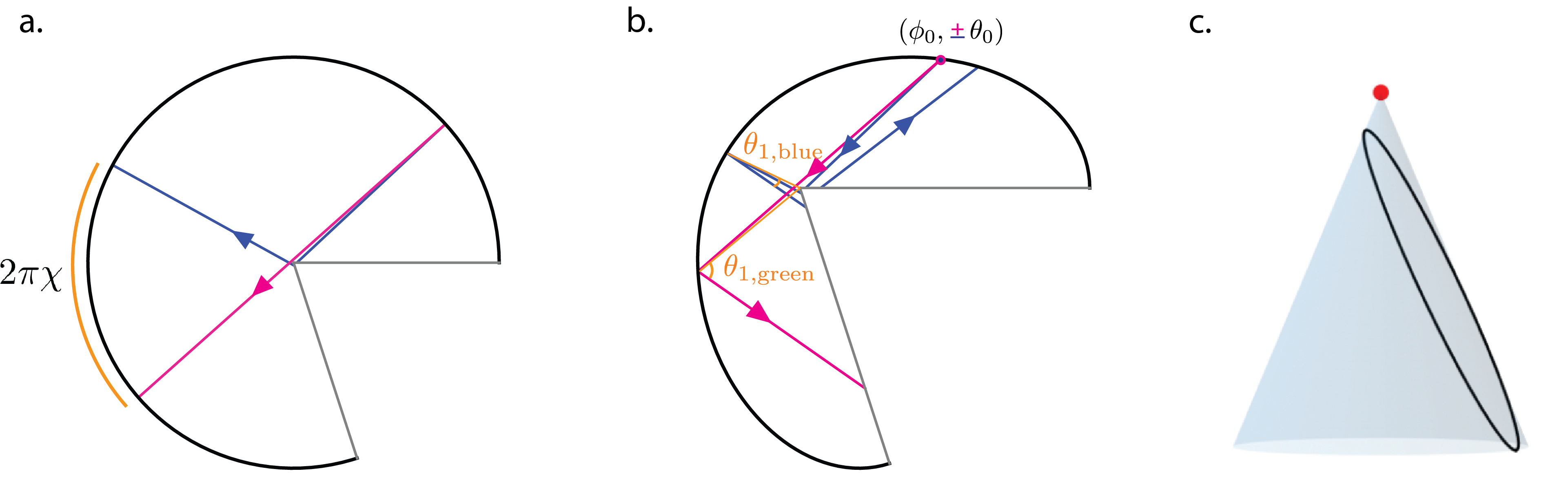}
    \caption{(a) The fate of two paths passing on different sides of the apex for a tilted cone with non-zero $\chi$ and zero $\gamma$. (b) The fate of two paths passing on different sides of the apex for a cone with non-zero $\chi$ and non-zero $\gamma$. $\theta_1$ is labeled for both the blue and pink path.(c) A cone with the base boundary in black for $\chi = 0.7$, $\gamma = 1.1$. We see that the apex is nearly coplanar with the base, and thus we expect the Poincare Map to be a relatively small perturbation on that of an ellipse.}.
    \label{fig:chaos appearence}
\end{figure}

Eq.\ref{eq:T(h)definition} reveals why we expect chaotic orbits appear when $\gamma\approx 0$ but not strictly equal to $0$. For $\gamma=0$, different trajectories with $H\approx 0$ will have different futures, even if their initial conditions are nearly the same. This conclusion follows because, trajectories which at any point have $H\approx 0$ that differ by a slight angle in their initial conditions will either go clockwise or counterclockwise around the apex, and will wind up at an angle of $2\pi\chi$ to each other, as shown in Fig.~\ref{fig:chaos appearence}(a). However, since for $\gamma =0$, $\tilde \gamma=0$ for all $\phi$, the set of trajectories that have $H_n \approx 0$ for any $n$ is a set of measure zero. 

However, even for very small $\gamma$, the set of initial $H_1$ for which some $H_n \approx 0$ increases as $H_n\neq H_{n-1}$, and is no longer measure zero. Thus there will be a larger set of trajectories which will have sensitive dependence on initial conditions. Note that $H_n\approx 0$ is equivalent to saying $\theta_n\approx 0$ where $\theta_n$ is defined as in Fig.~\ref{fig:poincareMap}. Thus in \Poincare space chaotic trajectories will appear near $\theta_n \approx 0$. We see this effect in Fig.~\ref{fig:phasediagram} and Fig.~\ref{fig:full phase diagram} for $\gamma \approx 0.1$ and $0.3$.

On the other hand, as shown in Fig.~\ref{fig:chaos appearence}(c), for large $\gamma$, near its maximum allowed value, the cone apex is only slightly out of plane of the elliptical boundary. At this stage we would expect the \Poincare space to start approaching that of an ellipse. So we would expect to see a return of 2-hourglass orbits, which we can also see for $\gamma \approx \gamma_{max}$ on Fig.~\ref{fig:phasediagram} and Fig.~\ref{fig:full phase diagram}.

\section{\Poincare maps for elliptical billiards}
\label{sec: appendix ellipse}

\begin{figure*}[t!]
    \centering
    \includegraphics[width = \linewidth]{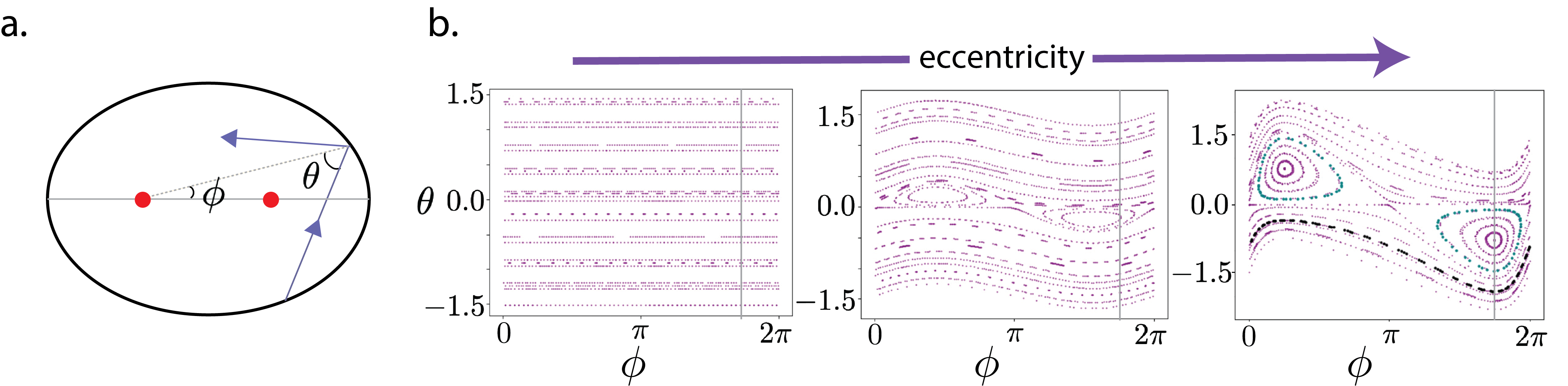}
    \caption{(a) Definition of $\phi$ and $\theta$ coordinates for the blue trajectory at its intersection with the ellipse boundary. (b) From left to right \Poincare maps in purple, in ($\phi, \theta$) space for ellipses of varying eccentricities (0,0.2,0.7) with initial conditions sampled along the vertical gray line. On the rightmost plot, a sequence of points corresponding to a rim trajectory is represented in black, and an hourglass trajectory is represented in teal. Note how both trajectories are qualitatively similar to their counterparts for a tilted cone with $\chi = 0.1$ and $\gamma =0.3$ as shown on Fig.~\ref{fig:3 types}(d). However, the more chaotic mixed trajectories characteristic of tilted cones are absent. } 
    \label{fig: ellipse Poincare}
\end{figure*}

As discussed in the main text, \Poincare maps analogous to the one used to represent trajectories on cones are a convenient way to visualize the space of trajectories in a dynamical system more generally. Since conical billiards have some qualitative similarities to elliptical billiards it is illustrative to compare their \Poincare maps. To define the angles $\phi$ and $\theta$ (analogous to the conical billiard coordinates in Fig.\ref{fig:poincareMap}(a)) for an elliptical billiard we choose one of the two focal points of the ellipse as the origin. When a trajectory approaches the boundary, we record the polar angle $\phi$ of impact and the angle, $\theta$, the trajectory makes with the line joining the intersection point to the focal point, see Fig.~\ref{fig: ellipse Poincare}(a). Here, the angle $\phi$ plays the role of a y-intercept and the angle $\theta$ a velocity direction or slope.  We then plot trajectories with multiple different initial conditions each as a sequence of points with coordinates ($\phi$,$\theta$) corresponding to bounces along the boundary. As expected for elliptical billiards, we see two qualitatively different types of trajectories; rim trajectories which show up as lines spanning the whole range of $\phi$ (in black on Fig.~\ref{fig: ellipse Poincare}) and hourglass trajectories which appear as pairs of loops in \Poincare space with one appearing at low $\phi$ and the other at high $\phi$. The hourglass trajectories correspond to the regions surrounding the two endpoints of the minor axis (top and bottom of the hourglass) spanned by hourglass trajectories.   Note that these two regions are separated by a homoclinic trajectory connecting the foci of the ellipse, not shown. Note the similarities of the \Poincare plots of these trajectories shown in Fig.~\ref{fig: ellipse Poincare} with those of conical billiards in Fig.~\ref{fig:3 types}(b). However, unlike for a tilted cone, there is no region of mixed or chaotic trajectories as this system is integrable~\cite{lynch2019integrable}.

\section{Coordinate transformation}
\label{sec: appendix Coordtransform}

In this section we will discuss the behavior of geodesics under the coordinate transformation used in the main text and shown on Fig.\ref{fig:coordinatesY} to transform a cone in three dimensions to an unrolled cone in the plane. In the absence of a tilt, we are led to a sector of a disk in two dimensions. As discussed in the main text, this transformation takes a point in $\mathcal{R}^3$ defined in cylindrical coordinates as $(\rho, \psi,z)$ to the the polar coordinate $(r,\phi) =(\frac{\rho}{\sin(\beta)}, \psi (1-\chi))$. We then impose boundary conditions such that $\phi = \phi +2\pi (1-\chi)$. 

\begin{figure}[t!]
    \centering
    \includegraphics[width = \columnwidth]{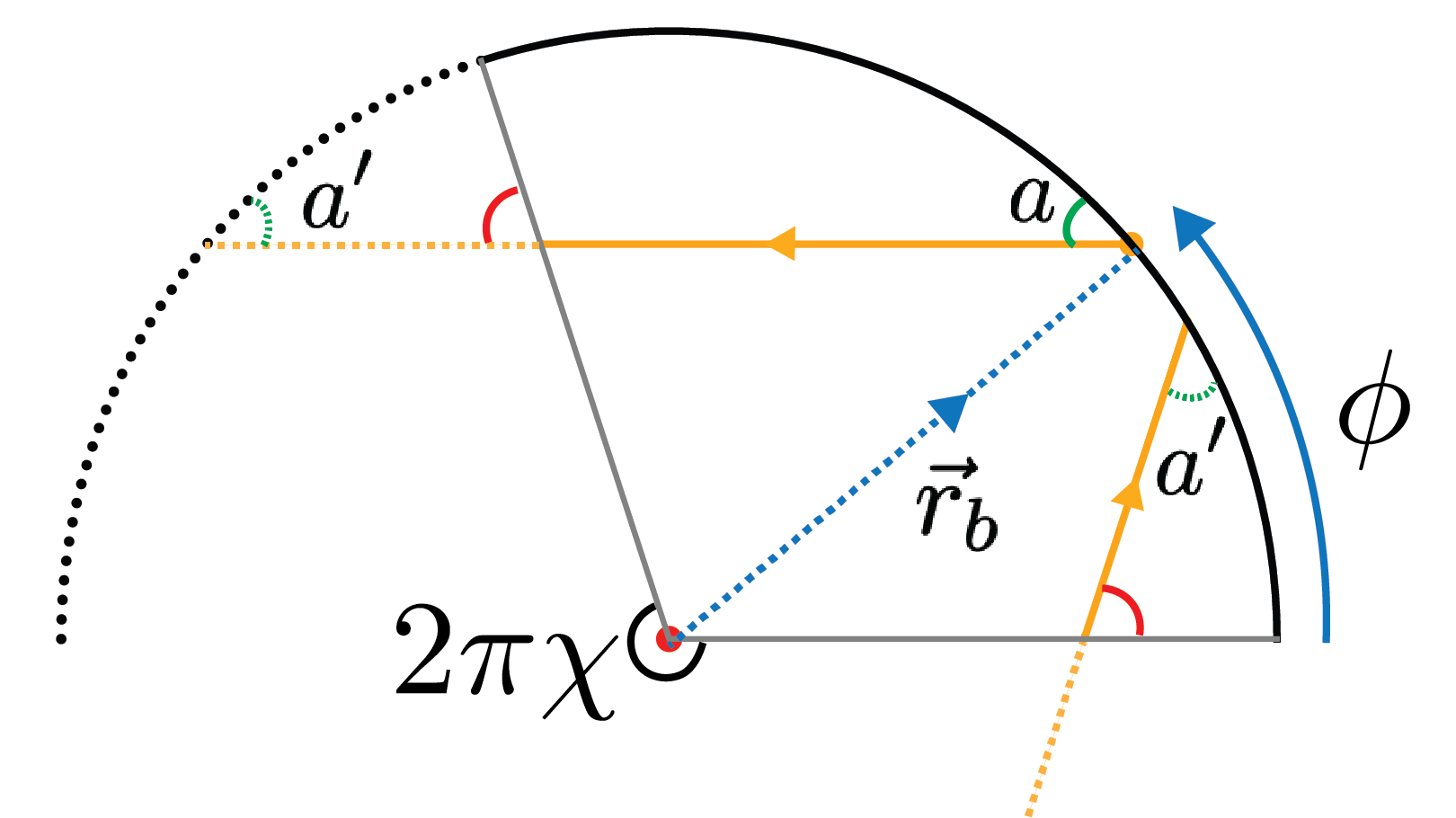}
    \caption{An unrolled cone with $\chi = 0.7, \gamma =0$ in unrolled coordinates. The two gray lines, corresponding to the cut leading from the base to the apex, are at $\phi = 0$ and $\phi = 2\pi(1-\chi)$. The dotted black curve is the continuation of the black boundary line under this identification. The orange line, initially making an angle $a$ with the cone base is then a geodesic. Note that the angle between the gray line and the geodesic remains the same as the orange geodesic crosses one gray line and emerges from the other, with the two angles marked in red identified. Under this transformation, the angle at which the geodesic intersects the boundary is $a'=a$, indicating a conserved quantity. In 2d space this corresponds to rotating the geodesic by $2\pi \chi$ in the clockwise direction.} 
    \label{fig:coord}
\end{figure}

Since geodesics are the shortest paths between two points, in unrolled coordinates, geodesics will be straight lines up until they intersect the cut of the cone (the two gray lines in Fig.~\ref{fig:coord}). When a geodesic, shown in orange for an untilted cone on Fig.~\ref{fig:coord}, intersects the cut of the cone at $\phi = 2\pi (1-\chi)$, shown in gray, it emerges on the other side of the cut at $\phi =0$, at the same radial coordinate but rotated clockwise by $2\pi \chi$. If the geodesic instead intersected the cut of the cone at $\phi = 0$ it would be rotated by $2\pi \chi$ counterclockwise. 

This rotation can be visualized by continuing the geodesic past the gray boundary line (shown as the orange dotted line on Fig.~\ref{fig:coord}. Then, for an untilted cone with $\gamma =0$, the geodesic is just a chord of a circle and thus the angles to the tangent of the circle at its two intersections, indicated by $a$ and $a'$ in Fig.~\ref{fig:coord} are equal. It follows that the distance along the black boundary between intersections with the geodesic is equal to the distance along the circumference of a circle spanned by a chord which intersects the boundary at the green angle $a$. This distance only depends on the green angle and does not depend on $\chi$ or on the azimuthal angle of the intersection between the geodesic and the boundary. Angles such as $a$ and $a'$ represent an additional conseved quantity (in addition to kinetic energy) for an untilted cone.

\section{Winding number for conical geodesics}\label{sec: appendix winding number}
Here we discuss the properties of conical geodesics. The analysis allows for understanding some of the behavior of periodic orbits for untilted cones with $\gamma = 0$. As discussed in the main text, these orbits occur when $\frac{2 (\pi/2 -\theta)}{2 \pi (1-\chi)}\in \mathbb{Q}$ where $\mathbb{Q}$ is the set of rational numbers. The trajectory then consists of a finite number of geodesic segments. 

Let us start by describing a single geodesic on the cone surface. Since periodic trajectories are made up of a finite number of geodesics, it is sufficient to understand the behavior of one geodesic to understand the behavior of periodic trajectories. 

Solving the geodesic equation on a cone~\cite{coxeter1969geometry} reveals the parametric equation for the most general geodesic on the surface of a cone using three dimensional cylindrical coordinates will be of the form
\begin{subequations}
\begin{align}
    \rho &= (1-\chi)r(s) \\ 
    \psi &= \psi_0+\frac{\tan^{-1}(\frac{\rho_0 \dot \rho_0 + s}{\rho_0 
    \sqrt{1-\dot \rho_0^2}}) + \tan^{-1}(\frac{\dot \rho_0}{\sqrt{1-\dot \rho_0^2}})}{1-\chi} \\
    z &= \sqrt{\chi(2-\chi)}r(s)    
\end{align}
\label{Eq.geodesic}
\end{subequations}
where $s$ is the arc-length along the geodesic, $\chi$ is related to the cone angle $\beta$ by $\chi = 1-\sin \beta$, 
\begin{align}\label{eq.r(s)}
    r(s) = \sqrt{(\rho_0 \dot \rho_0 + s)^2 + \rho_0^2 (1-\dot \rho_0^2)}
\end{align}
is the distance to the apex at each arclength point $s$, $\rho_0$ and $\psi_0$ are the initial radial and azimuthal position of the geodesic respectively, and $\dot \rho_0 = \frac{d\rho}{ds}|_{s=0}$ is the initial radial slope at $s=0$. 

It is interesting to ask what the maximum number of times a geodesic launched at the cone base, located at $\rho =\rho_{max}$ will wind around the cone apex before returning. It turns out that for an infinitely long cone (with the base infinitely far away), this number depends only on $\chi$ and not on the initial conditions for the geodesic. To show this, without loss of generality we can set $\dot \rho_0 = 0$. As discussed in Appendix~\ref{sec: appendix Coordtransform}, each geodesic will intersect the boundary at two locations which will correspond to maximum values of $\rho$ and will have a well defined minimum distance to the apex between those two locations and thus a minimum $\rho$ value at which  $\dot \rho_0 = 0$. Since the cone is azimuthally symmetric we can also without loss of generality set $\psi_0 = 0$. The expression for a geodesic (Eq.~\ref{Eq.geodesic}) then simplifies to
\begin{subequations}
\begin{align}
    \rho(s) &= (1-\chi)\sqrt{s^2 + \rho_0^2} \\
    \psi(s) &= \frac{\tan^{-1}(\frac{s}{\rho_0})}{1-\chi}  \\
    z(s) &= \sqrt{\chi(2-\chi)}\sqrt{s^2 + \rho_0^2}
\end{align}
\end{subequations}
with periodic boundary conditions with $\psi(s) \in (0, 2\pi)$ so that $\psi(s)$ is identified with $\psi(s)+2 n \pi$.

To determine a winding number $n$, we need to find an $s'$ for which $\psi(s')  = \psi(s)+2 n \pi$, which leads to
\begin{align}\label{eq.s'}
    &\tan^{-1}(\frac{s'}{\rho_0}) = 2 n\pi (1-\chi)  +\tan^{-1}(\frac{s}{\rho_0}) \nonumber\\ &\quad \implies s' = \rho_0\tan( 2 n\pi(1-\chi) +\tan^{-1}(\frac{s}{\rho_0})).
\end{align}
If we want to know how many times a geodesic makes a full circle around the apex, we choose $s = 0$, and find all values of $n$ for which $s'$ will have distinct values, and $\rho(s) < \rho_{max}$ where $\rho_{max}$. Thus we want

\begin{align}
     &2 n\pi(1-\chi) < \pi  \text{ and } (1-\chi)\sqrt{s^2 + \rho_0^2} <\rho_{max}:\nonumber\\  &n= \left \lfloor \frac{\abs{\pi/2 -\frac{\tan^{-1} \left ( \sqrt{\frac{\rho_{\max}^2 }{\rho_0^2}- 1}\right )}{1-\chi}}}{2\pi} \right \rfloor  \nonumber \\ & \qquad \quad+ \left \lfloor \frac{\abs{\pi/2 +\frac{\tan^{-1} \left ( \sqrt{\frac{\rho_{\max}^2 }{\rho_0^2}- 1}\right )}{1-\chi}}}{2\pi} \right \rfloor
\end{align}
Where $\lfloor x \rfloor$ is the floor function, i. e. the largest integer in $x$. 
If we allow $\frac{\rho_{max}}{\rho_0}\rightarrow \infty$ then the winding number becomes 
\begin{align}\label{eq.winding long cone}
     n = \left \lfloor \frac{1}{2(1-\chi)} \right \rfloor
\end{align}

\begin{figure}[ht!]
    \centering
    \includegraphics[width=\columnwidth]{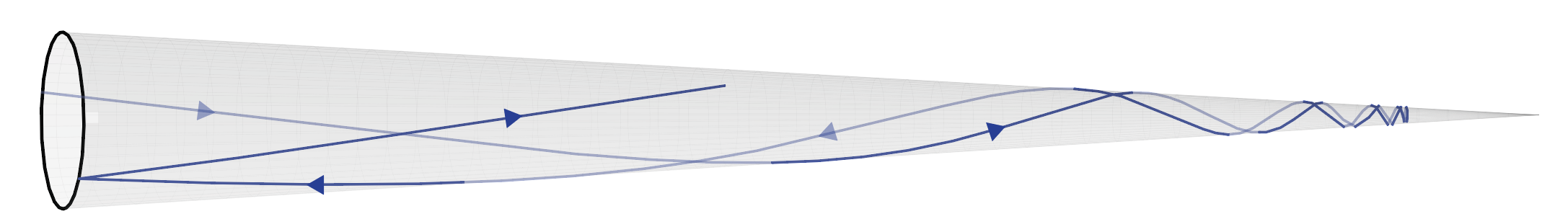}
    \caption{ Geodesic on a cone with $\chi = 0.95$ with one bounce off the base.  } 
    \label{fig: sharpcone winding}
\end{figure}

Note that this means that the winding number $n$ approaches infinity when for $\chi \rightarrow 1$. Thus, for a very sharp cone, a geodesic will wrap around the apex many times before winding back down the flank, as illustrated in Fig.~\ref{fig: sharpcone winding}. 

% For the following discussion we assume our cone base is very far from the apex compared to $\rho_0$, and use Eq.~\ref{eq.winding long cone}.
Upon inserting $s'$ from Eq.~\ref{eq.s'} back into Eq.~\ref{eq.r(s)}, we find that the geodesic intersects $\psi = 0$ at a sequence of points $r_n$ such that
\begin{align}
    r_n &= \rho_0 \sqrt{\tan(2 n \pi (1-\chi))^2 +1} \nonumber \\ &= \frac{\rho_0}{|\cos(2\pi n(1-\chi))|}  \nonumber \\ &=  \frac{\rho_0}{|\sin(2\pi n(1-\chi) + \pi/2)|}
\end{align}

which is the same functional form as a function of $n$ as we found for $x(\tilde \phi)$ as a function of $\tilde \phi$, for quasi periodic trajectories in the main text. For $\chi$ near $1$, using the same mathematical arguments as in the main text, we can show that $P(r_n)$ will scale the same way as we approach $r_n = \rho_0 (1-\chi)$ that it does for quasi periodic trajectories. 

\section{Measuring catacaustics}\label{sec: appendix caustic}
To obtain the results in Fig.~\ref{fig:caustic} for catacaustics on cones with varying degrees of tilt, we chose a rim trajectory on an unwrapped cone defined by a set of equally spaced points.
We then took a grid of 300x300 squares spanning the rectangle that the unrolled cone can be inscribed into.
We then calculated the number of points of the trajectory in each square, choosing a time discretization that allowed many points for each occupied square.

We defined the catacaustic to be the boundary between squares that have non-zero trajectory points in them and ones that have zero trajectory points. For a given boundary coordinate $\phi$ we determined the perpendicular line to the catacaustic and plotted the fraction of points in each box that the line crossed relative to the total number of points in all the boxes that the perpendicular crossed, as a function of the distance from the catacaustic. 

We only included those squares that had more than $0.2\%$ of the points crossed by the perpendicular.
We expect the averages to work less well near the boundary of the cone and the boundary of the catacaustic since the boxes would potentially intersect some of the region with no points. We avoided this edge effect by not including the first few points of the perpendicular near the caustic singularity or the last $2/5$ths of the perpendicular closest to the cone base. 
 
We then fitted the  sequence of points obtained with the above procedure to a power law
\begin{align}
    P(x) = \frac{b}{(x-a)^d}
\end{align}
Here the amplitude $b$ is taken to be such that the fit line intersects the average of the last ten points that the perpendicular line intersects. The parameters $a$ and $d$ describing the location and exponent of the catacaustic respectively, are fitting parameters.

\begin{figure*}[ht!]
    \centering
    \includegraphics[width=\linewidth]{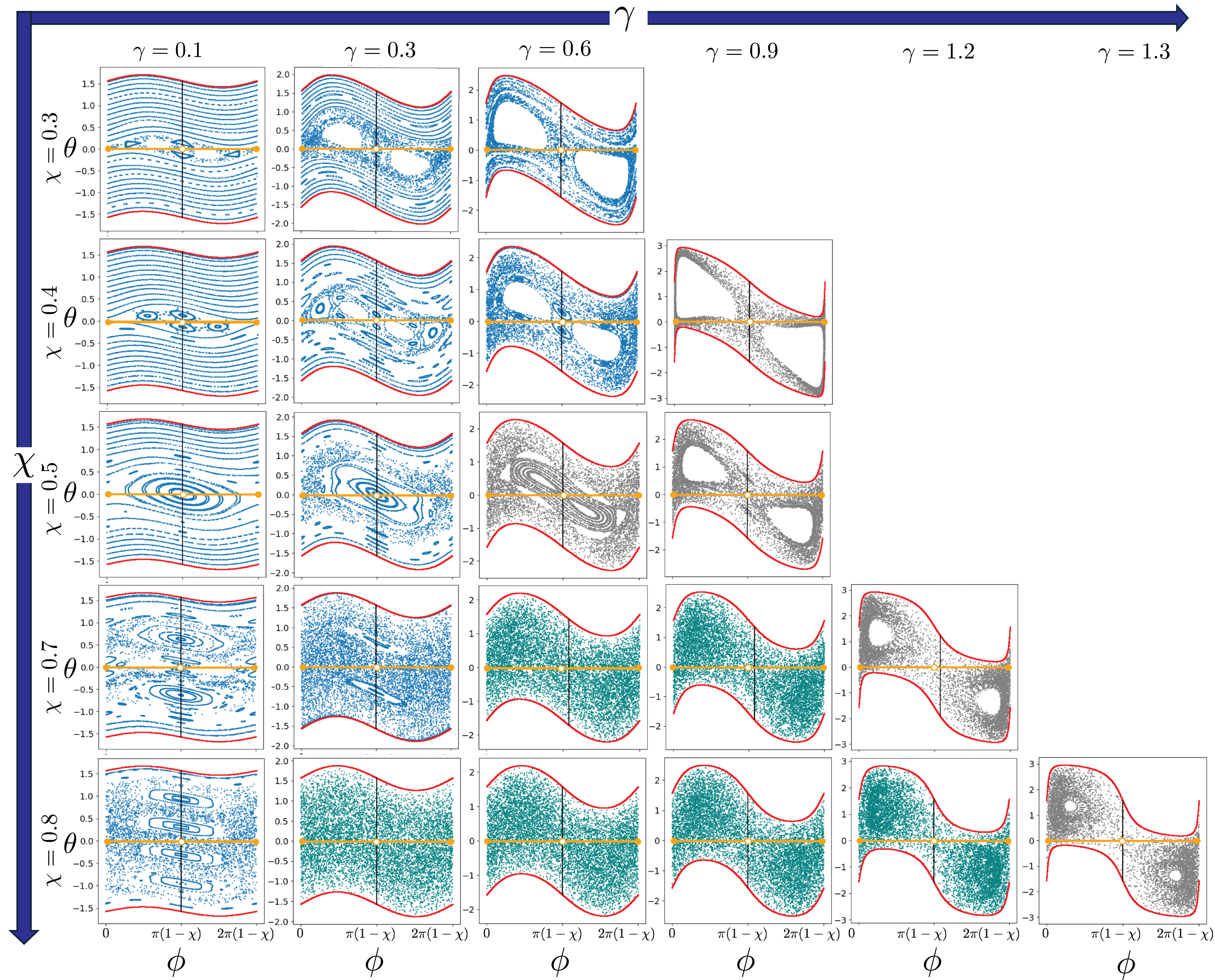}
    \caption{More fine grained \Poincare maps for varying $\chi$ and $\gamma$ for tilted cones. The faint horizontal yellow lines mark the points in \Poincare space that correspond to trajectories directly hitting the apex. The \Poincare maps are periodic in $\phi$. The wavy red lines mark the boundaries of the \Poincare space. A procedure to determine the equation for these lines is given in Appendix \ref{sec:Appendix thetamax}. The black vertical line marks the initial conditions that generate these \Poincare maps. The empty white spaces are regions that are not reached by trajectories beginning on those initial conditions. Blue plots are in region I, teal ones are in region IIA, and gray ones are in region IIB as defined in Sec.~\ref{sec:boundary} and Fig.~\ref{fig:boundary}. All \Poincare maps  in the blue region generate rim trajectories when periodicity in $\phi$ is taken into account. There are no rim trajectories on the gray and teal \Poincare maps, and all the gray plots and the blue plots for $\chi <0.5$ have a white region corresponding to two loop hourglass trajectories as discussed in the text. While being in region IIA of Fig.~\ref{fig:boundary} does not guarantee that trajectories are mixing and ergodic, we see here that many are.} 
    \label{fig:full phase diagram}
\end{figure*}

\begin{figure*}[ht!]
    \centering
    \includegraphics[width=\linewidth]{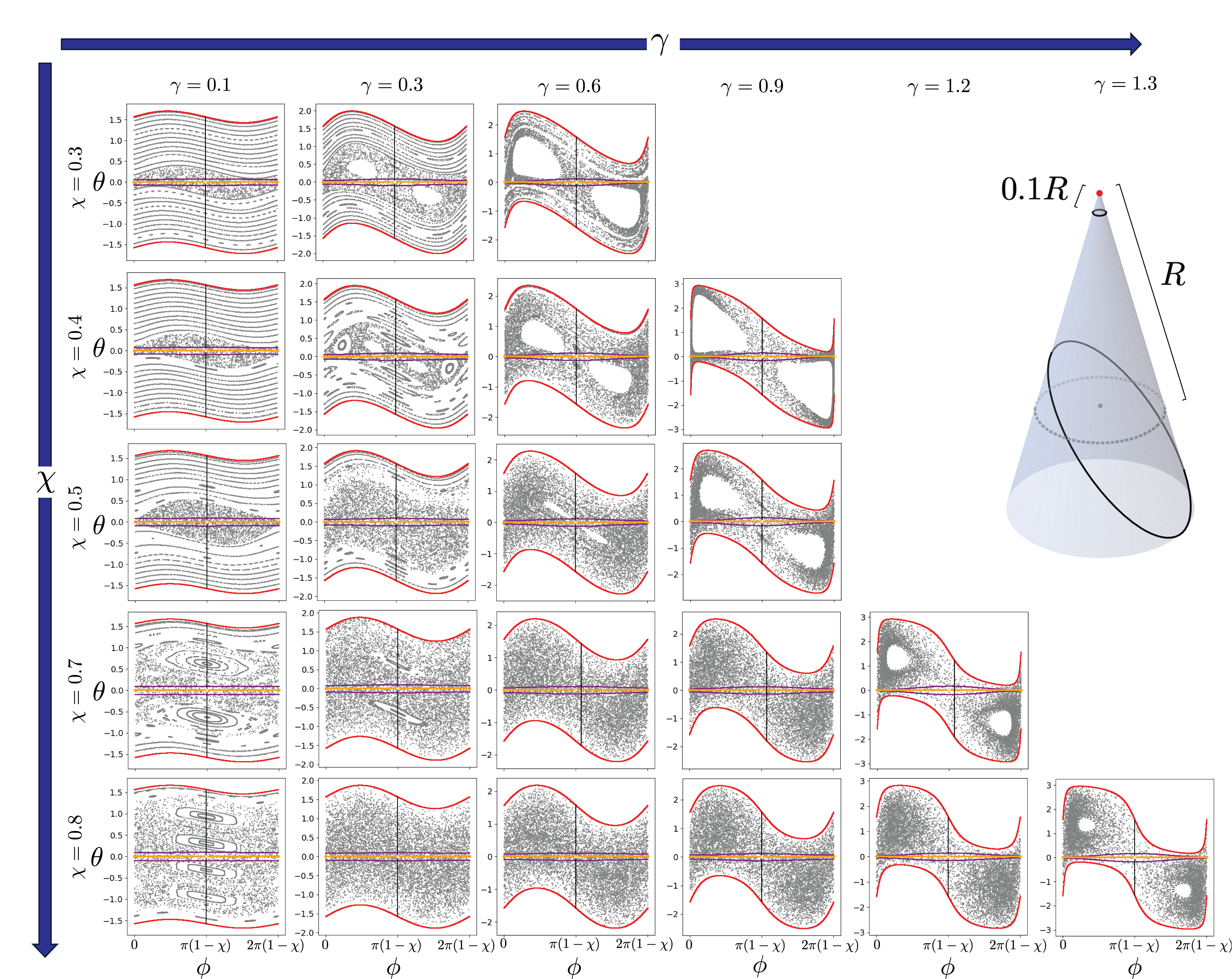}
    \caption{Analog of Fig.\ref{fig:full phase diagram} but with a second boundary at the top of the cone, defined by an intersection of the cone with a flat plane at a flank distance of $0.1*R$ (shown on the right side of the figure). Here R is defined in the same manner as in Fig.\ref{fig:intro}. Trajectories remain between the new upper boundary and the lower boundary at a tilt angle $\gamma$ (both in black). Like in Fig.\ref{fig:full phase diagram} the orange line marks the points on the \Poincare map which correspond to intersections with the apex, the red lines correspond to trajectories hugging the boundary, and the black line signifies where the initial conditions were sampled from. $\phi$, $\theta$ pairs which correspond to geodesic segments which intersect the upper boundary fall between the two purple lines. Note that  as one would expect, trajectories that did not fall in the region between purple lines remain unchanged from Fig.\ref{fig:full phase diagram}. Additionally,  note that introducing the upper cut increases the range of the mixed trajectories for a given $\chi$ and $\gamma$.  } 
    \label{fig:cutcone}
\end{figure*}
\section{Analytic form of the tilted cone boundary}\label{sec:Appendix coneboundary}
As described in the main text, see Fig.\ref{fig:setupFirst}, we describe our cone in three dimensions by taking a cone of flank length $R$ with apex at (0,0,0) and then rotating its base about the line $x = 0$, $z= R\cos(\beta)$ by an angle $\gamma$. We are interested in finding the equation for the flank distance to some point $O$ on the boundary, $r(\phi)$, where $\phi = \psi (1-\chi)$ is the polar angle of the unrolled cone as defined in Fig.~\ref{fig:setupFirst}(a). 

The point $O$ has cartesian coordinates $(O_x, O_y, O_z)$. Because it lies on the cone, we know that 
\begin{align}
    O_z = \frac{\sqrt{O_x^2 + O_y^2}}{\tan \beta} 
\end{align}
Similarly, because we know that $O$ lies on the plane that defines the base boundary, 
\begin{align}
    R\cos(\beta)-O_z = O_x \tan{\gamma}
\end{align}
Thus we are led to the relation
\begin{align}
    R\cos(\beta)- O_x \tan{\gamma} = \frac{\sqrt{O_x^2 + O_y^2}}{\tan \beta}
\end{align}

Upon noting that $\tan(\psi) = \frac{O_y}{O_x}$ by the definition of $\psi$, we find that
\begin{align}
    &R\cos(\beta)- O_x \tan{\gamma} = \frac{\sqrt{O_x^2 + O_x^2\tan^2(\psi)}}{\tan \beta}  \nonumber \\ & \quad \implies \frac{R\cos(\beta)}{O_x} - \tan(\gamma) = \frac{1}{-\cos(\psi)\tan(\beta)}
    \end{align}
    Thus we have
    \begin{align}
    O_x = \frac{R\cos(\beta) \tan(\beta)\cos(\psi)}{-\tan(\gamma)\cos(\psi) \tan(\beta) +1} 
    \end{align}
    which implies that
    \begin{align}
    O_z &= R-\frac{R\cos(\beta)\cos(\psi)\tan(\beta)\tan(\gamma)}{-\tan(\gamma)\cos(\psi) \tan(\beta) +1}  \nonumber \\ &= \frac{R\cos(\beta)}{1-\cos(\psi)\tan(\beta)\tan(\gamma)}
    \end{align}
    similar manipulations lead to
    \begin{align}
    r(\psi) &= \frac{R}{1-\cos(\psi)\tan(\beta)\tan(\gamma)}  \nonumber \\ &= \frac{R \cos(\gamma)\cos(\beta)}{\cos(\beta)\cos(\gamma)-\cos(\psi)\sin(\beta)\sin(\gamma)}
    \end{align}
    from which it follows that
    \begin{align}
    r(\phi) &= \frac{R \cos(\gamma)\cos(\beta)}{\cos(\beta)\cos(\gamma)-\cos(\frac{\phi}{1-\chi})\sin(\beta)\sin(\gamma)}  \nonumber \\ &= \frac{R \cos(\gamma)\sqrt{\chi(2-\chi)}}{\sqrt{\chi(2-\chi)}\cos(\gamma)-\cos(\frac{\phi}{1-\chi})(1-\chi)\sin(\gamma)}\label{eq:rphi}
\end{align}

We now want to determine when the boundary becomes partially concave, which occurs when the radius of curvature at the boundary given by ~\cite{coxeter1969geometry} 
\begin{align}
    r_{curve} = \frac{(r(\phi)^2 + (\frac{d r}{d\phi})^2)^{3/2}}{r^2 + 2(\frac{d r}{d\phi})^2 -r\frac{d^2 r}{d\phi^2} } 
    \end{align}
is set to infinity. Setting $r_{curve}$ to infinity gives
    \begin{align}
    r^2 + 2\Big(\frac{d r}{d\phi}\Big)^2 -r\frac{d^2 r}{d\phi^2}  = 0
\end{align}
After inserting this result into Eq.~\ref{eq:rphi} and solving, we find the critical tilt angle $\gamma_c$ above which the cone boundary becomes partially concave, 
\begin{align}
    \gamma_c = \tan^{-1} (\frac{\sin(\beta)}{\cos(\beta)}) = \beta,
\end{align}
as claimed in Eq.\ref{eq.phasediagramboundaries} of the main text.

\section{Analytic determination of $\theta_{\max}(\phi)$}\label{sec:Appendix thetamax}

We now determine the maximum allowed value of the slope variable, $\theta_{max}(\phi)$, for an unrolled cone, so that the geodesic it labels remains inside the cone, see Fig.\ref{fig:tiltedConeBound}. This condition means that $\theta_{max}(\phi)$ will be the angle between the boundary of the cone at $\phi$ and the vector $\vec r_b(\phi)=(-r(\phi)\cos(\phi), -r(\phi)\sin(\phi))$, see Fig.\ref{fig:tiltedConeBound}. Because the boundary is smooth, $\theta_{min}(\phi) = \theta_{\max}(\phi)-\pi$. 
As shown in Eq.~\ref{eq:rphi} in Appendix.~\ref{sec:Appendix coneboundary}, the boundary of a cone  with half-angle $\beta = \sin^{-1}(1-\chi)$ in unrolled coordinates is given by 
\begin{align}\label{eq. Appendix boundary}
    r(\phi) = \frac{R \cos(\gamma)\cos(\beta)}{\cos(\beta)\cos(\gamma)-\cos(\frac{\phi}{1-\chi})\sin(\beta)\sin(\gamma)}
\end{align}.
With the help of Eq.~\ref{eq. Appendix boundary}, we can determine the tangent to the boundary $\vec r(\phi) = [r(\phi) \cos \phi, r(\phi) \sin\phi]$ as $\vec r_T(\phi) =\frac{d \vec r(\phi)}{d\phi}$, and hence its slope as  
\begin{align}
    m =\frac{r_{T,y}}{r_{T,x}}= \frac{r(\phi) \cos(\phi) + \frac{d r(\phi)}{d\phi} \sin(\phi)}{- r(\phi) \sin(\phi) +\frac{d r(\phi)}{d\phi} \cos(\phi) } ,
\end{align}
where 
\begin{align}
  &\frac{dr(\phi)}{d \phi}  \nonumber \\ &\quad = \frac{-\cos (\gamma)\sin(\gamma) \sin(\frac{\phi}{1-\chi})}{(\sqrt{(2-\chi)\chi}\cos(\gamma) - (1-\chi) \cos(\frac{\phi}{1-\chi})\sin(\gamma))^2}  
\end{align}
where we have again used $\sin \beta = 1-\chi$ and $\cos \beta  = \sqrt{\chi (2-\chi)}$.
It follows that the tangent to the boundary takes the form,
\begin{align}
    \vec r_T(\phi) =  (r(\phi) \cos(\phi) + 1,r(\phi) \sin(\phi) + m )
\end{align}
We show both $\vec r_T(\phi)$ and $\vec r_b(\phi)$ on Fig.~\ref{fig:tiltedConeBound}. Without loss of generality we choose the positive $\theta$ direction to be when the path is moving clockwise, so that the cross product of $\vec r_T(\phi)$ and $\vec r_b$ to be positive.Thus, for example, $\theta_i>0$for the untilted cone in Fig.~\ref{fig:setup}(a). We can now find the angle between $\vec r_T(\phi)$ and $\vec r_b$ by taking their dot product, dividing by $|\vec r_b||\vec r_T|$ and multiplying by $\text{sgn}(\vec r_T(\phi) \cross \vec r_b$). This gives the equation for the top red line in Fig.~\ref{fig:phasediagram}. The bottom red line in Fig.~\ref{fig:phasediagram} is given by  $\theta_{max}(\phi) - \pi$.

\section{More detailed Phase diagram for tilted cones} \label{sec:Appendix full phase diagram}
A more densely populated version of Fig.\ref{fig:phasediagram} can be seen in Fig.\ref{fig:full phase diagram}. Additionally, Fig.\ref{fig:cutcone} shows an analogous image for a truncated cone to give us a sense of the effect of altering the singularity at the apex. We define the truncated cone to as the normal tilted cone but with an additional boundary at the height $h$ where $h$ is smaller than the maximum height of the tilted cone base. When trajectories hit this other boundary they bounce off with the same specular reflections as when they bounce off the tilted cone base. Note how the truncation creates more mixed trajectories near the orange line which corresponds to trajectories which would hit the apex if the truncation was not there.

\section{Uniform distribution in \Poincare space}\label{sec: appendix distribution}
\begin{figure}[t!]
    \centering
    \includegraphics[width=\columnwidth]{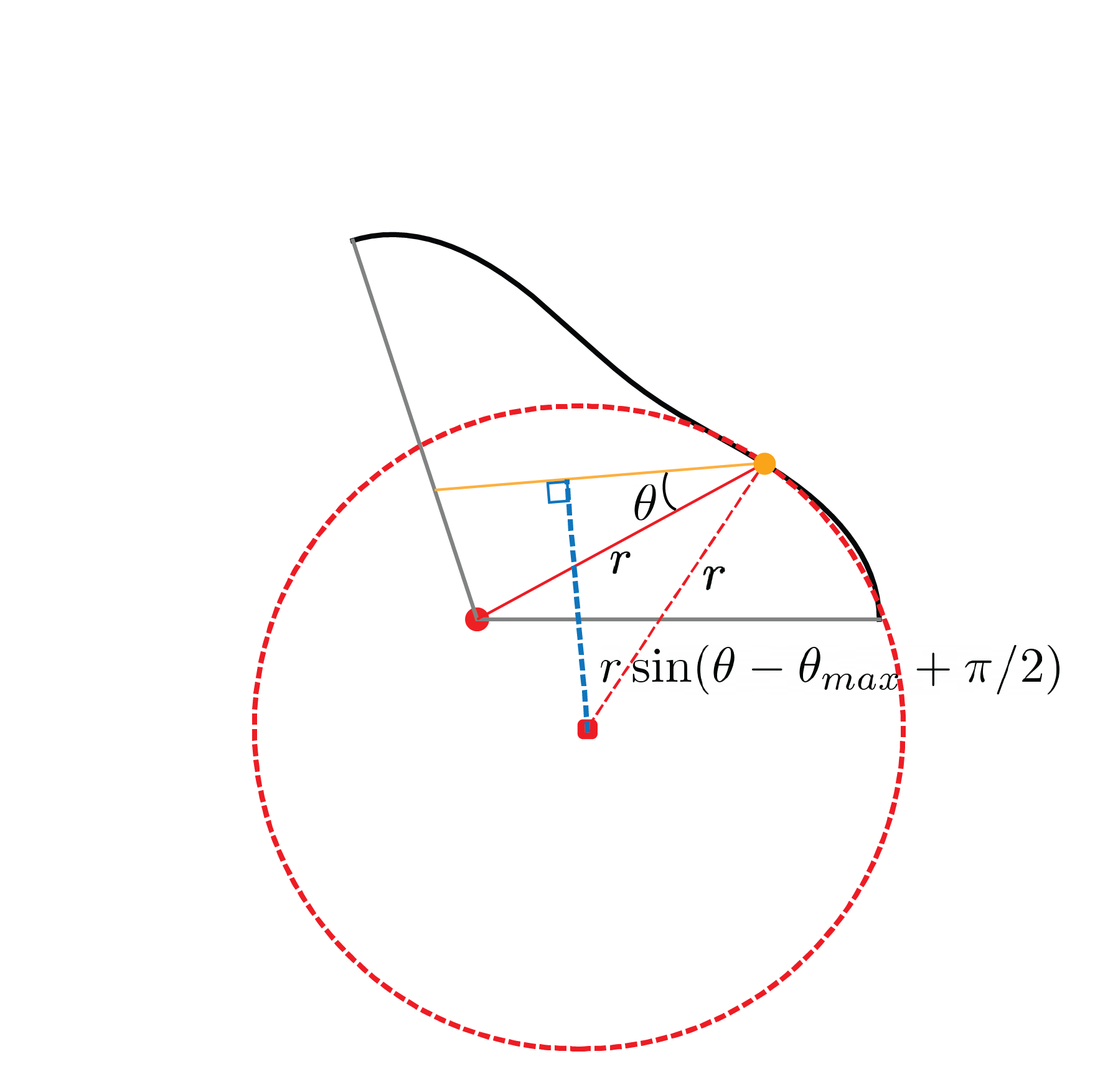}
    \caption{ Graphic demonstrating how the expected density of random trajectories on an unrolled cone was calculated to determine if the system is mixing. For the point marked in orange on the cone base at $\phi= \phi_0$ we determine the distance, $r$, from the apex to the point. We then draw a circle (dashed red line) of radius r which whose radius is perpendicular to the tangent to the boundary at the orange point. We choose the center of the circle to be on the same side of the boundary as the apex of the cone. A path that starts at point $(\phi_0,\theta)$ in \Poincare space will be perpendicular to the blue line from the dashed circle center. The blue line has length $r\sin(\theta-\theta_{max} +\pi/2)$. }.
    \label{fig:distributionCalculation}
\end{figure}

To determine what a randomly populated \Poincare space would look like, we ask how likely a randomly drawn geodesic on our unrolled cone is to go through a given boundary point at a given slope angle. To determine this probability, choose one point on the boundary (at $\phi=\phi_0$, marked as the orange point on Fig.\ref{fig:distributionCalculation}) and ask for the distribution of paths (orange line on Fig.\ref{fig:distributionCalculation}) that intersect the boundary at this point with different slopes $\theta$. We will use the definition of randomness inspired by the 'maximum ignorance principle' which states that we should aim to use a distribution of geodesics which is both scale and translation invariant~\cite{jaynes1973well,di2011bertrand}. On a circle, such a distribution can be achieved by choosing chords such that the perpendicular distance between them and the center of the circle is uniformly distributed~\cite{bertrand1889calcul,marinoff1994resolution}. Due to translation and scale invariance, the distribution on an unrolled tilted cone will be the same as that of a circle (in red on Fig.\ref{fig:distributionCalculation}), the boundary of which is parallel to the boundary of the cone at $\phi_0$. We need the boundary to be parallel to ensure that all chords that intersect the boundary at $\phi_0$ are captured. Due to scale invariance, we could in principle for each individual $\phi_0$ choose any radius for our circle. However, for any point along the boundary, all chords that originate on the cone that intersect it have a range of perpendicular distances from the apex from $0$ (for a chord going through the apex), to $r(\phi_0)$, the distance from the apex to the boundary at $\phi_0$ (for a chord that hugs the boundary at $\phi_0$). Thus, we would expect the total number of chords which intersect the boundary at $\phi_0$ to be proportional to $r(\phi_0)$. 

This reasoning reveals that at each $\phi_0$, the probability of finding a chord intersecting the boundary at $\phi_0$ at some slope $\theta$ is,

\begin{align}
     P(\theta,\phi_0) =\frac{1}{N} r(\phi_0) \cos(\theta -\theta_{max} +\pi/2)
\end{align} 
where $N$ is some normalization factor such that the total probability for all choices of $\phi_0$ is one.
In Fig.\ref{fig:9 chaos}(b), this is how the distribution of $\theta$ is determined for each given $\phi$. We then uniformly choose $\phi$ values at which to calculate the distribution of $\theta$ such that the density in \Poincare space of $\phi$ is on average the same as the average density in the $\theta$ direction (number of intersections at all $\theta$ divided by $\pi$) and plot the distributions of $\theta$ values for each $\phi$ to obtain the blue points on Fig. \ref{fig:9 chaos}(b). This is what we will take as the "random" distribution in \Poincare space that we will compare the distribution of $T^{-n}(A)$ to. Note that this distribution, while evenly distributed in real space is not evenly distributed in \Poincare space.

\bibliographystyle{unsrt}
\bibliography{Ergodicity}

\begin{thebibliography}{10}

\bibitem{sethna2021statistical}
James~P Sethna.
\newblock {\em Statistical mechanics: entropy, order parameters, and complexity}, volume~14.
\newblock Oxford University Press, USA, 2021.

\bibitem{adams2023ergodicity}
Terrence Adams and Anthony Quas.
\newblock Ergodicity and mixing properties.
\newblock In {\em Ergodic Theory}, pages 35--60. Springer, 2023.

\bibitem{wioland2013confinement}
Hugo Wioland, Francis~G Woodhouse, J{\"o}rn Dunkel, John~O Kessler, and Raymond~E Goldstein.
\newblock Confinement stabilizes a bacterial suspension into a spiral vortex.
\newblock {\em Physical review letters}, 110(26):268102, 2013.

\bibitem{ben2022disordered}
Ydan Ben~Dor, Sunghan Ro, Yariv Kafri, Mehran Kardar, and Julien Tailleur.
\newblock Disordered boundaries destroy bulk phase separation in scalar active matter.
\newblock {\em Physical Review E}, 105(4):044603, 2022.

\bibitem{keber2014topology}
Felix~C Keber, Etienne Loiseau, Tim Sanchez, Stephen~J DeCamp, Luca Giomi, Mark~J Bowick, M~Cristina Marchetti, Zvonimir Dogic, and Andreas~R Bausch.
\newblock Topology and dynamics of active nematic vesicles.
\newblock {\em Science}, 345(6201):1135--1139, 2014.

\bibitem{sun2023geometric}
Jessica~H Sun, Abigail Plummer, Grace~H Zhang, David~R Nelson, and Vinothan~N Manoharan.
\newblock Geometric frustration of hard-disk packings on cones.
\newblock {\em Physical Review E}, 108(5):054608, 2023.

\bibitem{bausch2003grain}
AR~Bausch, Mark~John Bowick, A~Cacciuto, AD~Dinsmore, MF~Hsu, DR~Nelson, MG~Nikolaides, A~Travesset, and DA~Weitz.
\newblock Grain boundary scars and spherical crystallography.
\newblock {\em Science}, 299(5613):1716--1718, 2003.

\bibitem{turner2010vortices}
Ari~M Turner, Vincenzo Vitelli, and David~R Nelson.
\newblock Vortices on curved surfaces.
\newblock {\em Reviews of Modern Physics}, 82(2):1301--1348, 2010.

\bibitem{zhang2022fractional}
Grace~H Zhang and David~R Nelson.
\newblock Fractional defect charges in liquid crystals with p-fold rotational symmetry on cones.
\newblock {\em Physical Review E}, 105(5):054703, 2022.

\bibitem{vafa2025defect}
Farzan Vafa, Grace~H Zhang, and David~R Nelson.
\newblock Defect ground states for liquid crystals on cones and hyperbolic cones.
\newblock {\em Journal of Physics A: Mathematical and Theoretical}, 58(22):225003, 2025.

\bibitem{vafa2024periodic}
Farzan Vafa, David~R Nelson, and Amin Doostmohammadi.
\newblock Periodic orbits, pair nucleation, and unbinding of active nematic defects on cones.
\newblock {\em Physical Review E}, 109(6):064606, 2024.

\bibitem{kourganoff2016uniform}
Micka{\"e}l Kourganoff.
\newblock Uniform hyperbolicity in nonflat billiards.
\newblock {\em arXiv preprint arXiv:1605.00290}, 2016.

\bibitem{tyc2022spherical}
Tom{\'a}{\v{s}} Tyc and Darek Cidlinsk{\`y}.
\newblock Spherical wedge billiard: From chaos to fractals and talbot carpets.
\newblock {\em Physical Review E}, 106(5):054202, 2022.

\bibitem{carmo2024mixing}
RB~do Carmo and T~Ara{\'u}jo Lima.
\newblock Mixing property of symmetrical polygonal billiards.
\newblock {\em Physical Review E}, 109(1):014224, 2024.

\bibitem{bunimovich1979ergodic}
Leonid~A Bunimovich.
\newblock On the ergodic properties of nowhere dispersing billiards.
\newblock {\em Communications in Mathematical Physics}, 65:295--312, 1979.

\bibitem{wojtkowski2020principles}
Maciej Wojtkowski.
\newblock {Principles for the design of billiards with nonvanishing Lyapunov exponents}.
\newblock In {\em Hamiltonian Dynamical Systems}, pages 531--554. CRC Press, 2020.

\bibitem{Tabachnikov1995}
Serge Tabachnikov.
\newblock {\em Billiards}, volume~1 of {\em Panoramas and Synthèses}.
\newblock Society for Industrial and Applied Mathematics (SIAM), Philadelphia, 1995.

\bibitem{lynch2019integrable}
Peter Lynch.
\newblock Integrable elliptic billiards and ballyards.
\newblock {\em European Journal of Physics}, 41(1):015005, 2019.

\bibitem{stachel2022motion}
Hellmuth Stachel.
\newblock On the motion of billiards in ellipses.
\newblock {\em European Journal of Mathematics}, 8(4):1602--1622, 2022.

\bibitem{stachel2021geometry}
Hellmuth Stachel.
\newblock {The geometry of billiards in ellipses and their Poncelet grids}.
\newblock {\em Journal of Geometry}, 112:1--29, 2021.

\bibitem{koiller1996static}
Jair Koiller, Roberto Markarian, Sylvie Oliffson~Kamphorst, and S{\^o}nia Pinto~de Carvalho.
\newblock Static and time-dependent perturbations of the classical elliptical billiard.
\newblock {\em Journal of Statistical Physics}, 83:127--143, 1996.

\bibitem{tabanov1994separatrices}
MB~Tabanov.
\newblock {Separatrices splitting for Birkhoff’s billiard in symmetric convex domain, closed to an ellipse}.
\newblock {\em Chaos: An Interdisciplinary Journal of Nonlinear Science}, 4(4):595--606, 1994.

\bibitem{dietz2022intermediate}
Barbara Dietz and Achim Richter.
\newblock Intermediate statistics in singular quarter-ellipse shaped microwave billiards.
\newblock {\em Journal of Physics A: Mathematical and Theoretical}, 55(31):314001, 2022.

\bibitem{lenz2007classical}
Florian Lenz, Fotis~K Diakonos, and Peter Schmelcher.
\newblock Classical dynamics of the time-dependent elliptical billiard.
\newblock {\em Physical Review E—Statistical, Nonlinear, and Soft Matter Physics}, 76(6):066213, 2007.

\bibitem{lenz2009evolutionary}
F~Lenz, C~Petri, FRN Koch, FK~Diakonos, and P~Schmelcher.
\newblock Evolutionary phase space in driven elliptical billiards.
\newblock {\em New Journal of Physics}, 11(8):083035, 2009.

\bibitem{stone2010chaotic}
A~Douglas Stone.
\newblock Chaotic billiard lasers.
\newblock {\em Nature}, 465(7299):696--697, 2010.

\bibitem{lazutkin1973existence}
Vladimir~F Lazutkin.
\newblock The existence of caustics for a billiard problem in a convex domain.
\newblock {\em Mathematics of the USSR-Izvestiya}, 7(1):185, 1973.

\bibitem{sinai1970dynamical}
Yakov~G Sinai.
\newblock Dynamical systems with elastic reflections.
\newblock {\em Russian Mathematical Surveys}, 25(2):137, 1970.

\bibitem{mcmullen2023billiards}
Curtis McMullen.
\newblock {Billiards and Teichm{\"u}ller curves}.
\newblock {\em Bulletin of the American Mathematical Society}, 60(2):195--250, 2023.

\bibitem{spagnolie2017microorganism}
Saverio~E Spagnolie, Colin Wahl, Joseph Lukasik, and Jean-Luc Thiffeault.
\newblock Microorganism billiards.
\newblock {\em Physica D: Nonlinear Phenomena}, 341:33--44, 2017.

\bibitem{dutta2020system}
Supriyo Dutta and Partha Guha.
\newblock A system of billiard and its application to information-theoretic entropy.
\newblock {\em arXiv preprint arXiv:2004.03444}, 2020.

\bibitem{albers2024billiards}
Thijs Albers, Stijn Delnoij, Nico Schramma, and Maziyar Jalaal.
\newblock Billiards with spatial memory.
\newblock {\em Physical Review Letters}, 132(15):157101, 2024.

\bibitem{badeau2024statistical}
Roland Badeau.
\newblock Statistical wave field theory.
\newblock {\em The Journal of the Acoustical Society of America}, 156(1):573--599, 2024.

\bibitem{kroetz2016dynamical}
Tiago Kroetz, H{\'e}rcules~A Oliveira, Jefferson~SE Portela, and Ricardo~L Viana.
\newblock Dynamical properties of the soft-wall elliptical billiard.
\newblock {\em Physical Review E}, 94(2):022218, 2016.

\bibitem{nagler2007leaking}
Jan Nagler, Moritz Krieger, Marco Linke, Johannes Sch{\"o}nke, and Jan Wiersig.
\newblock Leaking billiards.
\newblock {\em Physical Review E—Statistical, Nonlinear, and Soft Matter Physics}, 75(4):046204, 2007.

\bibitem{kudrolli1994signatures}
A~Kudrolli, S~Sridhar, Akhilesh Pandey, and Ramakrishna Ramaswamy.
\newblock Signatures of chaos in quantum billiards: Microwave experiments.
\newblock {\em Physical Review E}, 49(1):R11, 1994.

\bibitem{waalkens1997elliptic}
Holger Waalkens, Jan Wiersig, and Holger~R Dullin.
\newblock Elliptic quantum billiard.
\newblock {\em annals of physics}, 260(1):50--90, 1997.

\bibitem{blomquist2002geometry}
T~Blomquist, H~Schanze, IV~Zozoulenko, and H-J St{\"o}ckmann.
\newblock Geometry-dependent scattering through quantum billiards: Experiment and theory.
\newblock {\em Physical Review E}, 66(2):026217, 2002.

\bibitem{bertschinger2004real}
Nils Bertschinger and Thomas Natschl{\"a}ger.
\newblock Real-time computation at the edge of chaos in recurrent neural networks.
\newblock {\em Neural computation}, 16(7):1413--1436, 2004.

\bibitem{toyoizumi2011beyond}
Taro Toyoizumi and Larry~F Abbott.
\newblock Beyond the edge of chaos: Amplification and temporal integration by recurrent networks in the chaotic regime.
\newblock {\em Physical Review E—Statistical, Nonlinear, and Soft Matter Physics}, 84(5):051908, 2011.

\bibitem{chirikov2008chirikov}
Boris Chirikov and Dima Shepelyansky.
\newblock Chirikov standard map.
\newblock {\em Scholarpedia}, 3(3):3550, 2008.

\bibitem{weissert1997kolmogorov}
Thomas~P. Weissert.
\newblock {The Kolmogorov-Arnold-Moser Theorem:“Here comes the surprise”}.
\newblock {\em The Genesis of Simulation in Dynamics: Pursuing the Fermi-Pasta-Ulam Problem}, pages 51--82, 1997.

\bibitem{shenker1982critical}
Scott~J Shenker and Leo~P Kadanoff.
\newblock {Critical behavior of a KAM surface: I. Empirical results}.
\newblock {\em Journal of Statistical Physics}, 27(4):631--656, 1982.

\bibitem{strogatz2018nonlinear}
Steven~H Strogatz.
\newblock {\em Nonlinear dynamics and chaos: with applications to physics, biology, chemistry, and engineering}.
\newblock CRC press, 2018.

\bibitem{avendano2010caustics}
Maximino Avenda{\~n}o-Alejo, Luis Casta{\~n}eda, and Iv{\'a}n Moreno.
\newblock Caustics and wavefronts by multiple reflections in a circular surface.
\newblock {\em American Journal of Physics}, 78(11):1195--1198, 2010.

\bibitem{berry1981regularity}
Michael~V Berry.
\newblock Regularity and chaos in classical mechanics, illustrated by three deformations of a circular'billiard'.
\newblock {\em European Journal of Physics}, 2(2):91, 1981.

\bibitem{dumas2014kam}
H~Scott Dumas.
\newblock {\em {The: KAM Story, A Friendly Introduction To The Content, History, And Significance Of Classical Kolmogorov-arnold-moser Theory}}.
\newblock World Scientific Publishing Company, 2014.

\bibitem{grassberger1983measuring}
Peter Grassberger and Itamar Procaccia.
\newblock Measuring the strangeness of strange attractors.
\newblock {\em Physica D: nonlinear phenomena}, 9(1-2):189--208, 1983.

\bibitem{datseris2019estimating}
George Datseris, Lukas Hupe, and Ragnar Fleischmann.
\newblock {Estimating Lyapunov exponents in billiards}.
\newblock {\em Chaos: An Interdisciplinary Journal of Nonlinear Science}, 29(9), 2019.

\bibitem{wikiMix}
Mixing.
\newblock https://en.wikipedia.org/wiki/Mixing_(mathematics).
\newblock Accessed: 2024-11-10.

\bibitem{amir2013theory}
Ariel Amir, Jayson Paulose, and David~R Nelson.
\newblock Theory of interacting dislocations on cylinders.
\newblock {\em Physical Review E—Statistical, Nonlinear, and Soft Matter Physics}, 87(4):042314, 2013.

\bibitem{lepore2003diffraction}
Natasha Lepore.
\newblock {\em Diffraction and localization in quantum billiards}.
\newblock Harvard University, 2003.

\bibitem{ge2024direct}
Zhehao Ge, Anton~M Graf, Joonas Keski-Rahkonen, Sergey Slizovskiy, Peter Polizogopoulos, Takashi Taniguchi, Kenji Watanabe, Ryan Van~Haren, David Lederman, Vladimir~I Fal’ko, et~al.
\newblock Direct visualization of relativistic quantum scars in graphene quantum dots.
\newblock {\em Nature}, 635(8040):841--846, 2024.

\bibitem{coxeter1969geometry}
H.~S.~M. Coxeter.
\newblock {\em Introduction to Geometry}.
\newblock John Wiley \& Sons, New York, 2nd edition, 1969.
\newblock PDF version available.

\bibitem{jaynes1973well}
Edwin~T Jaynes.
\newblock The well-posed problem.
\newblock {\em Foundations of physics}, 3(4):477--492, 1973.

\bibitem{di2011bertrand}
P~Di~Porto, B~Crosignani, A~Ciattoni, and HC~Liu.
\newblock {Bertrand's paradox: a physical way out along the lines of Buffon's needle throwing experiment}.
\newblock {\em European journal of physics}, 32(3):819, 2011.

\bibitem{bertrand1889calcul}
Joseph Bertrand.
\newblock {\em Calcul des probabilit{\'e}s}.
\newblock Gauthier-Villars, 1889.

\bibitem{marinoff1994resolution}
Louis Marinoff.
\newblock {A resolution of Bertrand's paradox}.
\newblock {\em Philosophy of Science}, 61(1):1--24, 1994.

\end{thebibliography}
\end{document}